\documentclass[aps,prx,twocolumn,floatfix,superscriptaddress,showpacs,longbibliography]{revtex4-1}
\usepackage{xfrac}
\usepackage{amsfonts}
\usepackage{dsfont}
\usepackage{amsmath}
\usepackage{amssymb}
\usepackage{amsthm}
\usepackage{mathtools}
\usepackage{graphicx}
\usepackage[caption=false]{subfig}
\usepackage{floatrow}
\usepackage{color}
\usepackage{bbm}
\usepackage[normalem]{ulem}
\usepackage[hidelinks]{hyperref}
\hypersetup{
     colorlinks   = true,
     citecolor    = blue,
     linkcolor    = blue
}
\usepackage{mathptmx}
\usepackage{times}
\usepackage{enumerate}

\newcommand{\ignore}[1]{}
\newcommand{\nobibentry}[1]{{\let\nocite\ignore\bibentry{#1}}}

\newcommand{\bibfnamefont}[1]{#1}
\newcommand{\bibnamefont}[1]{#1}
\newcommand{\qav}[1]{\left<#1\right>}

\newcommand{\ket}[1]{\left\vert#1\right\rangle}

\def\ketbra#1#2{{\vert#1\rangle\!\langle#2\vert}}
\newcommand{\Andreu}[1]{{\color{black}#1}}

\DeclareMathAlphabet{\mathcal}{OMS}{cmsy}{m}{n}

\theoremstyle{definition}

\theoremstyle{plain}

\theoremstyle{plain}

\begin{document}

\title{Quantum systems correlated with a finite bath: nonequilibrium dynamics and thermodynamics}

\author{Andreu Riera-Campeny}
\affiliation{F\'{\i}sica Te\`{o}rica: Informaci\'{o} i Fen\`{o}mens Qu\`{a}ntics. Departament de F\'{\i}sica, Universitat Aut\`{o}noma de Barcelona, 08193 Bellaterra, Spain}
\author{Anna Sanpera}
\affiliation{F\'{\i}sica Te\`{o}rica: Informaci\'{o} i Fen\`{o}mens Qu\`{a}ntics. Departament de F\'{\i}sica, Universitat Aut\`{o}noma de Barcelona, 08193 Bellaterra, Spain}
\affiliation{ICREA, Psg. Llu\' is Companys 23, 08001 Barcelona, Spain.}
\author{Philipp Strasberg}
\affiliation{F\'{\i}sica Te\`{o}rica: Informaci\'{o} i Fen\`{o}mens Qu\`{a}ntics. Departament de F\'{\i}sica, Universitat Aut\`{o}noma de Barcelona, 08193 Bellaterra, Spain}

\begin{abstract}

Describing open quantum systems far from equilibrium is challenging, in particular when the environment is \textit{mesoscopic}, when it develops nonequilibrium features during the evolution, or when memory effects cannot be disregarded. Here, we derive a master equation that explicitly accounts for system-bath correlations and includes, at a coarse-grained level, a dynamically evolving bath. It applies to a wide variety of environments, for instance,  those which can be described by Random Matrix Theory or the Eigenstate Thermalization Hypothesis. We obtain a local detailed balance condition which, interestingly, does not forbid the emergence of stable negative temperature states in unison with the definition of temperature through the Boltzmann entropy. We benchmark the master equation against the exact evolution and observe a very good agreement in a situation where the conventional Born-Markov-secular master equation breaks down. Interestingly, the present description of the dynamics is robust and it remains accurate even if some of the assumptions are relaxed. Even though our master equation describes a dynamically evolving bath \textit{not} described by a Gibbs state, we provide a consistent nonequilibrium thermodynamic framework and derive the first and second law as well as the Clausius inequality. Our work paves the way for studying a variety of nanoscale quantum technologies including engines, refrigerators, or heat pumps beyond the conventionally employed assumption of a static thermal bath.
\end{abstract}		

\date{\today}
\maketitle

\section{Introduction}

To understand the potential of future quantum technologies, it is essential to develop an efficient description of microscopic systems far from equilibrium. An important tool to describe the nonequilibrium dynamics of small systems in contact with an external environment are quantum master equations \cite{Breuer2002, Schaller2014, deVega2017}. Master equations have the advantage that they apply to a large class of open systems, are intuitive, and often allow for further analytical progress in the description. Unfortunately, master equations often rely on the assumption that the environment is large, thermal, memoryless, and weakly coupled to the system. Therefore, they quickly break down for many interesting applications \cite{Breuer2002,Schaller2014,deVega2017}. 

Here, we reconsider a class of master equations, first proposed in \cite{Esposito2003a}, which are general, intuitive, and analytically tractable, but overcome to some extent the assumption of a large, thermal, and memoryless environment. We refer to them as the extended microcanonical master equation (EMME). The idea is to additionally keep track of the bath dynamics at a coarse-grained level and include to some degree system-bath correlations. This approach was previously formalized using correlated projection operator techniques \cite{Budini2005, Breuer2006, Budini2006, Breuer2007} and it has been shown to significantly improve standard perturbative master equations \cite{Breuer2006, Esposito2003b, Breuer2007, Fischer2007}. However, it has not yet become a widespread tool. We believe the reason is that the general physical properties of this class of master equations have been not yet investigated and applications of it remained restricted to specifically tailored models. It is our goal to overcome these limitations in the present paper.

Interestingly, we show that the EMME does not only provide an efficient way to describe the non-Markovian dynamics of open quantum systems, but it also connects to a plethora of actively discussed topics in nonequilibrium statistical mechanics. In the following, we summarize our main results, which also serves as an outline for the rest of the paper.

In Sec.~\ref{sec:derivation}, we derive the EMME using three different methods, all leading to the same structure and phenomenology. Among them, one method uses Random Matrix Theory (RMT) and another invokes the Eigenstate Thermalization Hypothesis (ETH). The fact that we obtain the same equation using different methods indicates that our master equation has a clear degree of universility since, in principle, it can be applied to many open quantum systems.

In Sec.~\ref{sec:general_props}, we observe that the EMME preserves the total (system plus bath) coarse-grained energy and we derive local detailed balance. Remarkably, the local detailed balance condition does not forbid the emergence of stable negative temperatures defined according to the Boltzmann entropy.

We devote Sec.~\ref{sec:case_study} to test numerically our analytical results. We benchmark the EMME against the frequently employed Born-Markov-secular (BMS) master equation and against the exact dynamics. To that end, we consider a spin system randomly coupled to a finite environment for which the EMME shows a very good agreement with the exact dynamics.

In Sec.~\ref{sec:emergent_thermo}, we introduce a consistent nonequilibrium thermodynamic framework that includes slowly driven systems. Using this framework, we obtain the first and second law of thermodynamics. Moreover, we connect the first and the second law with the Clausius inequality by introducing an effective nonequilibrium temperature.  

In Sec.~\ref{sec:multiple_baths}, we extend the aforementioned results to the case of multiple environments.

In Sec.~\ref{sec:comparison}, we compare the EMME with other master equation approaches and present our conclusions.

\Andreu{Finally, to keep the presentation focused, generalizations and additional results are shifted to the Appendix.}
 
\section{The extended microcanonical master equation} \label{sec:derivation}

\subsection{General idea and final result}

One of the central goals of open quantum systems theory is to derive a closed evolution equation for the \textit{relevant} degrees of freedom. Such equation can be formally obtained using projection operator techniques \cite{Breuer2002, deVega2017}. Projection operator techniques are based on the definition of a projection superoperator $\mathcal{P}$, and its complementary $\mathcal{Q} = \mathcal{I}-\mathcal{P}$ (where $\mathcal{I}$ is the identity map), that divide the Hilbert space into relevant ($\mathcal{P}$) and irrelevant ($\mathcal{Q}$) degrees of freedom. Because $\mathcal{P}$ and $\mathcal{Q}$ are orthogonal projectors, they satisfy $\mathcal{P}^2 = \mathcal{P}$, $\mathcal{Q}^2 = \mathcal{Q}$, and $\mathcal{QP} = \mathcal{PQ} = 0$, and are otherwise quite arbitrary. Given the state $\rho$ of an isolated system, the use of projection techniques provides a closed equation for the dynamics of the relevant part $\mathcal{P}\rho = \mathcal{P}[\rho]$, achieved by formally integrating out the dynamics of the irrelevant part $\mathcal{Q}[\rho]$ \cite{Breuer2002,deVega2017}. 

We consider an isolated system (the universe) composed of the system $S$ and the environment (or bath) $B$. The isolated system undergoes unitary dynamics generated by the Hamiltonian $\text{H} = \text{H}_S + \text{H}_\text{int} + \text{H}_B$, where $\text{H}_S$ and $\text{H}_B$ contain only system and bath degrees of freedom respectively while $\text{H}_\text{int}$ represents the interaction energy between the system and the bath.

 In the interaction picture with respect to $\text{H}_0 = \text{H}_S + \text{H}_B$, the evolution of the isolated system is generated by the von Neumann equation ($\hbar=1$)
\begin{align}
\partial_t \tilde{\rho}(t) = -i [\tilde{\text{H}}_{\text{int}}(t),\tilde{\rho}(t)] \coloneqq \mathcal{L}(t)[\tilde{\rho}(t)], 
\end{align}
where the tilde denotes operators in the interaction picture, e.g., $\tilde{\rho}(t) = \exp(i \text{H}_0 t) \rho(0) \exp(-i\text{H}_0 t)$. Under the assumptions of (i)  weak-coupling and (ii) an initial state contained in the relevant part $\mathcal{P}[\rho(0)] = \rho(0)$, the dynamics of the relevant degrees of freedom are described by the well-known \Andreu{second-order time-convolutionless (or finite-time Redfield)} master equation \cite{Breuer2002, deVega2017}
\begin{align}
\partial_t \mathcal{P}\tilde{\rho}(t) =& \mathcal{PL}(t)[\mathcal{P}\tilde{\rho}(t)] \nonumber\\
&+ \int_0^t dt' \mathcal{P}\mathcal{L}(t)\mathcal{Q}  \mathcal{L}(t') [\mathcal{P}\tilde{\rho}(t)]\label{eq:redfield_equation},
\end{align}
where we have disregarded terms of $\mathcal{O}(\text{H}_\text{int}^3)$. Dropping assumption (ii) would lead to an extra non-homogeneous term in Eq.~\eqref{eq:redfield_equation} that, typically, is only relevant for the transient dynamics. 

It is worth noting that the derivation of Eq.~\eqref{eq:redfield_equation} makes no use of the explicit form of the projection superoperator $\mathcal{P}$. For later comparison, we introduce the projection superoperator $\mathcal{P}_{\text{Born}}$ that leads to the standard BMS master equation ($k_B =1$)
\begin{align}
\mathcal{P}_{\text{Born}}[\rho] \coloneqq \rho_S \otimes \frac{e^{-\text{H}_B/T_\text{can}} }{Z_B},\label{eq:born_projector}
\end{align}
where $\rho_S \coloneqq \text{tr}_B[\rho]$, $T_\text{can}$ is the canonical temperature of the reference state of the bath, and $Z_B \coloneqq \text{tr}[\exp(-\text{H}_B/T_\text{can})]$ is the partition function. We emphasize that $\mathcal{P}_\text{Born}$ is defined with respect to a fixed Gibbs state of the bath, which is uncorrelated with the system.

\Andreu{In some physical situations, however, the system-bath interaction causes the bath to evolve and develop correlations with the system.} In order to better approximate this situation, we instead consider the following classically correlated projection superoperator
\begin{align}
\mathcal{P}[\rho] \coloneqq  \sum_{E} \rho_S(E) \otimes \frac{\Pi_E}{V_E},\label{eq:our_projector}
\end{align}
where all the terms deserve a comment. First, the \textit{macroscopic} energies $\{E\}$ are a set of coarse-grained bath energies. To be precise, consider the spectral decomposition of the bath Hamiltonian $\text{H}_B = \sum_{E_i}  E_i \ketbra{E_i}{E_i}$, where the set of \textit{microscopic} energies $\{E_i\}$ is ordered according to $E_i \leq E_j$ if $i<j$. In contrast, we define the set $\{E\}$ of macroscopic energies by dividing the spectrum of the bath into non-overlapping energy windows $E_\delta \coloneqq [E-\delta/2,E+\delta/2)$ of width $\delta$. Accordingly, we introduce the projectors $\Pi_E \coloneqq \sum_{E_i\in E_\delta} \ketbra{E_i}{E_i}$ corresponding to the different energies $E$ that can be distinguished by macroscopic measurements. We also introduce the volume $V_E \coloneqq \text{tr}[\Pi_E]$, which represents the number of microstates in the macrostate $E$. Finally, \Andreu{$\rho_S(E)\coloneqq\text{tr}_B[\rho\Pi_E]$} is the \Andreu{unnormalized} conditional state of the system when the bath is found in the macroscopic state $E$, and its trace gives the probability $p(E) \coloneqq \text{tr}[\rho_S(E)]$ of the bath being in that macrostate $E$. Then, the reduced state of the system can be obtained as $\rho_S = \sum_E \rho_S(E)$ which is normalized since $\sum_E p(E) = 1$. For the time being, we focus on the case where $\text{H}_B$ represents a single bath, leaving the extension to multiple environments for Sec.~\ref{sec:multiple_baths}. 

\Andreu{Our goal is describing the dynamics of open quantum systems that interact and build up correlations with a finite bath. Hence, we first define precisely what a finite bath actually is. First, the term \textit{finite} refers to an environment with a finite dimension whose state can \textit{not} be approximated by a time-independent reference state. Second, the term \textit{bath} implies that such system should exhibit bath-like properties, which are ultimately related to a large number of microstates. In particular, the coarse-graining procedure should ensure that in each energy window $E_\delta$ there are enough microscopic energies $E_i$. In fact, as already recognized by Boltzmann, the aforementioned coarse-graining procedure is crucial to reconcile the underlying reversible quantum mechanical description with the irreversible macroscopic world and permits a simplified dynamical description. The same coarse-graining procedure was also used by von Neumann \cite{vonNeumann1929_ger} (see \cite{vonNeumann1929_en} for the English translation) almost a century ago.}

To fix further notation, we introduce the system Hamiltonian $\text{H}_S = \sum_k \varepsilon_k \ketbra{k}{k}$ and we fix the interaction $\text{H}_\text{int} = \lambda\text{S}\otimes \text{B}_{\text{int}}$ where $\lambda$ is an energy scale. The general expressions for multiple coupling operators (i.e., $\text{H}_\text{int} = \sum_\alpha \text{S}^\alpha\otimes\text{B}^\alpha$) can be found in the Appendix. 

Under the conditions spelled out above, our central object of study is a master equation describing the time evolution of $\rho_S(E)$. If we use the conventional Markov and secular approximation\cite{Breuer2002, Schaller2014, deVega2017} it reads 
\begin{align}
\partial_t  \rho_S(E) =& -i [\text{H}'_{S}(E),\rho_S(E)]\nonumber\\
&+\sum_{\omega} \left(\frac{\gamma(E,E-\omega)}{V_{E-\omega}} \text{S}_{\omega} \rho_S(E-\omega)\text{S}^\dagger_{\omega} \right.\nonumber\\
& -\left.\frac{\gamma(E+\omega,E)}{2 V_E}  \left\{ \rho_S(E),\text{S}^{\dagger}_{\omega} \text{S}_{\omega}\right\} \right),\label{eq:emme}
\end{align}
where $\omega$ sums over all possible system transition frequencies. Furthermore, we have introduced the dissipation rates
\begin{align}
\gamma(E,E')&\delta_{E',E+\omega} \coloneqq\int_\mathbb{R} d\tau \text{tr}_B[\tilde{\text{B}}^{\dagger}(-\tau) \Pi_E \text{B}\Pi_{E'}]e^{i \omega\tau},\label{eq:factorization_condition}
\end{align}
the operators $\text{S}_\omega \coloneqq \sum_{kq} \qav{k|\text{S}|q}\ketbra{k}{q}\delta_{\varepsilon_q-\varepsilon_k,\omega}$, and the modified Hamiltonian $\text{H}'_{S}(E)$ that commutes with the bare system Hamiltonian $\text{H}_S$. All of them are defined precisely below.

It is important to emphasize two general features of our EMME. First, one can show that Eq.~\eqref{eq:emme} fits into the general form investigated by Breuer~\cite{Breuer2007}, who shows that it preserves the trace of $\rho_S$ and complete positivity of $\rho_S(E)$ at all times. Second, although we derived Eq.~\eqref{eq:emme} using the Markov and secular approximation, which implies that $\rho_S(E)$ evolves in a Markovian manner, the reduced system state $\rho_S$ does \emph{not}. Therefore, the EMME is able to capture non-Markovian system dynamics.

We now provide a step-by-step derivation of Eq.~\eqref{eq:emme} in Sec.~\ref{subsec:detailed_derivation}. Particular care is required when evaluating the bath correlation function. In Sec.~\ref{subsec:correlation_function}, we use three different methods to arrive at the same conclusion. Further mathematical details are shifted to App.~\ref{app:details}. Readers not interested in the details of the derivation can skip the rest of this section and continue reading in Sec.~\ref{sec:general_props}, where we start to focus in detail on the physics predicted by the EMME. 

\subsection{Detailed derivation}\label{subsec:detailed_derivation}

In this subsection, we give a detailed derivation of the EMME, which corresponds to finding explicit expressions for the first and second order terms in Eq.~\eqref{eq:redfield_equation}. To this aim, it will prove useful to decompose $\text{H}_\text{int}$ into block diagonal and off-diagonal parts
\begin{align}
\text{H}_\text{int} = \sum_E \delta \text{H}(E)\otimes \Pi_E + \text{V},\label{eq:decomposition}
\end{align}
where we have implicitly defined 
\begin{align}
&\delta \text{H}(E) \coloneqq \lambda  \qav{\text{B}_{\text{int}}}_E \text{S}, \nonumber\\
&\text{V} \coloneqq \lambda \text{S}\otimes\text{B} = \lambda \text{S}\otimes(\text{B}_{\text{int}} -\sum_{E} \qav{\text{B}_{\text{int}} }_E \Pi_E).\label{eq:explicit_decomposition}
\end{align}
Here, $\qav{\cdots}_E \coloneqq \text{tr}_B[\cdots \Pi_E/V_E]$ denotes the microcanonical average at energy $E$. Note that the operator $\text{B}$ has the important property $\qav{\text{B}}_E=0$, which we use below. 

Using Eq.~\eqref{eq:decomposition}, the first order term in Eq.~\eqref{eq:redfield_equation} reads
\begin{align}
\mathcal{PL}(t)[\mathcal{P}\tilde{\rho}(t)] &= -i \sum_E [ \tilde{\delta\text{H}}(E;t)\Pi_E, \mathcal{P}\tilde{\rho}(t)]. \label{eq:plp_explicit}
\end{align}
In standard projection operator techniques that employ $\mathcal{P}_{\text{Born}}$, the first order term in Eq.~\eqref{eq:redfield_equation} can be set to zero without loss of generality by including its contribution in the system Hamiltonian $\text{H}_S$ \cite{Breuer2002,deVega2017,Schaller2014}. This is no longer possible for the projection in Eq.~\eqref{eq:our_projector}. The second order term is obtained using similar manipulations as
\begin{align}
\mathcal{P}\mathcal{L}(t)\mathcal{Q} &\mathcal{L}(t') [\mathcal{P}\tilde{\rho}(t)] \nonumber\\
&= - \sum_E \text{tr}_B \left\{ \Pi_E [\tilde{\text{V}}(t),[\tilde{\text{V}}(t'),\mathcal{P}\tilde{\rho}(t)]]\right\} \otimes \frac{\Pi_E}{V_E}.\label{eq:plqlp_explicit}
\end{align}

The evolution equation for each component $\tilde{\rho}_S(E;t)$ is then found by inserting the expressions~\eqref{eq:plp_explicit} and ~\eqref{eq:plqlp_explicit} in Eq.~\eqref{eq:redfield_equation} and making use of our correlated projector $\mathcal{P}$ in Eq.~\eqref{eq:our_projector}. It yields\begin{align}
&\partial_t \tilde{\rho}_S(E;t) = -i [\tilde{\delta\text{H}}(E;t),\tilde{\rho}_S(E;t)] \nonumber\\
 &+\sum_{E'} \int_0^t dt' \text{tr}_B \left\{ \Pi_E [\tilde{\text{V}}(t),[\tilde{\rho}_S(E';t)\otimes \frac{\Pi_{E'}}{V_{E'}},\tilde{\text{V}}(t')]]\right\}.\label{eq:emme_second_order}
\end{align}
Next, we introduce the microcanonical bath correlation function
\begin{align}
C_B(E,E';t'-t) \coloneqq \lambda^2 \text{tr}\qav{\text{B}^{\dagger}(t')\Pi_E \text{B}(t)}_{E'},
\end{align}
which can be explicitly computed as 
\begin{align}
C_B(E,E';-\tau) = \sum_{E_i\in E_\delta}\sum_{E_j\in E'_\delta} \frac{\lambda^2}{V_{E'}} |\langle E_i|\text{B}|E_ j\rangle|^2e^{i(E_i-E_j)\tau}.
\label{eq:correlation_function}
\end{align}
\Andreu{We also introduce the decomposition $\tilde{\text{S}}(t) = \sum_\omega\text{S}_\omega \exp(-i\omega t)$, where $\text{S}_{\omega} \coloneqq\sum_{kq} \qav{k|\text{S}|q}\ketbra{k}{q}\delta_{\varepsilon_q-\varepsilon_k , \omega}$. Using both expressions into Eq.~\eqref{eq:emme_second_order} one arrives to
\begin{align}
\partial_t \tilde{\rho}_S(E;t) =& -i [\tilde{\delta\text{H}}(E;t),\tilde{\rho}_S(E;t)] \nonumber\\
 &\hspace{-1.5cm}+\sum_{E'}\sum_{\omega\omega'} \int_0^t dt' e^{i(\omega' t'-\omega t)} \left(C_B(E,E';t'-t) \text{S}_\omega\tilde{\rho}_S(E',t)\text{S}_{\omega'}^\dagger \right.\nonumber\\
& \left.- C_B(E',E;t'-t) \tilde{\rho}_S(E;t)\text{S}_{\omega'}^\dagger\text{S}_{\omega}\right)+\text{h.c.}. \label{eq:emme_second_order_2}
\end{align}
Equation} \eqref{eq:emme_second_order_2} is the finite-time Redfield version of the EMME (in the interaction picture), which is ready for numerical implementation and gives improved results for transient times (see Sec.~\ref{sec:case_study}). It is, however, still hard to work with Eq.~\eqref{eq:emme_second_order_2} analytically. Therefore, we employ the standard Markov and secular approximations \cite{Breuer2002, Schaller2014, deVega2017}, which, nonetheless, give different results from the standard BMS master equation due to the different choice of the projection superoperator $\mathcal{P}$ in Eq.~\eqref{eq:our_projector}. 

\Andreu{The Markov approximation relies on the fact that the microcanonical bath correlation function decays rapidly to zero. To understand the range of validity of the this approximation, we use the following time-scale argument. Let us denote by $\tau_B$ the correlation time of the bath defined such that $C_B(E,E';-\tau) \approx 0$ for all $\tau \geq \tau_B$. Of course, $\tau_B$ is a function of the energy width $\delta$, i.e. $\tau_B = \tau_B(\delta)$, and depends on the particular coarse-graining procedure. If the bath energies are fine-grained (i.e., $\delta\to 0$), the correlation function oscillates at frequency $E_i-E_j$ and never decays. In such a case, $\tau_B\to\infty$ and the Markovian approximation breaks down. Instead, for a sufficiently large $\delta$ many-frequencies $E_i-E_j$ contribute to Eq.~\eqref{eq:correlation_function} and the correlation function rapidly decays. Then, one can safely extend the upper limit of the time integrals in Eq.~\eqref{eq:emme_second_order_2} to infinity and the Markov approximation holds.
In this sense, a finite coarse-graining $\delta$ is necessary to reconcile the reversible microscopic description with the irreversible macroscopic world.

However, for finite baths the correlation function never decays exactly to zero and it can exhibit recurrences for sufficiently long times. Then, the validity of the Markov assumption relies on the fact that the typical Poincar\'{e} recurrence time is exceedingly large (see for instance \cite{Reimann2008, Venuti2015}), and one is often interested in time-scales of evolution much smaller than the recurrence time.}

Our second approximation, which is called secular approximation, consists in averaging out the rapidly oscillating terms in the interaction picture. Formally, this is done by introducing the time average
\begin{align}
\overline{\text{O}} \coloneqq \lim_{T\to\infty} \frac{1}{T} \int_0^T dt'\tilde{\text{O}}(t'),
\end{align}
which, used in Eq.~\eqref{eq:emme_second_order_2} (after taking the limit $t\to\infty$ in the upper limit of the integral) selects the components $\omega = \omega'$.

Introducing the one sided Fourier transform
\begin{align}
\Gamma(E,E';\omega) \coloneqq V_{E'} \int_0^\infty d\tau C_B(E,E';-\tau) e^{i \omega \tau},
\end{align}
the resulting equation after both approximations can be compactly written as
\begin{align}
\partial_t \tilde{\rho}_S(E;t) =& -i [\overline{\delta\text{H}}(E),\tilde{\rho}_S(E;t)]+\nonumber\\
&+\sum_{E'}\sum_{\omega}\left(\frac{\Gamma(E,E';-\omega)}{V_{E'}}  \text{S}_{\omega} \tilde{\rho}_S(E';t)\text{S}^{\dagger}_{\omega}\right. \nonumber\\
& -\left. \frac{\Gamma(E',E;-\omega)}{V_E} \tilde{\rho}_S(E;t) \text{S}^{\dagger}_{\omega} \text{S}_{\omega} \right)+ \text{h.c.},\label{eq:lindblad_emme_1}
\end{align}
where the change of variables $\tau=t-t'$ has been performed. Also, $\overline{\delta\text{H}}(E)$ is an energy dependent Hamiltonian shift that commutes with the system Hamiltonian $\text{H}_S$ and has the explicit expression $\overline{\delta\text{H}}(E) = \sum_{k} \langle \text{B}_{\text{int}}\rangle_E\langle k| \text{S}|	k\rangle \ketbra{k}{k} $.

To compare the EMME with the conventional BMS master equation, it is convenient to decompose the function $\Gamma(E,E';\omega)$ into its real and imaginary parts as
\begin{align}
&\text{A}(E,E';\omega) \coloneqq \frac{1}{2i} (\Gamma(E,E';\omega) -\Gamma(E,E';\omega)^*),\nonumber\\
&\gamma(E,E';\omega) \coloneqq \Gamma(E,E';\omega) +\Gamma(E,E';\omega)^*. \label{eq:dissipation_rates}
\end{align}
from where it follows that 
\begin{align}
\gamma(E,E';\omega) =  V_{E'}\int_\mathbb{R} d\tau C_B(E,E';-\tau) e^{i\omega\tau}.\label{eq:closed_dissipation}
\end{align}

Below, from Subsubsec.~\ref{subsec:first} to Subsubsec.~\ref{subsec:eth}, we show that the function $\gamma(E,E';\omega)$ is generically peaked around $E' = E + \omega$ and we are allowed to factorize
\begin{align}
\gamma(E,E';\omega) = \gamma(E,E')\delta_{E',E+\omega}.
\end{align}

We also introduce the energy-dependent Lamb-shift Hamiltonian
\begin{align}
\text{H}_{\text{LS}}(E) \coloneqq -\sum_{E'} \sum_{\omega}\frac{\text{A}(E',E,-\omega)}{V_E}\text{S}^{\dagger}_\omega \text{S}_\omega,
\end{align}
which, also commutes with the system Hamiltonian $\text{H}_S$. Then, the modified Hamiltonian
\begin{align}
\text{H}'_{S}(E) \coloneqq \text{H}_S+\overline{\delta\text{H}}(E)+ \text{H}_{\text{LS}}(E),
\end{align}
commutes with $\text{H}_S$ and corresponds to an $E$ dependent shift of the system energies. 

Finally, after moving to the Schr\"odinger picture, we obtain our first main result: the EMME within the Markov and secular approximation shown in Eq.~\eqref{eq:emme}. 

The exact computation of $C_B(E,E';-\tau)$, or the associated $\gamma(E,E';\omega)$, depends on the fine structure of the bath energy levels as well as the exact form of the coupling operators $\text{B}$. In relevant physical situations, neither the fine structure of the bath energy levels nor the exact form of the coupling operator are typically available. For this reason, it is important to find approximate methods to compute the correlation function that only depend on the coarse structure of the bath energy levels. In the following subsubsections, we present three methods to obtain the functions $\gamma(E,E';\omega)$: the first one neglects part of the internal bath dynamics, the second one uses RMT, and the third invokes the ETH. Moreover, we further connect those three methods in App.~\ref{app:connection}. 

\subsection{Evaluation of the bath correlation function}\label{subsec:correlation_function}

\subsubsection{Heuristic approach}\label{subsec:first}

\Andreu{The idea behind the heuristic approach is to assume that there exists a coarse-graining $\delta$ such that $\delta \tau_B(\delta) \ll 1$. In that case, for the relevant times $\tau\leq\tau_B$ one can expand 
\begin{align}
e^{i(E_i-E_j)\tau} \approx e^{i(E-E')\tau}+\mathcal{O}(\delta\tau_B).
\end{align}
Essentially, this corresponds to replacing the energy differences $E_i-E_j\mapsto E-E'$ when $E_i \in E_\delta$ and $E_j \in E'_\delta$ in Eq.~\eqref{eq:correlation_function}, as it was previously considered in Ref.~\cite{Strasberg2019}. Then, the correlation function yields
\begin{align}
C_B&(E,E';-\tau) \approx \lambda^2 \text{tr}[\text{B}^{
\dagger} \Pi_E \text{B} \Pi_{E'}]  \frac{e^{i(E-E')\tau}}{V_{E'}}. \label{eq:first_correlation}
\end{align}
Note that the case $\delta\to 0$ was studied in Refs.~\cite{Esposito2003a, Esposito2007}. With the help of Eq.~\eqref{eq:closed_dissipation} and Eq.~\eqref{eq:first_correlation} (see App.~\ref{app:details_first}), one obtains the dissipation rates
\begin{align}
\gamma_\text{heuristic}(E,E') = \frac{2\pi \lambda^2}{\delta} \text{tr}[\text{B}^{\dagger} \Pi_E \text{B} \Pi_{E'}] .
\label{eq:first_rates}
\end{align}
Hence, the procedure above provides an additional interpretation to the parameter $\delta$. Of course, given a bath, it is not clear whether such a coarse-graining $\delta$ exists and for this reason we refer to this approach as heuristic. On the other hand, the present evaluation of the correlation function does not rely on any explicit assumption on the bath coupling operator $\text{B}$. Alternative methods to evaluate $C_B(E,E';-\tau)$ which are based on assumptions about the bath coupling operator $\text{B}$ are the content of the next two subsections. }

\subsubsection{Random matrix coupling with dense environment}\label{subsec:stochastic}

\Andreu{In many physical situations, the interaction Hamiltonian $\text{H}_\text{int}$ might be too complicated to be obtained with \textit{ab initio} methods. Hence, in the same spirit of the heuristic approach, our aim is to evaluate the correlation function without fully specifying the bath coupling operator $\text{B}$. One possibility is offered by RMT (for a review on the topic see for instance \cite{Beenakker1997}), which has been widely used in many physical contexts due to its universality. In particular, it has been already used to describe the decay of quantum systems in contact with complex environments, see e.g., \cite{Breuer2006,Gelbart1972, Mello1988, Pereyra1991, Cohen2000, Esposito2003b, Lebowitz2004, Cohen2004}. In general, a random matrix ansatz seems to work well for strongly non-integralbe systems (see for instance Ref.~\cite{Nation2019}).

Our approach is based on extracting $\text{B}$ from a random matrix ensemble and compute the corresponding correlation function. In principle, two different members of the ensemble can give rise to a very different dynamics. The essence of RMT relies on the fact that, interestingly, this is often not the case and the fine-structure of the coupling operators is only important in exceptional cases. In particular, it has been shown that not only the RMT approach gives the correct mean value when compared with the predictions of statistical mechanics, but also the variance between the two is very small (see, e.g. Ref.~\cite{Reimann2015}). Thus, almost all members of the random matrix ensemble give the same prediction. Hence, despite performing the random matrix ensemble average theoretically, no such average is implied experimentally.} 

Here, we take the RMT approach and consider that the coupling to the environment is done via the random matrix
\begin{align}
\text{B}= \sum_{E\neq E'}\sum_{E_i\in E_\delta}\sum_{E_j \in E'_\delta} \left[b(E,E') + c(E_i,E_j)\right]\ketbra{E_i}{E_j},\label{eq:stochastic_coupling}
\end{align}
where $b(E,E')$ are deterministic functions of the macroscopic energies, and $c(E_i,E_j)$ are random numbers. We consider $c(E_i,E_j)$ to be i.i.d. complex random variables with zero mean, i.e., $\mathbb{E}[c(E_i,E_j)]=0$, and variance $a^2$, i.e., $\mathbb{E}[c(E_i,E_j) c(E_{i'},E_{j'})] = a^2 \delta_{E_i,E_i'}\delta_{E_j,E_j'}$. Using the random coupling of Eq.~\eqref{eq:stochastic_coupling}, and averaging over the correlation function in Eq.~\eqref{eq:correlation_function} (see App.~\ref{app:details_stoch} for details), one obtains the complex dissipation rates 
\begin{align}
\gamma_\text{rmt}(E,E&') = \frac{2\pi\lambda^2}{\delta}V_E V_{E'} (|b(E,E')|^2 + a^2).\label{eq:stoch_rates}
\end{align}

\subsubsection{The Eigenstate Thermalization Hypothesis}\label{subsec:eth}

The ETH is an \textit{ansatz} for the matrix elements of a local observable in the energy eigenbasis of a quantum many body system (see \cite{Deutsch1991,Srednicki1994,dAlessio2016,Deutsch2018} ). It has been successfully used to study equilibration and thermalization in a variety of isolated quantum systems. Yet, its exact range of validity is still under debate, but there is a common consensus that it applies to many body systems whose classical counterpart is chaotic (although not exclussively, see \cite{dAlessio2016} and references therein). Here, we use the ETH to make progress in computing the bath correlation function $C_B(E,E';-\tau)$, thereby linking the field of equilibration and thermalization in isolated many body systems to the field of open quantum systems. The ETH can be formulated as follows: the matrix elements of a local observable $\text{O}$ in the energy eigenbasis of a non-integrable quantum many body system obey the following ansatz:
\begin{align}
\text{O}_{ij} = O(E_{ij}) \delta_{ij} + \sqrt{\frac{1}{V_{E_{ij}}}} f(E_{ij},\Omega_{ij}) R_{ij},
\end{align}
where the mean energy $E_{ij} = (E_i + E_j)/2$ and the energy difference $\Omega_{ij} = E_i-E_j$ have been introduced. All elements of the above equation deserve a comment. The functions $O(E_{ij})$ and $f(E_{ij},\Omega_{ij})$ are smooth functions of their arguments. Moreover, the function $f$ must decay as $|\Omega_{ij}|$ grows and has the symmetry property $f(E_{ij},-\Omega_{ij})= f^*(E_{ij},\Omega_{ij})$. Finally, the numbers $R_{ij} = R_{ji}^*$ have zero mean and unit variance, and vary erratically with $i$ and $j$. These erratically varying $R_{ij}$ allow us to effectively use arguments from random matrix theory without the need to actually perform any ensemble average. 

The main insight arises from the fact that the open system $S$ couples locally (through its boundary) to the bath $B$ via the operators $\text{B}$. Since the ETH holds for local observables of a quantum many-body system, we can make progress on the computation of the bath correlation function $C_B(E,E';-\tau)$ using the ETH ansatz for the operators $\text{B}$. Then, introducing $\bar{E} = (E+E')/2$, one arrives to the complex dissipation rates (see App.~\ref{app:details_eth})
\begin{align}
\gamma_{\text{eth}}(E,E') = \frac{2\pi\lambda^2}{\delta} V_{E}V_{E'} \frac{|f(\bar{E},E-E')|^2}{V_{\bar{E}}}.
\end{align}
\section{Energy conservation, equilibrium states, and mutual information}\label{sec:general_props}

After proving that the EMME has certain degree of universality, we devote this section to investigate its properties. We start by noting that the populations $p(\varepsilon_k,E) = \qav{k|\rho_S(E)|k}$ evolve autonomously under the rate equation 
\begin{align}
\partial_t p(\varepsilon_k,E) = &\sum_{q} \left( \frac{W_{kq}(E,E+\omega_{kq})}{V_{E+\omega_{kq}}} p(\varepsilon_q,E+\omega_{kq})\right. \nonumber\\
&\qquad \left. - \frac{W_{qk}(E+\omega_{kq},E)}{V_E} p(\varepsilon_k,E) \right),\label{eq:emme_populations}
\end{align}
where we have defined $\omega_{kq} = \varepsilon_k - \varepsilon_q$ and the transition rates 
\begin{align}
W_{kq}(E,E') \coloneqq \gamma(E,E')|\qav{k|\text{S}|q}|^2.\label{eq:transition_rates}
\end{align} 
In App.~\ref{app:general_props}, we prove the positivity of the transition rates $W_{kq}(E,E') \geq 0$ as well as the symmetry $W_{kq}(E,E') = W_{qk}(E',E)$.

\subsection{Strict energy conservation}\label{subsec:strict_energy_conservation}

A crucial property of Eq.~\eqref{eq:emme} is that the coarse-grained total energy of the system and bath composite is preserved under the evolution. In fact, the statement is stronger since not only the average value of the total energy is preserved, but also the associated probability distribution. To be precise, we introduce the coarse-grained total energy $E_{\text{tot}} \coloneqq \varepsilon_k + E$ associated to the system having energy $\varepsilon_k$ and the bath being at the energy window $E$. We denote the corresponding probability of being in the energy shell $E_\text{tot}$ as $P(E_{\text{tot}}) \coloneqq \sum_{k} p(\varepsilon_k,E_{\text{tot}}-\varepsilon_k)$. From Eq.~\eqref{eq:emme_populations}, it follows that 
\begin{align}
\partial_t P (E_{\text{tot}}) = 0,
\end{align}
and therefore, the probability of being in the energy shell $E_{\text{tot}}$ is a conserved quantity of the evolution. In particular, its average $U \coloneqq \sum_{k,E} P(E_\text{tot}) E_\text{tot}$ fulfills $dU/dt =0$. Hence, the two variables $\varepsilon_k$ and $E$ are not independent and, given $E_\text{tot}$, one can obtain $\varepsilon_k$ from $E$ or \textit{vice versa}.

\subsection{Equilibrium states and local detailed balance}\label{subsec:steady-state}

From Eq.~\eqref{eq:emme_populations}, the rate to jump from the state $(\varepsilon_q,E')$ to the state $(\varepsilon_k,E)$ is $W_{kq}(E,E')/V_{E'}$. Using the symmetry property of the transition rates $W_{kq}(E,E') = W_{qk}(E',E)$, we find that the ratio of the rates to jump from the state $(\varepsilon_k,E)$ to the state $(\varepsilon_q,E+\omega_{kq})$ reduces to 
\begin{equation}
\frac{V_E}{V_{E+\omega_{kq}}} = e^{-[\mathcal{S}_\text{mic}(E+\omega_{kq})-\mathcal{S}_\text{mic}(E)]},\label{eq:local_detailed_balance}
\end{equation}
where we have introduced the microcanonical (or Boltzmann) entropy $\mathcal{S}_{\text{mic}}(E) = \log V_E$. Equation~\eqref{eq:local_detailed_balance} constitutes the local detailed balance condition for the EMME. Note that the local detailed balance condition in Eq.~\eqref{eq:local_detailed_balance} matches the one obtained within the context of classical Markov dynamics in phase space \cite{Maes2003}. 

At equilibrium all probability flows are balanced and the local detailed balance condition implies 
\begin{equation}
\frac{p_\text{eq}(\varepsilon_k,E)}{p_\text{eq}(\varepsilon_q, E+\omega_{kq})} = e^{-[\mathcal{S}_\text{mic}(E+\omega_{kq})-\mathcal{S}_\text{mic}(E)]},\label{eq:entropy_volumes}
\end{equation}
for the equilibrium probabilities. Using the total energy $E_{\text{tot}}$ introduced above, we can write the steady-state condition in a more symmetric manner as
\begin{align}
\frac{p_\text{eq}(\varepsilon_k, E_{\text{tot}}-\varepsilon_k)}{p_\text{eq}(\varepsilon_q,E_{\text{tot}}-\varepsilon_q)} = \frac{V_{E_{\text{tot}}-\varepsilon_k}}{V_{E_{\text{tot}}-\varepsilon_q}}\label{eq:steady_state}.
\end{align}
Since $P(E_{\text{tot}})$ is constant, the final energy distribution for each of the probabilities $p(\varepsilon_k,E_\text{tot})$ is fixed to have the equilibrium value
\begin{align}
p_\text{eq}(\varepsilon_k,E_{\text{tot}}-\varepsilon_k) = P(E_{\text{tot}})\frac{V_{E_{\text{tot}}-\varepsilon_k}}{\sum_{q} V_{E_{\text{tot}}-\varepsilon_q}},\label{eq:general_ss}
\end{align}
which was first noted in \cite{Esposito2007}. In summary, the steady-state condition in Eq.~\eqref{eq:steady_state} implies that the system explores equiprobably all the available phase space given the macroscopic constraint that the total energy equals $E_\text{tot}$.

One may wonder whether the steady state of the EMME deviates from the steady state of the conventionally used BMS master equation. From Eq.~\eqref{eq:entropy_volumes}, introducing the definition of the microcanonical (or Boltzmann) temperature $dE =T_{\text{mic}}(E) d\mathcal{S}_{\text{mic}}(E)$, we obtain for small $\omega_{kq}$ 
\begin{align}
\frac{p_{\text{eq}}(\varepsilon_k,E)}{p_{\text{eq}}(\varepsilon_q,E+\omega_{kq})} = e^{-\omega_{kq}/T_{\text{mic}}(E) }.\label{eq:steady_thermal}
\end{align}
This still involves the joint probability distributions of system and bath. If the bath energies are restricted to a microscopically large but macroscopically small energy range, we can assume that $T_\text{mic}$ depends only very slowly on $E$ such that it is a constant to first order. In that scenario, the population ratio in Eq.~\eqref{eq:steady_thermal} is independent from the bath energy $E$. Assuming an unbounded bath spectrum, one can multiply Eq.~\eqref{eq:steady_thermal} by $p(\varepsilon_q,E+\omega_{kq})$, and then sum over $E$, to obtain 
\begin{align}
\frac{p_{\text{eq}}(\varepsilon_k)}{p_{\text{eq}}(\varepsilon_q)} = e^{-\omega_{kq}/T_\text{mic}} \left(1 - \sum_{E=0}^{\omega_{kq}} \frac{p(\varepsilon_{q},E)}{p(\varepsilon_q)}\right),\label{eq:equivalence_of_ensembles}
\end{align}
where we have taken, without loss of generality, $\omega_{kq}>0$. The second term within brackets in Eq.~\eqref{eq:equivalence_of_ensembles} is a correction that appears due to the strict energy conservation condition $\partial_t P(E_\text{tot})=0$. Namely, for $E<\omega_{kq}$ the process $\varepsilon_q \mapsto \varepsilon_k$ cannot occur because there is no bath transition that can supply the energy deficit $\omega_{kq}$. In general, the contribution of the correction term will be small as long as the bath has initially a sufficiently high energy.

Instead, the conventional BMS master equation derived using the projector $\mathcal{P}_\text{Born}$ in Eq.~\eqref{eq:born_projector} predicts the steady state
\begin{align}
\frac{p_{\text{eq}}(\varepsilon_k)}{p_{\text{eq}}(\varepsilon_q)} = e^{-\omega_{kq}/T_{\text{can}}}, \label{eq:gibbs_eq}
\end{align}
where $T_{\text{can}}$ is fixed by the choice of the projector in Eq.~\eqref{eq:born_projector}. At first glance, the steady-state probabilities in Eq.~\eqref{eq:equivalence_of_ensembles} and Eq.~\eqref{eq:gibbs_eq} are similar. Indeed, if the backaction of the system on the bath is negligible and if the equivalence of ensembles holds, then $T_\text{can} = T_\text{mic}$, but we remark that $T_\text{mic}\neq T_\text{can}$ in general. This is best illustrated by the extreme case of negative temperature ($T_\text{mic} < 0$) steady states, for which the energy populations increase with energy. Those negative temperature states arise when, at least locally, the volume terms of the bath decrease with energy $V_E < V_{E'}$ for $E>E'$. \Andreu{Then, the equivalence of ensembles clearly breaks down, as there is typically no Gibbs state which approximates the true state of the bath.}

We remark that, following the observation of negative temperature states \cite{Braun2013}, the question whether negative temperatures are thermodynamically consistent has attracted much attention recently, with arguments presented against \cite{RomeroRochin2013,Dunkel2014,Campisi2015} or in favour \cite{Eitan2017,Swendsen2018,Schneider2014} of it. The debate remained, however, mostly on an abstract and axiomatic level. We here contribute to this fundamental question by numerically observing the emergence of stable population inverted steady states of an open system (see below). By \textit{stable} we mean that all initial states of the open system tend to this population inverted state in the long-time limit (unless additional symmetries are present preventing the existence of a unique steady state). This result is also supported by the exact numerical integration of the full Schr\"odinger equation (see second row of Fig.~\ref{fig:twobands}). These states match a Gibbs distribution with negative temperature \emph{if} one uses the Boltzmann entropy to define temperature. In Sec.~\ref{sec:emergent_thermo} we will also formulate nonequilibrium first and second law for the EMME. Thus, within our framework we can deal with negative temperature states without observing the emergence of any thermodynamic inconsistencies.

\subsection{System-bath correlations}

In the literature, the use of the uncorrelated projection superoperator $\mathcal{P}_\text{Born}$ is often justified by invoking the weak coupling assumption. However, as we show in Sec.~\ref{sec:emergent_thermo}, this is not true and even at the weak coupling limit strong system-bath correlations can build up (see also Ref.~\cite{Mitchison2018}). The EMME allows us to access part of these correlations and also to quantify them. We proceed as follows.

In order to quantify the (possibly quantum) system-bath correlations, we introduce the always positive quantum mutual information
\begin{align}
\mathcal{I}^{S:B}[\rho] \coloneqq \text{tr}[\rho(\log\rho - \log(\rho_S\otimes\rho_B))]\geq 0.
\end{align}
Then, a high value of the mutual information indicates that the uncorrelated projector $\mathcal{P}_\text{Born}$ fails to capture the correlated nature of the system-bath dynamics and, therefore, \Andreu{the correctness of the BMS description is not guaranteed.}

 In absence of quantum correlations, the quantum mutual information is bounded from above by $\mathcal{I}^{S:B} \leq \log d_S$, where $d_S$ is the dimension of the system Hilbert space (naturally, we assume the dimension of the bath $d_B$ larger than the system dimension $d_S$). Thus, in our context a value $\mathcal{I}^{S:B} \lesssim \log d_S$ corresponds to strong system-bath correlations. 

The EMME, however, only keeps track of part of the full system-bath dynamics and, ultimately, has information about the classical probability distribution $p(\varepsilon_k,E)$, with $p(\varepsilon_k,E) = \text{tr}[\rho \ketbra{k}{k} \otimes \Pi_E/V_E]$. The system-bath correlations included in $p(\varepsilon_k,E)$ are quantified using the coarse-grained mutual information
\begin{align}
\mathcal{I}^{S:B}_\text{cg}(\mathbf{p}) \coloneqq \sum_{k,E} p(\varepsilon_k, E) \log \left(\frac{p(\varepsilon_k,E)}{p(\varepsilon_k)p(E)}\right)\geq 0,\label{eq:mutual_information_obs}
\end{align}
where $\mathbf{p}$ is the vector with components $p(\varepsilon_k,E)$. It can be shown (see App.~\ref{app:mutual_information}), that 
\begin{align}
\mathcal{I}^{S:B}[\rho] \geq \mathcal{I}^{S:B}_\text{cg}(\mathbf{p}) \geq 0,\label{eq:mutual_information_bound}
\end{align}
where the first inequality becomes an equality for the state $\rho = \sum_{k,E} p(\varepsilon_k,E)\ketbra{k}{k}\otimes\Pi_E/V_E$. 

Physically speaking, the reason why we can observe strong system-bath correlations with the EMME arises from the fact that the total energy $E_\text{tot}$ is conserved under the evolution, which constraints the values that the bath energy $E$ can take given a system energy $\varepsilon_k$. Hence, this constrained dynamics can give rise to high system-bath correlations as we numerically observe in Sec.~\ref{sec:emergent_thermo} (see Fig.~\ref{fig:mutual_information}).

\section{Example: spin coupled to a structured environment}\label{sec:case_study}
\floatsetup[figure]{style=plain,subcapbesideposition=top}
\begin{figure}
{\includegraphics[width=.8\textwidth]{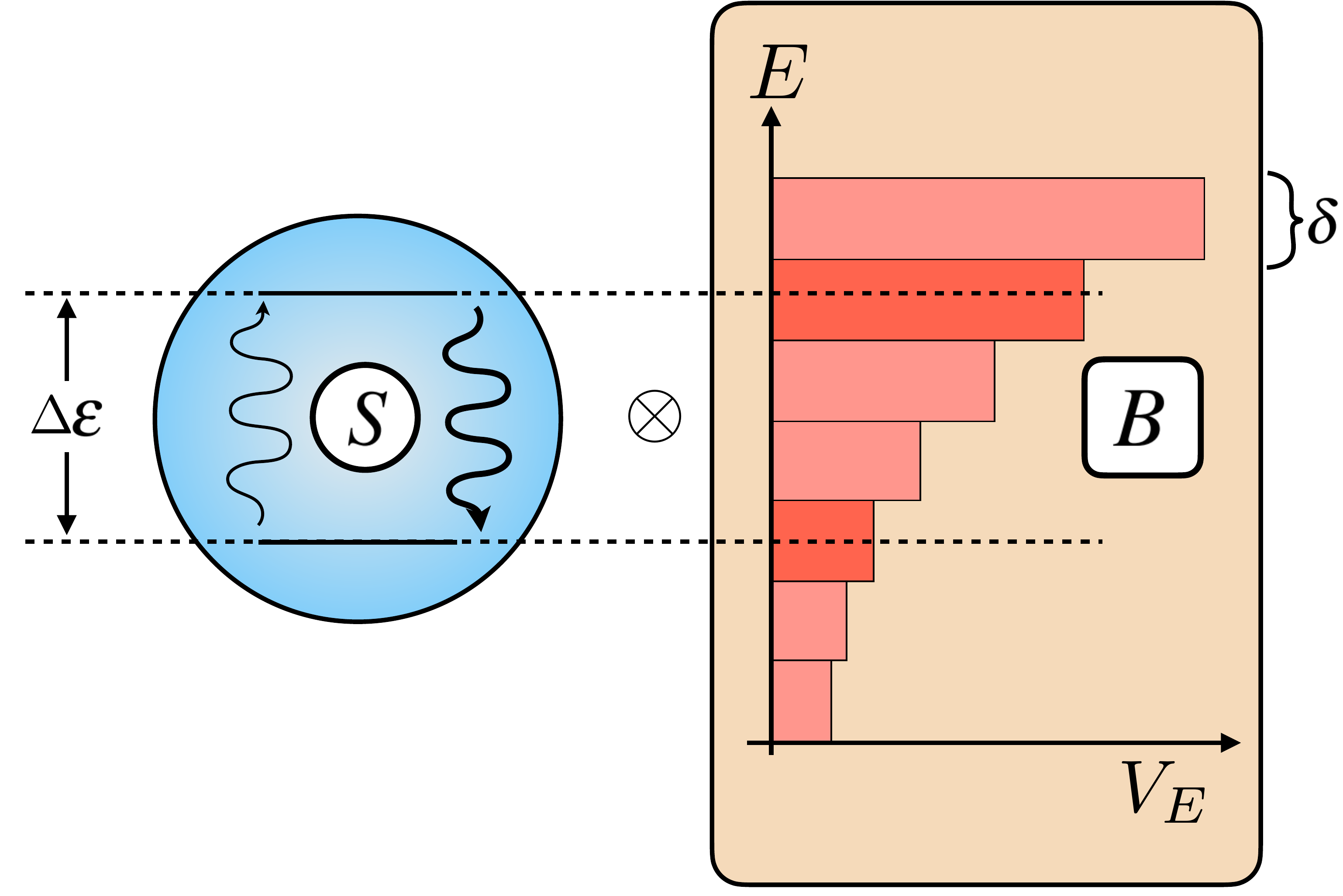}}
  \caption{Scheme of the model of a spin coupled to a structured environment.\label{fig:scheme_model}} 
\end{figure}

\floatsetup[figure]{style=plain,subcapbesideposition=top}
\begin{figure*}
{\includegraphics[width=\textwidth]{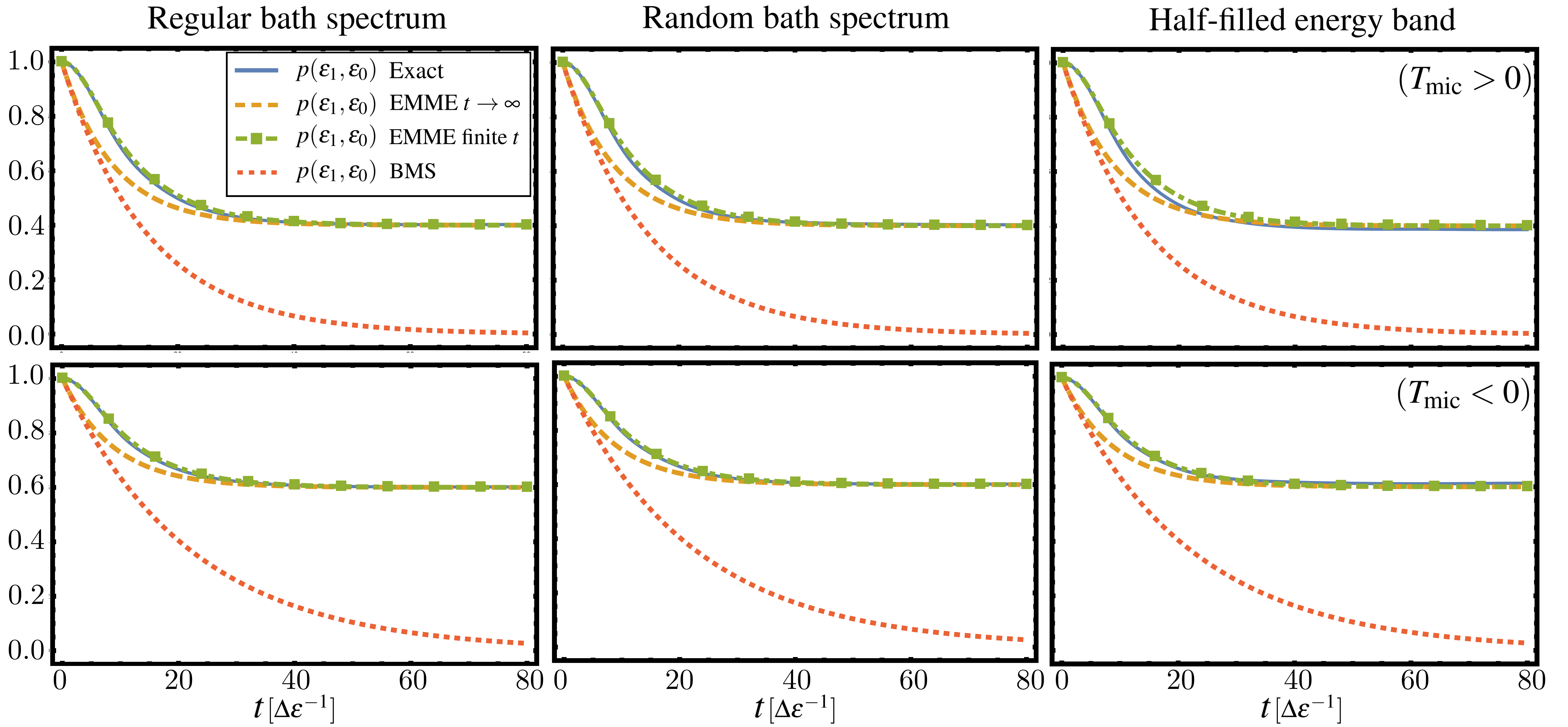}}\\
\caption{Comparison of the evolution of the joint probability $p(\varepsilon_1,E=\varepsilon_0)$ for a spin system coupled to a two bands environment with initial state $\rho(0) = \ketbra{1}{1}\otimes \Pi_{\varepsilon_0}/V_{\varepsilon_0}$, and $b(E,E')=0$ for all $E$ and $E'$. The results are obtained with the following three different methods: exact evolution solving the Schr\"odinger equation (blue solid line), solution using the EMME with the Markov approximation $t\to\infty$ (orange dashed line), solution using the finite-time Redfield version of the EMME (green dot-dashed line with square markers), and finally the solution using the standard BMS master equation (red dotted line). First column: regular bath with equidistant energy levels. Second column: randomly distributed energy levels. Third column: randomly distributed energy levels and the initial state of the bath is taken to be only half filled, i.e. a mixed stated of the $V_{\varepsilon_0}/2$ levels with lower energy.  First row: $V_{\varepsilon_0} =  400$ and $V_{\varepsilon_1} = 600$. Second row: $V_{\varepsilon_0} =  600$ and $V_{\varepsilon_1} = 400$. Parameters: $\lambda = 3\times 10^{-3}$, $\delta = 0.5$, $\Delta\varepsilon = 1$, $a=1$.\label{fig:twobands}}
\end{figure*}

To numerically check the validity of various results derived in Sec.~\ref{sec:general_props} and Sec.~\ref{sec:emergent_thermo}, we consider here an extension of the model studied in \cite{Breuer2006}. First, we derive the EMME for this model and then we compare its prediction with the exact integration of the Schr\"odinger equation. The model consists in a single spin with two energy levels $\ket{0}$ and $\ket{1}$ coupled to a finite environment (see Fig.~\ref{fig:scheme_model}). For convenience, we introduce the raising and lowering operators $\sigma_+ = \ketbra{1}{0} = \sigma_{-}^\dagger$. The bare system and bath Hamiltonians are
\begin{align}
&\text{H}_S = \varepsilon_0 \ketbra{0}{0}+ \varepsilon_1 \ketbra{1}{1},\nonumber\\
&\text{H}_B = \sum_{E_i} E_i \ketbra{E_i}{E_i}.\label{eq:qubit_ham}
\end{align}
We consider the interaction $\text{H}_\text{int} = \text{V} = \lambda (\sigma_++\sigma_-) \otimes \text{B}$, and model the bath coupling operator using the RMT approach of Subsec.~\ref{subsec:stochastic}, in such a way that $\text{B}$ is given by Eq.~\eqref{eq:stochastic_coupling}. This choice leads to $\delta \text{H}(E) = 0$ for all energies $E$. Recall that $c(E_i,E_j)$ are i.i.d. complex random numbers with zero mean and variance $a^2$. For simplicity, we assume that $\varepsilon_1-\varepsilon_0 = \Delta \varepsilon = n\delta$ with $n\in \mathbb{N}$, which ensures the existence of an energy $E'$ such that $E' = E +\Delta \varepsilon$.

Remarkably, it is relatively simple to obtain the finite-time Redfield version of the EMME analytically for this system. As explained in App.~\ref{app:time-integrals}, this is achieved by conveniently introducing the function
\begin{align}
\zeta(t) \coloneqq \frac{\delta}{\pi} \int_0^t \frac{\sin^2(\delta\tau/2)}{(\delta\tau/2)^2} d\tau,\label{eq:finite_t}
\end{align}
which fulfills $\zeta(t\to\infty) = 1$. Then, the finite-time Redfield of the EMME takes the same for as Eq.~\eqref{eq:emme} but the dissipation rates are multiplied by the time-envelope $\zeta(t)$. For our particular example, this procedure leads to
\begin{align}
&\partial_t \rho_S(E) = -i [\text{H}'_{S}(E),\rho_S(E)]\nonumber\\
&+\zeta(t)\gamma(E,E+\Delta \varepsilon)\left(\frac{\sigma_+ \rho_S(E+\Delta\varepsilon) \sigma_-}{V_{E+\Delta \varepsilon}} -\frac{\{ \rho_S(E),\ketbra{1}{1}\}}{2V_E}\right) \nonumber\\
&+\zeta(t)\gamma(E,E-\Delta \varepsilon)\left(\frac{\sigma_- \rho_S(E-\Delta\varepsilon)\sigma_+}{V_{E-\Delta \varepsilon}} -\frac{\{ \rho_S(E),\ketbra{0}{0}\}}{2V_E} \right). \label{eq:qubit_explicit}
\end{align}
where we have introduced $\gamma(E,E')  = 2\pi \lambda^2 (|b(E,E')|^2 + a^2)V_E V_{E'}/\delta$. The function $\gamma(E,E')$ has the property $\gamma(E,E') = \gamma(E',E)$, and it vanishes if either $E$ or $E'$ do not exist. Physically speaking, the second line of Eq.~\eqref{eq:qubit_explicit} represents a process in which a quantum of bath energy excites the system, while the third line represents the opposite process in which the system gets de-excited. Equation~\eqref{eq:qubit_explicit}, corresponds to the finite-time Redfield equation for this particular model, and we provide its analytical solution in App.~\ref{app:spin}.

In order to benchmark the EMME, we proceed to investigate numerically a particular case of the above described model. Namely, we consider an environment of only two energy windows of macroscopic energies $E=\{ \varepsilon_0,\varepsilon_0+\Delta\varepsilon\}$, with width $\delta = \Delta \varepsilon/2$. The bath coupling operator $\text{B}$ takes the form in Eq.~\eqref{eq:stochastic_coupling} where we have set \Andreu{for simplicity $b(E,E') = 0$ and $a^2=1$}. Also, we fix the initial state to $\rho(0) = \ketbra{1}{1}\otimes \Pi_{\varepsilon_0}/V_{\varepsilon_0}$. Note that, for this particular choice, even if we would consider more bands in the environment, their populations would not change with time. Finally, in the spirit of RMT, the numerics is done with a single realization of the bath coupling operator B.

In order to challenge the assumptions done in the derivation, we consider the following three scenarios: (i) \textit{Regular bath spectrum}: the energy levels of the bath are equidistantly distributed (as in Ref.~\cite{Breuer2006}); (ii) \textit{Random bath spectrum}: the energy levels of the bath are randomly distributed within each energy window; and (iii) \textit{Half-filled energy window}: not only the bath energy levels are randomly distributed, but also the initial state has only half of the energy window occupied. Hence, we have $\mathcal{P}[\rho(0)]\neq\rho(0)$, but we still use the EMME from Eq.~\eqref{eq:emme} neglecting any inhomogeneous contribution.

To be precise, the regular bath spectrum has energies $E_{i_k} = \varepsilon_k -\delta/2 + i_{k} \delta/V_{\varepsilon_k}$ where $i_{k} \in \{0,1,\cdots,V_{\varepsilon_k}-1\}$ for $k=0,1$; the random bath spectrum has energies sampled from a flat distribution in the energy window $[\varepsilon_k-\delta/2,\varepsilon_k +\delta/2)$ for $k =0,1$; and the initial state of the half-filled energy window is taken to be $\rho(0) = \ketbra{1}{1}\otimes \sum_{i=1}^{V_{\varepsilon_0}/2} \ketbra{E_i}{E_i}$. Note that the situations described in (i)--(iii) increasingly challenge the assumptions done in the derivation of the EMME. We numerically compare them in Fig.~\ref{fig:twobands} (one for each column).

In the first row of Fig.~\ref{fig:twobands}, we set $V_{\varepsilon_0} = 400 < V_{\varepsilon_1} = 600$ and observe that the prediction of Eq.~\eqref{eq:qubit_explicit} (green dot-dashed line with square markers) agrees very well with the exact result (solid blue line) even for Random bath spectrum or the Half-filled window. Instead, for this model, the standard BMS master equation (red dotted line) fails to capture the dynamics. If one sets $\zeta(t) \mapsto 1$ (orange dashed line), as opposed to keeping the time dependent dissipation rates $\zeta(t)$ (green dot-dashed line with square markers), the equation fails to describe the dynamics only at short time-scales, but correctly predicts the steady-state. This short-time behavior is ultimately a consequence of the failure of the Fermi golden rule and has a universal character \cite{Braun2001}. 

In the second row of Fig.~\ref{fig:twobands}, we exchange the volumes of the bands such that $V_{\varepsilon_0} = 600 > V_{\varepsilon_1} = 400$. In all three scenarios, the EMME describes accurately the dynamics and predicts correctly the steady state. In this case, the equilibrium state shows population inversion and, in agreement with the discussion in Subsec.~\ref{subsec:steady-state}, it can be described by a negative temperature state $T_\text{mic} < 0$. 

In App.~\ref{app:EMME_failure}, we challenge the EMME even further by considering weaker and stronger coupling strengths $\lambda$, as well as smaller volumes $V_E$ for the energy windows of the bath. Even though the EMME is not able to always reproduce the dynamics accurately, for large volumes $V_E\gtrsim 100$, it does typically give the right time-scale of decay and a good approximation for the steady-state populations.

\section{Non equilibrium thermodynamics }\label{sec:emergent_thermo}

In the first part of the present article we have derived the EMME starting from a microscopic description of the system and the bath. One motivation to derive such a master equation is its potential to describe small quantum devices including heat engines, refrigerators or heat pumps. If the bath is finite, operating those small quantum devices can cause the bath to develop nonequilibrium features during the evolution and then, the standard approach relying on a large bath in a Gibbs state cannot be applied. In this respect, it is important to obtain a consistent (nonequilibrium) thermodynamic interpretation of the dynamics. 

Our master equation can describe three sources of non-equilibrium: (i) A non-thermal initial state $\rho(0)$ of the system and bath composite as already considered in the numerical simulations above. (ii) A time-dependent system energy spectrum $\epsilon_k(\lambda_t)$, where $\lambda_t$ represents a sufficiently slow driving protocol ($\dot{\lambda}_{t} \ll1$) such that one can directly replace $\varepsilon_k \mapsto \varepsilon_k (\lambda_t)$ in Eq.~\eqref{eq:emme}. And (iii) the system $S$ being in contact with multiple environments (which we discuss in Sec.~\ref{sec:multiple_baths}). We devote this second part of the article to connect the nonequilibrium quantum dynamics of the EMME with the laws of thermodynamics.

\subsection{The first law of thermodynamics}

We start with the definition of the internal energy of the universe
\begin{align}
U(t)& \coloneqq \sum_{k,E} (\varepsilon_k(\lambda_t) + E) p(\varepsilon_k,E).
\end{align}
By the first law of thermodynamics, its change can only be due to the mechanical work done on the system. Then, using $\partial_t P (E_{\text{tot}}) = 0$, we obtain
\begin{align}
\frac{d}{dt} U = \dot{W} = \sum_{k,E}( \partial_t \varepsilon_k(\lambda_t))p(\varepsilon_k, E).\label{eq:first_law_a}
\end{align}

We also introduce the internal energy of the system
\begin{align}
U_S(t) &\coloneqq \sum_{k,E} \varepsilon_k(\lambda_t) p(\varepsilon_k,E).
\end{align}
Since the system $S$ is in contact with a bath, the change in its internal energy is now due to work \textit{and} heat
\begin{align}
\frac{d}{dt}U_S = \dot{W} + \dot{Q}. \label{eq:first_law_b}
\end{align}
Then, the heat flux is found to be
\begin{align}
\dot{Q} \coloneqq -\sum_{k,E} E \partial_t p(\varepsilon_k, E) = \sum_{k,E} \varepsilon_k(\lambda_t) \partial_t p(\varepsilon_k,E),
\end{align}
where the second equality follows again from $\partial_t P (E_{\text{tot}}) = 0$.

\subsection{The second law of thermodynamics}

The second law of thermodynamics states that a change in the thermodynamic entropy of the universe is always non-negative. However, there is no general consensus on the microscopic definition of the thermodynamic entropy. Here, we use the recently (re)discovered observational entropy \cite{Safranek2019_1, Safranek2019_2, Schindler2020, Strasberg2020}
\begin{align}
\mathcal{S}_{\text{obs}}(\mathbf{p}) \coloneqq \sum_{k,E} p(\varepsilon_k,E) \left(- \log p(\varepsilon_k,E)+\log V_E \right), \label{eq:observational_entropy}
\end{align}
which also appears in the work of von Neumann and Wigner~ \cite{vonNeumann1929_en}. Here, we use $\mathbf{p}$ to denote the vector of probabilities $p(\varepsilon_k,E)$ and we also write $\mathcal{S}_{\text{obs}}(t)$ instead of $\mathcal{S}_{\text{obs}}(\mathbf{p}(t))$ when its meaning is clear from the context. We note that the observational entropy $\mathcal{S}_\text{obs}$ coincides with the well-known von Neumann entropy
\begin{align}
\mathcal{S}_{\text{vN}}[\rho] \coloneqq -\text{tr}[\rho\log\rho], \label{eq:vonNeumann_entropy}
\end{align}
when $\rho$ is diagonal and fulfills $\qav{k,E_i|\rho|k,E_i} = p(\varepsilon_k,E)/V_E$ $\forall E_i\in E_\delta$. Finally, we introduce the coarse-grained relative entropy 
\begin{align}
\mathcal{D}_{\text{cg}}(\mathbf{p}||\mathbf{q}) \coloneqq \sum_{k,E} p(\varepsilon_k,E) \log \frac{p(\varepsilon_k,E)}{q(\varepsilon_k,E)}\geq 0,\label{eq:relative_entropy_cl}
\end{align}
for two arbitrary vectors of probabilities $\mathbf{p}$ and $\mathbf{q}$. 

Equipped with those information theoretical quantities, our aim is to derive a second law in terms of the observational entropy $\mathcal{S}_{\text{obs}}$ and, hence, we define the entropy production rate
\begin{align}
\dot{\Sigma} \coloneqq \frac{d}{dt}\mathcal{S}_\text{obs}(t).
\end{align}
In contrast to Ref. \cite{Strasberg2020}, where only the integrated change $\Delta \mathcal{S}_{\text{obs}}(t) = \mathcal{S}_\text{obs}(t) - \mathcal{S}_\text{obs}(0)$ in observational entropy was shown to be positive, we derive here the stronger result that the entropy production \emph{rate} is always positive $\dot{\Sigma}\geq 0$. To prove the positivity of $\dot{\Sigma}$, it is convenient to introduce the joint Gibbs distribution at temperature $T$
\begin{align}
p_T(\varepsilon_k(\lambda_t),E) \coloneqq \frac{V_E\exp(-(\varepsilon_k(\lambda_t) + E)/T)}{Z_S(\lambda_t)\, Z_{B}},
\end{align}
where $Z_S(\lambda_t) \coloneqq \sum_k \exp(-\varepsilon_k(\lambda_t)/T)$ and $Z_B \coloneqq \sum_E V_E \exp(- E/T)$ are the partition functions. Note that $\mathbf{p}_T(\lambda_t)$ is a stationary distribution since it fulfills the steady-state condition in Eq.~\eqref{eq:steady_state}. Below, we show that it is possible to recast the entropy production rate as
\begin{align}
\dot{\Sigma} =- \left. \partial_t\right|_{\lambda_t} \mathcal{D}_{\text{cg}}(\mathbf{p}(t)||\mathbf{p}_T(\lambda_{t})), \label{eq:entropy_prod}
\end{align}
\Andreu{where the symbol $\partial_x|_{y}$ stands for the partial derivative with respect to $x$ while keeping $y$ fixed.} The result above then implies that the entropy production rate is non-negative:
\begin{align}
\frac{d}{dt} \mathcal{S}_{\text{obs}}(t) = \dot{\Sigma} \geq 0. \label{eq:second_law_rate}
\end{align}
This follows from two facts: First, $\textbf{p}_T$ is an equilibrium state of the dynamics and, second, the dynamics are Markovian, which implies that we can use monotonicity of the relative entropy. If any of these two assumptions is violated, negative entropy production rates can appear although $\Sigma(t) = \Delta \mathcal{S}_\text{obs}(t)$ remains positive \cite{Strasberg2019_b}. 

The proof is as follows: we start writing the classical relative entropy in Eq.~\eqref{eq:entropy_prod} in terms of the observational entropy as
\begin{align}
\mathcal{D}_{\text{cg}}(\mathbf{p}(t)||\mathbf{p}_T(\lambda_{t})) = &- \mathcal{S}_{\text{obs}}(t) +T^{-1}U(t) \nonumber\\
&+ \log Z_S(\lambda_t) + \log Z_B.
\end{align}
Then, using the chain rule $d/dt = \partial_t + \dot{\lambda}_{t}\partial_{\lambda_t}$ we obtain
\begin{align}
\left. \partial_t\right|_{\lambda_t} \mathcal{D}_{\text{cg}}(\mathbf{p}(t)||\mathbf{p}_T(\lambda_{t})) = \frac{d}{dt} \mathcal{S}_{\text{obs}}(t) - \frac{1}{T}\left(\frac{dU }{dt} - \dot{W}\right).\label{eq:second_law_derivation}
\end{align}
Finally, we note that the rightmost term in Eq.~\eqref{eq:second_law_derivation} vanishes upon the use of the first law in Eq.~\eqref{eq:first_law_a}. Hence, we obtain the second law of thermodynamics in Eq.~\eqref{eq:second_law_rate} which implies a positive entropy production rate. Note that the second law $\dot\Sigma(t) \geq 0$ derived here holds as long as $\mathcal{P}[\rho(0)] = \rho(0)$ and, in particular, we did not invoke at any time the assumption of thermal equilibrium for the bath.

\subsection{Connecting the first and second law}

In the above subsections, we have derived independently the first and second law of thermodynamics and, at this point, they appear rather disconnected. In standard phenomenological thermodynamics, the first and second law are related through the well-known Clausius inequality. Namely, if during a transformation the bath is well approximated at all times $t$ by an equilibrium state at temperature $T_B(t)$, then the Clausius inequality reads
\begin{align}
\Delta \mathcal{S}^{S}(t) - \int_0^t dt' \frac{\dot{Q}(t')}{T_B(t')} \geq 0,\label{eq:clausius_phenomenological}
\end{align}
where $\mathcal{S}^S$ denotes the system thermodynamic entropy. Equation~\eqref{eq:clausius_phenomenological} connects the entropic changes with the heat flux into the system $\dot{Q}$ when the environment is found at temperature $T_B$. In particular, a large environment with an infinite heat capacity would keep its temperature constant through the process (i.e., $T_B(t)\approx T_B(0)$). Then, the Clausius inequality simplifies to
\begin{align}
\Delta \mathcal{S}^{S}(t) - \frac{Q(t)}{T_B(0)} \geq 0,\label{eq:clausius_constant_T}
\end{align}
where $Q(t) = \int_0^t dt' \dot{Q}$. Equation~\eqref{eq:clausius_constant_T} is conventionally considered the second law in quantum thermodynamics. Importantly, Eq.~\eqref{eq:clausius_constant_T} should be regarded as a consequence of the second law only under the conditions spelled out above. 

In our description, during a transformation $\rho(0)\mapsto\rho(t)$, the environment generically goes through several nonequilibrium states for which a temperature $T_B$ may not even be defined. To establish a connection with the Clausius inequality, we define an effective nonequilibrium temperature $T_B^*(t)$ by demanding that the actual bath energy matches the one of a fictitious canonical ensemble at that temperature. In equations, $T_B^*(t)$ is determined by solving 
\begin{align}
\sum_{E} E  \frac{V_E e^{-E/T_B^*(t)}}{Z_B(t)} = \sum_{k,E} E p(\varepsilon_k,E;t) = U_B(t).
\end{align}
Operationally, $T_B^*(t)$ corresponds to the temperature of a \textit{super-bath} that, if weakly coupled to the bath $B$, would give rise to a total vanishing heat current between the bath and the super-bath.

\floatsetup[figure]{style=plain,subcapbesideposition=top}
\begin{figure}
{\includegraphics[width=\textwidth]{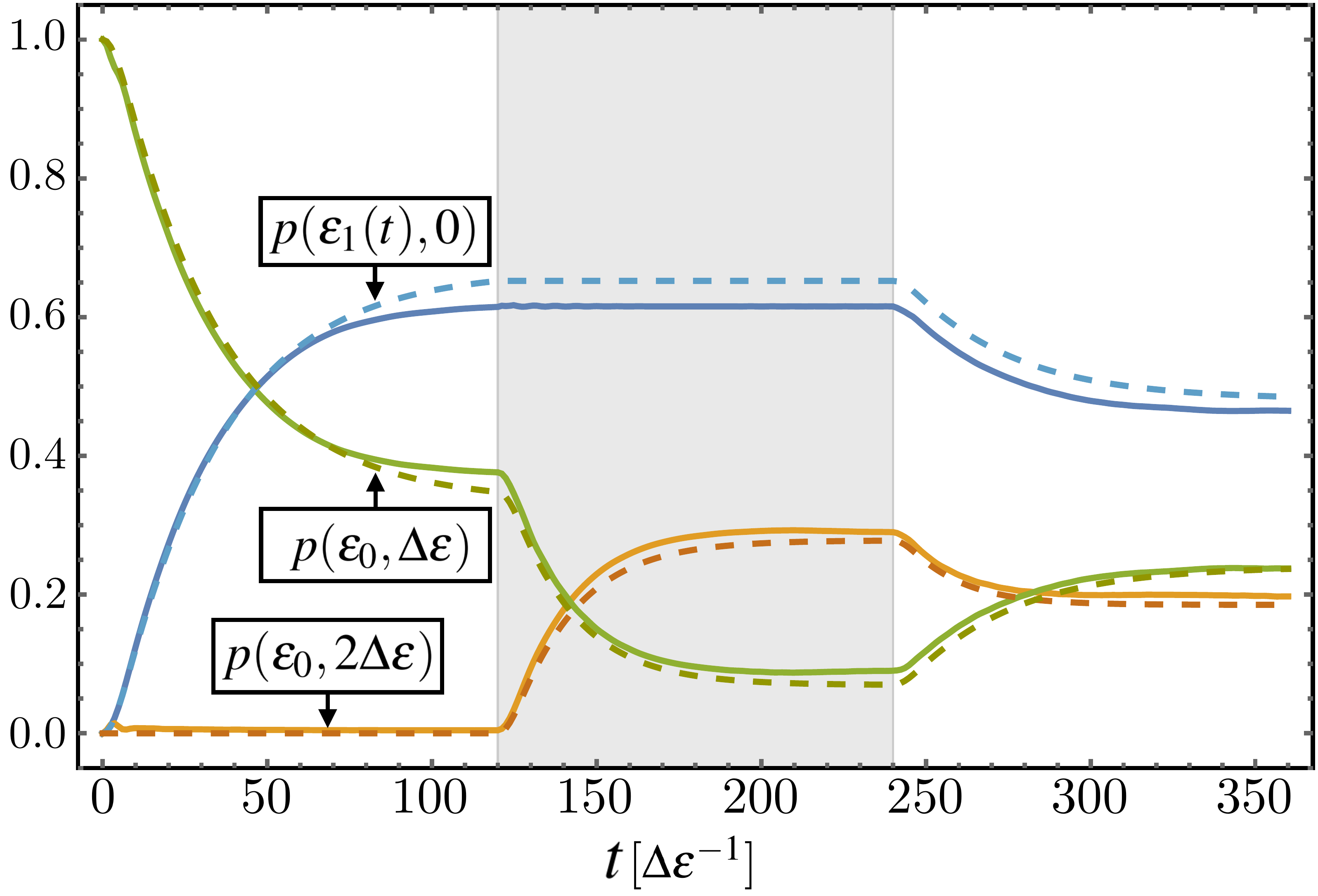}}
  \caption{Comparison of the exact dynamics (solid lines) with the dynamics predicted by the EMME (dashed lines) for a spin with energies $\varepsilon_0 = 0$ and $\varepsilon_1(\lambda_t) = \Delta\varepsilon$ (white background) and $\varepsilon_1(\lambda_t) = 2\Delta\varepsilon$ (shadowed background). See main text for details. Parameters: $\delta = 0.5$, $t_\star = 120\Delta \varepsilon^{-1}$, $b(E,E') = 0$ $\forall E,E'$, and $a=1$.\label{fig:quenched_dynamics}} 
\end{figure}

As emphasized above, the Clausius inequality follows from the second law Eq.~\eqref{eq:second_law_rate} only under additional approximations. To derive it rigorously, we assume the following:  (i) $\rho(0)$ is a product state of system and bath, i.e. $\rho(0) = \rho_S(0) \otimes \rho_B(0)$, and (ii) the initial bath state $\rho_B(0)$ is a Gibbs state of the bath at temperature $T(0)$. 

We start by noting that, for a Gibbs state at an arbitrary temperature $T$, the following differential relation holds:
\begin{align}
dU_B = T d\mathcal{S}^B_{\text{obs}}(\mathbf{p}_{T}). \label{eq:differential_relation}
\end{align}
Here, we use the superscript $B$ to indicate that $\mathcal{S}_{\text{obs}}^B$ corresponds to the observational entropy of the bath alone. Next, we note the following identity
\begin{align}
\Delta \mathcal{S}_{\text{obs}}^B (t)& = \mathcal{S}_{\text{obs}}^B (t)-\mathcal{S}_{\text{obs}}^B (\mathbf{p}_{T_B^*(t)})\nonumber\\
& +\mathcal{S}_{\text{obs}}^B (\mathbf{p}_{T_B^*(t)})-\mathcal{S}_{\text{obs}}^B (0). \label{eq:entropy_identity}
\end{align}
The difference of the first two terms of the above equation is negative since, by construction, $\mathbf{p}(t)$ and $\mathbf{p}_{T_B^*(t)}$ have the same energy and the Gibbs state maximizes the entropy. Using the differential relation in Eq.~\eqref{eq:differential_relation}, the last two terms of Eq.~\eqref{eq:entropy_identity} can be cast as
\begin{align}
\mathcal{S}_{\text{obs}}^B (\mathbf{p}_{T_B^*(t)})-\mathcal{S}_{\text{obs}}^B (\mathbf{p}_{T_B(0)})  = -\int_0^t dt' \frac{\dot{Q}(t')}{T_B^*(t')}.\label{eq:integral_change}
\end{align}
Moreover, using the coarse-grained mutual information Eq.~\eqref{eq:mutual_information_obs} as well as the initial product state assumption, we obtain the following relation
\begin{align}
\Delta \mathcal{S}_{\text{obs}}^S(t) + \Delta \mathcal{S}_{\text{obs}}^B(t) = \Delta \mathcal{S}_{\text{obs}} (t) + \mathcal{I}^{S:B}_\text{cg}(\mathbf{p}(t)) \geq 0.\label{eq:local_positivity}
\end{align}
Finally, putting together Eq.~\eqref{eq:entropy_identity}, Eq.~\eqref{eq:integral_change} and Eq.~\eqref{eq:local_positivity} the following chain of inequalities is found
\begin{align}
\Delta \mathcal{S}_{\text{obs}}^S(t) - \int_0^t dt' \frac{\dot{Q}(t')}{T_B^*(t')} &\geq \Delta \mathcal{S}_{\text{obs}}^S(t) + \Delta \mathcal{S}_{\text{obs}}^B(t) \nonumber\\
&\geq \Delta \mathcal{S}_{\text{obs}}(t) \geq 0.\label{eq:clausius_inequality}
\end{align} 
which proves the aforementioned Clausius inequality and connects the first and second law. We remark that the first inequality in Eq.~\eqref{eq:clausius_inequality} becomes an exact equality whenever the bath does not develop any noticeable nonequilibrium features. 

\subsection{Testing the results numerically}

\floatsetup[figure]{style=plain,subcapbesideposition=top}
\begin{figure}
{\includegraphics[width=\textwidth]{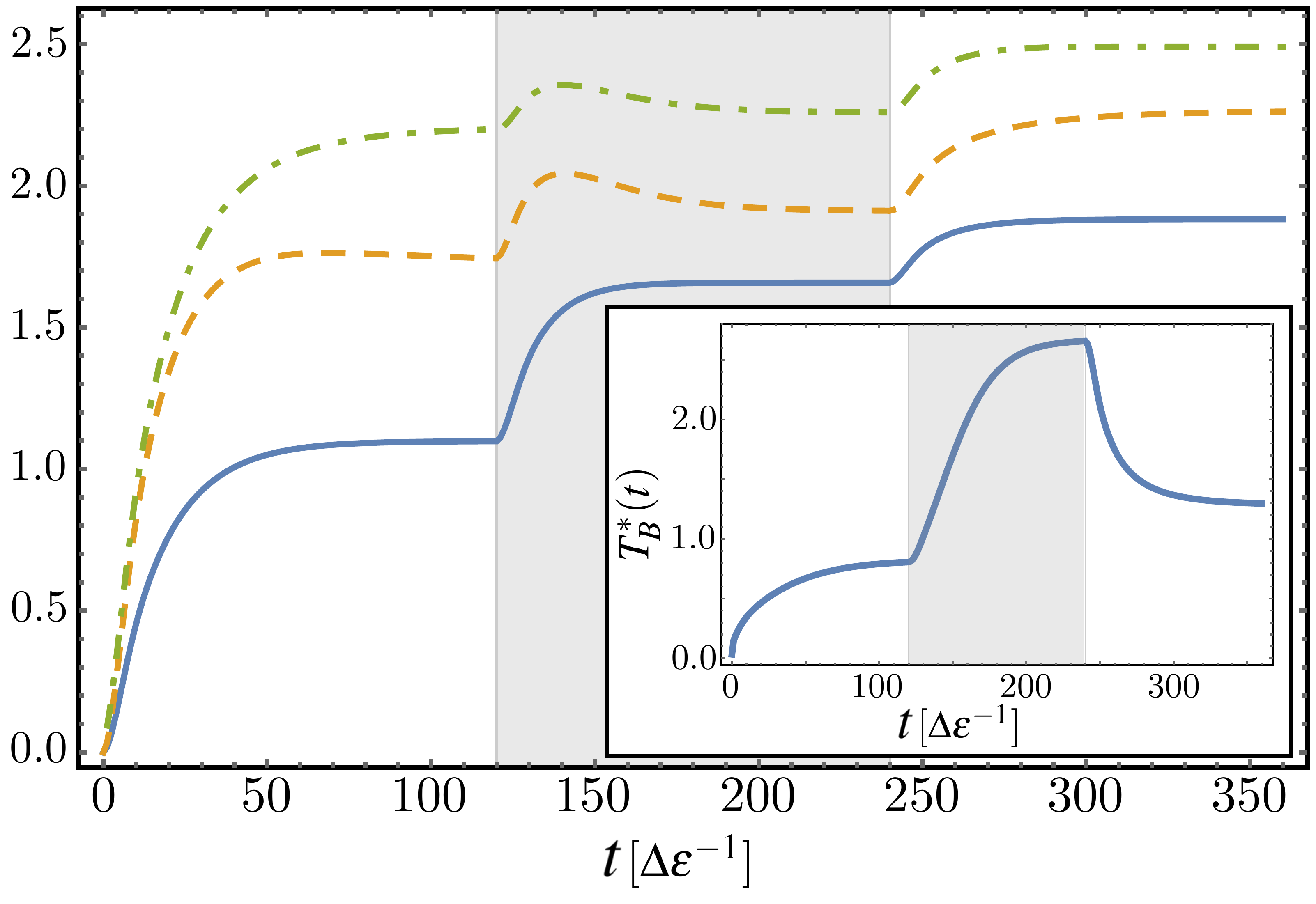}}
  \caption{Numerical proof of the Clausius inequality: $\Delta\mathcal{S}_\text{obs}(t)$ (blue solid line), $\Delta\mathcal{S}^S_\text{obs}(t)+\Delta\mathcal{S}^B_\text{obs}(t)$ (orange dashed line) and $\Delta\mathcal{S}^S_\text{obs}(t)-\int_0^t dt' \dot{Q}(t')/T_B^*(t')$ (green dot-dashed line) computed using the EMME of a for a spin with energies $\varepsilon_0 = 0$ and $\varepsilon_1(t) = \Delta\varepsilon$ (white background) and $\varepsilon_1(t) = 2\Delta\varepsilon$ (shadowed background). See main text for details. Inset: Corresponding non-equilbrium temperature of the bath $T_B^*(t)$ ($k_B=1$) for the same protocol. Parameters: $\delta = 0.5$, $t_\star = 120\Delta \varepsilon^{-1}$, $b(E,E') = 0$ $\forall E,E'$, and $a=1$.\label{fig:clausius_inequality}}
\end{figure}

To conclude this section, we test numerically the results derived above. We consider again the same spin system described in Sec.~\ref{sec:case_study} but now we allow the energies $\varepsilon_k(\lambda_t)$ to depend parametrically on time. For concreteness, we leave the energy $\varepsilon_0 = 0$ constant and quench periodically the energy of the excited state as
\begin{align}
\varepsilon_1(\lambda_t) = \left\{\begin{matrix}
\Delta \varepsilon & t\in [0,t_\star)\quad \\
2\Delta\varepsilon & t\in [t_\star,2 t_\star),
\end{matrix} \right.,
\end{align}
where $2t_\star$ is the period, and $\Delta\varepsilon$ a fixed energy splitting. Since we are quenching back an forth the energy of the excited state, three energy windows of the environment are now explored. We consider their associated macroscopic energies to be $E \in \{0,\Delta\varepsilon,2\Delta\varepsilon\}$. 

In Fig.~\ref{fig:quenched_dynamics}, we compare the exact dynamics with the prediction of the EMME. The volume terms of the bands are set to $V_E\in\{100,200,400\}$ and we chose the initial state $\rho(0) = \ketbra{1}{1}\otimes \Pi_{E=0}/V_{E=0}$. We observe that the EMME is also able to reproduce accurately the dynamics when the system energy levels are periodically quenched. This justifies in retrospective our claim above that we can replace the static system energies $\varepsilon_k$ with time-dependent energies $ \varepsilon_k(\lambda_t)$ as long as $\lambda_t$ varies slowly compared to the relaxation time of the bath.

\floatsetup[figure]{style=plain,subcapbesideposition=top}
\begin{figure}
{\includegraphics[width=\textwidth]{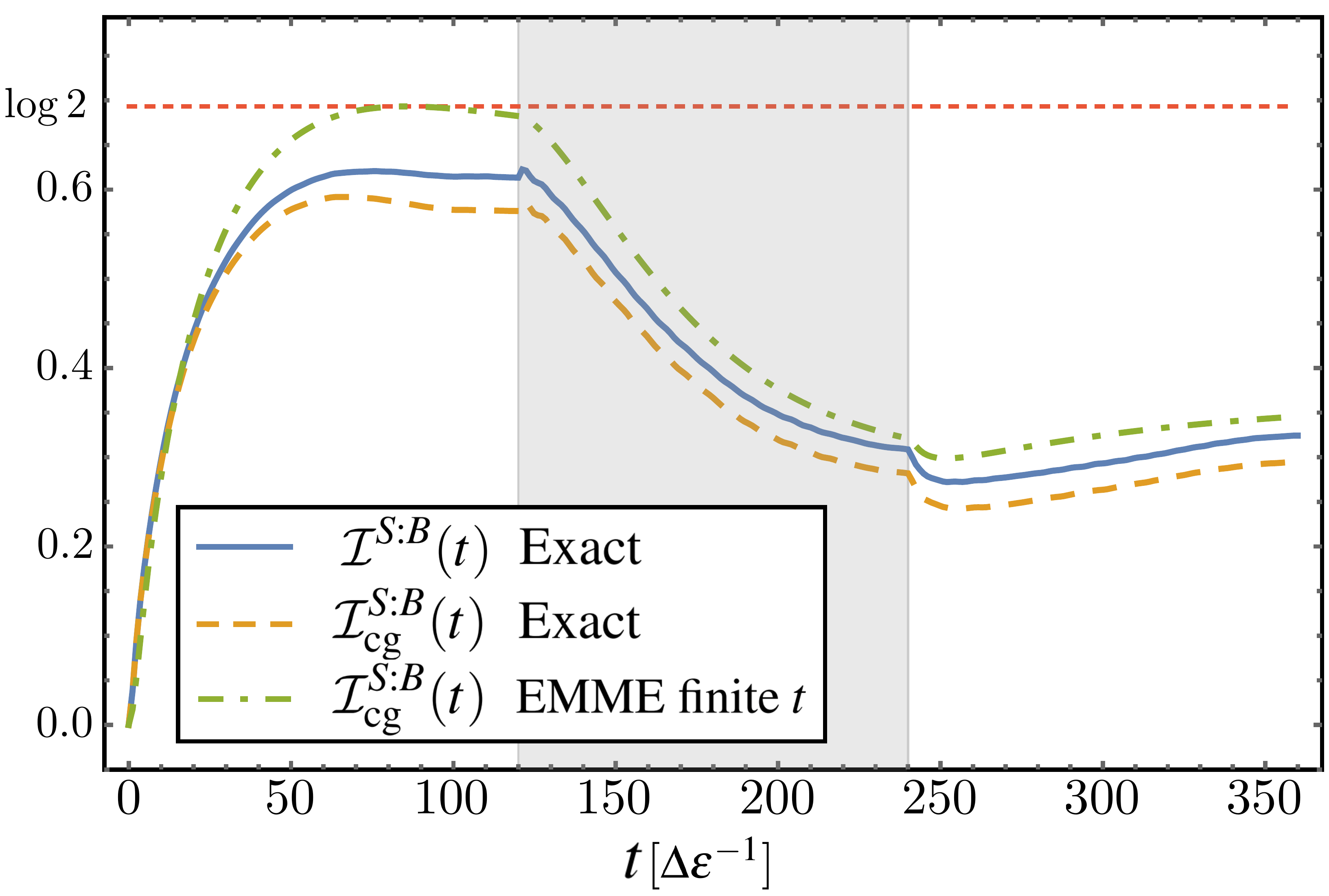}}
  \caption{Evolution of the mutual information comparing the exact solution of the quantum mutual information (blue solid line) with the exact coarse-grained mutual information (orange dashed line) and the approximation obtained using the EMME (green dot-dashed lines) for a spin with energies $\varepsilon_0 = 0$ and $\varepsilon_1(\lambda_t) = \Delta\varepsilon$ (white background) and $\varepsilon_1(\lambda_t) = 2\Delta\varepsilon$ (shadowed background). See details in the main text. The initial state is chosen $\rho(0) = \ketbra{1}{1}\otimes \Pi_{E=0}/V_{E=0}$. Parameters: $\delta = 0.5$, $t_\star = 120\Delta \varepsilon^{-1}$, $b(E,E') = 0$ $\forall E,E'$, and $a=1$.\label{fig:mutual_information}}
\end{figure}

During the evolution the system and bath develop nonequilibrium features, but the Clausius inequality and our second law in Eq.\eqref{eq:second_law_rate} remain valid at all times as demonstrated in Fig.~\ref{fig:clausius_inequality}. In the inset, we show the corresponding evolution of the non-equilibrium temperature of the bath $T^*_B(t)$ which, due to the bath finite heat capacity, cannot be approximated by a constant value. The fact that we start with a zero effective temperature $T_B^*(0) = 0$ is a result of our choice for the initial state $\rho(0)$ and a consequence of the bath model, where we neglect any energy levels below the lowest band that participates in the dynamics.

The difference between the dash-dotted green curve and the dashed orange curve in Fig.~\ref{fig:clausius_inequality} is a nonequilibrium effect resulting from a bath state deviating from an ideal thermal state. The difference between the solid blue curve and the dashed orange curve in Fig.~\ref{fig:clausius_inequality} is, instead, a result of the (neglected) classical system-bath correlations as measured by the coarse-grained mutual information in Eq.~\eqref{eq:mutual_information_obs}. To investigate the latter, we numerically compute the evolution of the mutual information. In Fig.~\ref{fig:mutual_information}, we show that $\mathcal{I}^{S:B}$ always upper bounds $\mathcal{I}^{S:B}_\text{cg}$, which increases close to their maximum value $\log 2$. Since the mutual information can grow close to its maximum value, the system-bath correlations are not negligible showing that \Andreu{the weak-coupling approximation does not justify the use of an uncorrelated reference state of system and bath}. Also, the EMME provides a good approximation of the real value of $\mathcal{I}^{S:B}_\text{cg}$, but is prediction can violate the bound in Eq.~\eqref{eq:mutual_information_bound}. The reason that the EMME overestimates the system-bath correlation is a consequence of the strict energy conservation derived in Subsec.~\ref{subsec:strict_energy_conservation}, which is never exactly satisfied for any finite coarse-graining and any finite system-bath coupling strength.

\section{Generalization to multiple environments}\label{sec:multiple_baths}

In the previous sections, we have studied the case of a quantum system $S$ in contact with a single heat bath $B$. Similar findings hold for the case of multiple environments and, for completeness, we outline in this section the generalization of the main results to multiple environments. Its detailed study including the treatment of various interesting applications is left for future work.

We consider multiple baths labeled by $\nu = 1,\cdots, n$, with Hamiltonian $\text{H}_{B_\nu}$ coupled to the system with an interaction $\text{H}_{\text{int},\nu} = \lambda \text{S}_{\nu} \otimes \text{B}_{\text{int},\nu}$ (again, more general expressions can be found in the Appendix). Denoting $E_{\nu,i}$ the eigenenergies of the $\nu^{th}$ bath Hamiltonian, we proceed to coarse-grain the energies into energy windows $E_{\nu,\delta_{\nu}} = [E_\nu-\delta_{\nu}/2,E_\nu+\delta_{\nu}/2)$ centered around the energy $E_\nu$. Then, we can define a projection operator
\begin{align}
\mathcal{P}[\rho] \coloneqq  \sum_{\mathbf{E}} \rho_S(\mathbf{E}) \otimes \frac{\Pi_\mathbf{E}}{V_\mathbf{E}},
\end{align}
where the vector $\mathbf{E} = (E_1,\cdots,E_n)$, the projector $\Pi_\mathbf{E} = \Pi_{E_1}\otimes\cdots\otimes\Pi_{E_n}$, and the joint volume $V_\mathbf{E} = V_{E_1}\cdots V_{E_n}$ have been introduced. All the steps followed in Sec.~\ref{sec:derivation} are valid under the replacement $E\mapsto\mathbf{E}$ with the important remark that we obtain an additive structure for the EMME, i.e., there are no crossed terms with $\nu \neq \nu'$(see App.~\ref{app:multiple_baths}). For the multiple bath scenario, the bath correlation function is given by
\begin{align}
C_{B_\nu}(\mathbf{E},\mathbf{E'};-\tau) \coloneqq \lambda^2 \qav{\tilde{\text{B}}^{\dagger}_\nu (-\tau) \Pi_\mathbf{E} \text{B}_\nu}_{\mathbf{E}'}.
\end{align}
Note that $C_{B_\nu}(\mathbf{E},\mathbf{E'};-\tau)$ is very sparse, since the vectors $\mathbf{E}$ and $\mathbf{E}'$ can only be different in the $\nu^{th}$ component. Therefore, it is possible to redefine complex decay rates $\gamma_\nu (E_\nu, E'_\nu;\omega)$ that depend only in the $\nu^{th}$ component of the energy vectors $\mathbf{E}$ and $\mathbf{E}'$. For the same reasons that we gave in Sec.~\ref{sec:derivation}, it is justified to factor $\gamma_\nu(E_\nu, E'_\nu;\omega) = \gamma_\nu(E_\nu, E'_\nu) \delta_{E'_\nu,E_\nu+\omega}$ yielding the EMME
\begin{align}
\partial_t  \rho_S(\textbf{E}) &= -i [\text{H}'_{S}(\mathbf{E}),\rho_S(\textbf{E})]\nonumber\\
&+\sum_\nu\sum_{\omega} \left(\frac{\gamma_\nu(E_\nu,E_\nu-\omega)}{V_{E_\nu-\omega}} \text{S}_{\nu,\omega} \rho_S(\textbf{E}+\omega \textbf{e}_\nu)\text{S}^{\dagger}_{\nu,\omega} \right.\nonumber\\
& -\left.\frac{\gamma_\nu(E_\nu+\omega,E_\nu)}{2 V_{E_\nu}}  \left\{ \rho_S(\textbf{E}),\text{S}^{\dagger}_{\nu,\omega} \text{S}_{\nu,\omega}\right\} \right),\label{eq:emme_multiple_bath}
\end{align}
where $\textbf{e}_\nu$ is the unit vector along the $\nu^{th}$ component.

From Eq.~\eqref{eq:emme_multiple_bath}, it is possible to obtain the conservation of the probability $\partial_t P(E_{\text{tot}}) = 0$, with $P(E_{\text{tot}}) = \sum_{k,\mathbf{E}} p(\varepsilon_k, \mathbf{E}) \delta_{E_{\text{tot}},\varepsilon_k +\sum_\nu E_\nu}$, as well as a  steady-state condition similar to Eq.~\eqref{eq:steady_state} (further details are given in App.~\ref{app:multiple_baths}).

The thermodynamic behavior studied in Sec.~\ref{sec:emergent_thermo} also extends to multiple environments. Introducing $\dot{Q}_\nu = \sum_{k,\mathbf{E}} E_\nu  \partial_t p(\varepsilon_k, \textbf{E})$, the first law takes the form
\begin{align}
\frac{d}{dt} U_S = \sum_\nu \dot{Q}_\nu + \dot{W}.
\end{align}
To obtain the second law, we first need to introduce the multiple-bath observational entropy 
\begin{align}
\mathcal{S}_{\text{obs}}(\mathbf{p}) = \sum_{k,\textbf{E}} p(\varepsilon_k, \textbf{E}) (-\log p(\varepsilon_k, \textbf{E}) + \log V_\textbf{E}),
\end{align}
for which it is possible to show that $d \mathcal{S}_{\text{obs}}(t)/dt \geq 0$. The proof uses the additive structure of the EMME, and then proceeds analogously to the single bath case. Finally, Clausius inequality is similarly found as 
\begin{align}
\Delta \mathcal{S}^S_{\text{obs}}(t) - \sum_\nu \int_0^t dt' \frac{\dot{Q}_\nu(t')}{T^*_\nu(t')}\geq \Delta \mathcal{S}_{\text{obs}}(t)\geq 0.
\end{align}

\section{Outlook and comparison with other master equations}\label{sec:comparison}

We conclude by comparing the EMME with other master equation approaches. In the previous sections we have already discussed the EMME in comparison with the popular Redfield and BMS (or ``quantum optical'') master equation \cite{Breuer2002,deVega2017,Schaller2014}, whose dynamic and thermodynamic predictions can differ significantly from the EMME. Note that, similar in spirit to the EMME, the BMS master equation is sometimes refined by equipping it with an additional counting field, which keeps track of the changes in bath energy \cite{Esposito2009,Schaller2014}. However, these counting field master equations make no further use of this information to obtain a more accurate system dynamics: after integrating out the counting field, the reduced dynamics of the system is still given by the standard BMS master equation. Therefore, we here focus only on a comparison with more advanced master equations going beyond this standard approach. 

Clearly, one way to obtain improved results is to use the standard projection operators $\mathcal{P}_\text{Born}$ as for the BMS master equation, but to go beyond second order in the interaction Hamiltonian. However, this quickly becomes cumbersome and, as Ref. \cite{Breuer2006} has shown, even the fourth-order master equation does not necessarily improve the accuracy, still giving \emph{qualitatively} wrong results in comparison with the EMME and the exact solution. A more sophisticated idea in comparison to simply ``cranking up" the perturbative hierarchy is to apply different approximation techniques to the memory kernel in the Nakajima-Zwanzig equation based on, e.g., semiclassical simulations of the bath dynamics \cite{Kelly2013,Kelly2015}. While being non-perturbative and more accurate, this approximation has been so far mostly used for numerical case-by-case studies and it seems hard to obtain general insights from it. 

Another approach, which still resides in the standard picture by tracing out the entire bath and keeping only information about the system, is to combine second order master equations and polaron transformations \cite{Brandes2005}. In this approach one first maps a strongly coupled system-bath Hamiltonian to a weakly coupled one by using the polaron transformation and afterwards combines it with standard perturbative master equations. This allows to treat strong coupling, as demonstrated, e.g., in Refs.~\cite{Segal2006, Schaller2013, Gelbwaser2015, Wang2015}, but it does not overcome the Markovian approximation and essentially treats the bath as being in (conditional) equilibrium throughout. Nevertheless, combining polaron transformations, which work well only for particular system-bath models, with the EMME seems to be a promising avenue for future research to treat strongly coupled systems in a more accurate way. 

Finally, another master equation approach, which shares some similarities with our approach by explicitly treating parts of the bath degrees of freedom, is based on Markovian embedding strategies \cite{Hughes2009a, Martinazzo2011,Woods2014}. By redefining the system-bath partition and applying a master equation to an enlarged but weakly coupled and Markovian system, as done using, e.g., the reaction coordinate master equation \cite{Hughes2009b, IlesSmith2014, Strasberg2016, Newman2017, Strasberg2018}, or other formally exact but more involved master equations \cite{Tamascelli2018, Brenes2020}, numerically accurate results can be obtained while retaining at the same time detailed information about system-bath correlations and (parts of) the bath degrees of freedom. We have not yet benchmarked our master equation with these techniques, but we expect the latter to be more accurate. On the downside, these Markovian embedding strategies, as well as the aforementioned master equations using polaron transformations or semiclassical simulations of the memory kernel, all rely on the paradigmatic Caldeira-Leggett model. We believe it is a significant advantage that the EMME applies in principle to \emph{every} system-bath model. 

Thus, to summarize, the EMME opens up the possibility to treat a variety of interesting nonequilibrium situations, including, e.g., finite heat baths, spin environments, nonlinear system-bath interactions, impurities in quantum many body systems, etc., in a dynamically more accurate way beyond the restrictive \Andreu{static bath} and Markovian approximation and with an intuitive and consistent thermodynamic interpretation. Perhaps in combination with other techniques, such as the ones just mentioned, we are convinced that it provides an efficient, flexible and intuitive tool for future research in quantum nanotechnologies.

\section*{Aknowledgements}

We thank Massimiliano Esposito and Kavan Modi for stimulating discussions on this and related topics. We acknowledge financial support from the Spanish MINECO/ AEI FIS2016-80681-P, PID2019-107609GB-I00, from the Catalan Government: projects CIRIT 2017-SGR-1127, AGAUR FI-2018-B01134, and QuantumCAT 001-P-001644 (RIS3CAT comunitats), co-financed by the European Regional Development Fund (FEDER), and also from the German Research fundation DFG (project STR 1505/2-1).

\appendix
\begin{center}
{ \textbf{APPENDIX}}
\end{center}

\Andreu{For the sake of the explanation, we have considered a single system-bath coupling operator throughout the main text. For completeness, we spell out here the most general form of the EMME for multiple environments and multiple coupling operators, corresponding to the interaction $\text{H}_{\text{int},\nu} = \lambda \sum_{\alpha_\nu} \text{S}_\nu^{\alpha_\nu}\otimes\text{B}_\nu^{\alpha_\nu}$. In this case, the EMME yields
\begin{align}
&\partial_t  \rho_S(\textbf{E}) = -i [\text{H}'_{S}(\mathbf{E}),\rho_S(\textbf{E})]\nonumber\\
&+\sum_\nu\sum_{\alpha_\nu \alpha'_\nu}\sum_{\omega} \left(\frac{\gamma_\nu^{{\alpha_\nu}{\alpha'_\nu}}(E_\nu,E_\nu-\omega)}{V_{E_\nu-\omega}} \text{S}^{\alpha_\nu}_{\nu,\omega} \rho_S(\textbf{E}+\omega \textbf{e}_\nu)\text{S}^{{\alpha_\nu'}\dagger}_{\nu,\omega} \right.\nonumber\\
&\qquad \quad -\left.\frac{\gamma^{{\alpha_\nu}{\alpha'_\nu}}_\nu(E_\nu+\omega,E_\nu)}{2 V_{E_\nu}}  \left\{ \rho_S(\textbf{E}),\text{S}^{{\alpha'_\nu}\dagger}_{\nu,\omega} \text{S}^{\alpha_\nu}_{\nu,\omega}\right\} \right).\label{eq:emme_general}
\end{align}
Equation~\eqref{eq:emme_general} reduces to Eq.~\eqref{eq:emme_multiple_bath} for $\text{H}_{\text{int},\nu} = \lambda \text{S}_\nu\otimes\text{B}_\nu$ (for all $\nu$) and to Eq.~\eqref{eq:emme} for a single bath index $\nu$. 
}

\section{Details on the derivation of the EMME}\label{app:details}

In this appendix we provide details on the derivation of the EMME, the computation of the dissipation rates in the three approaches listed in the main text, and a comparison between them. For the sake of generality, we consider the interaction $\text{H}_\text{int} = \lambda\sum_\alpha \text{S}^\alpha\otimes\text{B}_\text{int}^\alpha$, which appears as a double index $\alpha\alpha'$ in the expression of the correlation functions (i.e., $C_B^{\alpha\alpha'}(E,E';\omega)$) and the corresponding dissipation rates (i.e., $\gamma^{\alpha\alpha'}(E,E')$). In the three methods, the aim is to compute the complex dissipation rates $\gamma^{\alpha\alpha'}(E,E')$ starting from the correlation function
\begin{align}
C_B^{\alpha\alpha'}(E,E';-\tau) =  \sum_{E_i\in E_\delta}\sum_{E_j\in E'_\delta} \frac{\lambda^2 }{V_{E'}} \text{B}^{\alpha'*}_{ij}\text{B}^{\alpha}_{ij}e^{i(E_i-E_j)\tau}. \label{eq:correlation_appendix}
\end{align}

\subsection{Details on the heuristic approach}\label{app:details_first}

As explained in the main text, this approach is based on the substitution $E_i-E_j \mapsto E-E'$ in Eq.~\eqref{eq:correlation_appendix}. Then, using Eq.~\eqref{eq:closed_dissipation}, one obtains
\begin{align}
\gamma^{\alpha\alpha'}_{\text{heuristic}}(E,E';\omega) &= \lambda^2 \text{tr}_B[\text{B}^{\alpha'\dagger}\Pi_E \text{B}^{\alpha}\Pi_{E'}] \nonumber\\
 &\times \int_\mathbb{R}d\tau e^{i(\omega +E-E')\tau}.
\end{align}
The macroscopic energies of the bath can always be expressed as $E = n\delta$ for some $n \in \mathbb{N}$. Then, the time integrals give rise to
\begin{align}
\int_\mathbb{R} d(\delta\tau) e^{\pm i( (\omega/\delta) +n -n')(\delta\tau)} \approx 2\pi \delta'_{E', E + \omega_{kq}}.\label{eq:slow_delta}
\end{align}
where the modified Kronecker delta should be interpreted as the function:
\begin{align}
\delta'_{E',E+\omega_{kq}} = \left\{\begin{matrix}
1 & \text{ if } E' \text{ s.t. } |E'-E-\omega| \leq \delta\\
0 &\text{else}
\end{matrix} \right.,
\end{align}
and then, one can directly identify the expression of $\gamma^{\alpha\alpha'}_{\text{heuristic}}(E,E')$ in Eq.~\eqref{eq:first_rates}.

\subsection{Details random matrix approach}\label{app:details_stoch}

We consider bath coupling operators of the form
\begin{align}
\text{B}^\alpha&=\nonumber\\
& \sum_{E\neq E'}\sum_{E_i\in E_\delta}\sum_{E_j \in E'_\delta} \left[b^\alpha(E,E') + c^\alpha(E_i,E_j)\right]\ketbra{E_i}{E_j},
\end{align}
where $b^\alpha(E,E')$ are deterministic functions of the macroscopic energies and $c^\alpha(E_i,E_j)$ are i.i.d. complex random numbers with zero mean and variance $\mathbb{E}[c^\alpha(E_i,E_j)c^{\alpha'}(E'_i,E'_j)] = a^2 \delta_{\alpha\alpha'}\delta_{E_i E_i'}\delta_{E_j E_j'}$. Then, the ensemble averaged bath correlation function yields
\begin{align}
\mathbb{E}[C_B^{\alpha\alpha'}(E,E';-\tau)] = \frac{\lambda^2}{V_{E'}} &\left( b^{\alpha' *}(E,E')b^{\alpha}(E,E')  + a^2\delta_{\alpha\alpha'}\right)\nonumber\\ &\times \sum_{E_i\in E_\delta}\sum_{E_j\in E'_\delta} e^{i E_i \tau} e^{-i E_j\tau}.\label{eq:stochastic_corr}
\end{align}

To compute the double sum term in Eq.~\eqref{eq:stochastic_corr} we introduce the density of states $g(E) = \partial_E \sum_{E_i} \Theta(E-E_i)$ and assume: (i) the bath is dense enough to justify $\sum_{E_i} \mapsto \int g(e)de$ and (ii) $g(e)$ is approximately constant in each energy window. Then,
\begin{align}
\sum_{E_i\in E_\delta}\sum_{E_j\in E'_\delta} e^{i (E_i -E_j)\tau}\approx V_E V_{E'} e^{i(E-E')\tau}\frac{\sin^2(\delta\tau/2)}{(\delta\tau/2)^2},
\end{align}
where we have used the relation $g(E)\delta = V_E$. The time-dependent properties of the correlation function are then described by the function
\begin{align}
h(\tau) = \frac{\delta}{2\pi} \frac{\sin^2(\delta\tau/2)}{(\delta\tau/2)^2}
\end{align}
As we show in App.~\ref{app:time-integrals}, the Fourier transform of $h(\tau)$ is strongly peaked around the origin, and it allows to approximate
\begin{align}
\int_\mathbb{R} d\tau h(\tau) e^{i\Omega\tau}=\hat{h}(\Omega) \approx \delta_{\Omega,0},
\end{align}
where, in the computation of the complex dissipation rates, $\Omega =\omega+E-E'$. Then, using Eq.~\eqref{eq:closed_dissipation}, we obtain the complex dissipation rates
\begin{align}
\gamma^{\alpha\alpha'}_\text{rmt}(E&,E') \nonumber\\
&= \frac{2\pi\lambda^2}{\delta}(b^*_{\alpha'}(E,E')b_{\alpha}(E,E') + a^2 \delta_{\alpha\alpha'})V_EV_{E'}. \label{eq:dissipation_rates_stoch}
\end{align}

\subsection{Details ETH approach}\label{app:details_eth}

In order to exploit the ETH to compute the correlation function in Eq.~\eqref{eq:correlation_appendix}, two issues arise: First, there is no guarantee that the operators $\text{B}^\alpha$ are Hermitian and, second, it is not clear how the ETH should be modified when one considers correlation between different observables. The first issue is easily solved by noting that any operator can be decomposed as $\text{O} = \text{O}_+ + i\text{O}_-$, where $\text{O}_+$ and  $\text{O}_-$ are Hermitian. Therefore,
\begin{align}
\text{V} = \lambda \sum_\alpha \text{S}^\alpha\otimes\text{B}^\alpha = \lambda \sum_{\alpha} \left( \text{S}_+^\alpha \otimes \text{B}^\alpha_+ - \text{S}^\alpha_- \otimes \text{B}_-^\alpha\right).
\end{align}
Then we can assume without loss of generality $\text{S}^\alpha$ and $\text{B}^\alpha$ to be Hermitian. Then, using the ETH ansatz we find 
\begin{align}
\text{B}^{\alpha}_{ij} = B^\alpha(E_{ij}) \delta_{ij} + \sqrt{\frac{1}{V_{E_{ij}}}} f^\alpha(E_{ij},\Omega_{ij}) R^{\alpha}_{ij}.\label{eq:eth_ansatz}
\end{align}
Because the definition of $\text{B}^\alpha$ is such that $\qav{\text{B}^\alpha}_E=0$ for all $E$, the first term of the ansatz can be set to zero. Then, the microcanonical bath correlation function yields
\begin{align}
&C_B^{\alpha\alpha'}(E,E';-\tau) =\nonumber\\
& \frac{\lambda^2}{V_{E'}}\sum_{E_i,E_j}\frac{f^{\alpha'*}(E_{ij},\Omega_{ij})f^\alpha(E_{ij},\Omega_{ij})}{V_{E_{ij}}} R^{\alpha'*}_{ij}R^{\alpha}_{ij}e^{i (E_i - E_j)\tau}.\label{eq:eth_corr}
\end{align}
Now, we would like to use the statistical properties of the erratically varying random numbers $R^\alpha_{ij}$. It is clear that $R^{\alpha}_{ij}$ should have zero mean as before. However, note that $R^{\alpha}_{ij}$ cannot be uncorrelated for different $\alpha$ or, otherwise, two point correlation of two generic different observables would vanish \cite{dAlessio2016}. We avoid the second issue by leaving without specify the correlation $\overline{R^{\alpha'*}_{ij}R^{\alpha}_{ij}}$, where here the overline denotes the average in the spirit of the ETH.

We proceed as follows: On one hand, the numbers $R^{\alpha}_{ij}$ change erratically with $i$ and $j$ and even $j\mapsto j+1$ can abruptly change its value. On the other hand, the function $f^\alpha(E,\Omega)$ is a smooth function of its arguments and, for a dense enough bath, the substitution $j\mapsto j+1$ will give rise to a perturbative correction. Therefore, it is justified to substitute in Eq.~\eqref{eq:eth_corr} $R^{\alpha'*}_{ij}R^{\alpha}_{ij}\mapsto\overline{R^{\alpha'*}_{ij}R^{\alpha}_{ij}}$. Then, we introduce the function
\begin{align}
F^{\alpha\alpha'}(E_{ij},\Omega_{ij}) \coloneqq f^{\alpha'*}(E_{ij},\Omega_{ij})f^{\alpha}(E_{ij},\Omega_{ij}) \overline{R_{ij}^{\alpha'*} R_{ij}^{\alpha}},\label{eq:bigF_function}
\end{align}
which, consistently with the ETH ansatz, is a smooth function of its arguments, decays with $|\Omega|\to \infty$, and has the symmetry property 
\begin{align}
(F^{\alpha\alpha'}(E,\Omega))^* = F^{\alpha'\alpha}(E,\Omega) = F^{\alpha\alpha'}(E,-\Omega).
\end{align}
For a sufficiently regular and dense environment, it is justified to approximate $E_{ij} \approx \bar{E} = (E+E')/2$ and replace $\sum_{E_i} \mapsto \int de g(e)$. Assuming that $g(e)$ is constant within each energy window we arrive at 
\begin{align}
C_B^{\alpha\alpha'}(E,E';&-\tau) =\lambda^2 \frac{g(E) g(E')}{V_{E'}} \nonumber\\
& \times \iint de\, de' \frac{F^{\alpha\alpha'}(\bar{E},e-e')}{V_{\bar{E}}} e^{i (e-e')\tau},\label{eq:corr_eth}
\end{align}
Finally, using Eq.~\eqref{eq:closed_dissipation}, the integral over time of the correlation function gives a factor $2\pi \delta(\omega + e-e')$ and leads to the expression
\begin{align}
\gamma^{\alpha\alpha'}_{\text{eth}}(E,E') = \frac{2\pi\lambda^2}{\delta} V_E V_{E'} \frac{F^{\alpha\alpha'}(\bar{E},E-E')}{V_{\bar{E}}}.
\end{align}

\subsection{Connection between the three approaches}\label{app:connection}

Finally, we investigate under which circumstances the three derivations given above give rise to the same decay rates. To this end, we rewrite them in the alternative form
\begin{align}
&\gamma^{\alpha\alpha'}_{\text{heuristic}}(E,E') = \frac{2\pi\lambda^2}{\delta} V_E V_{E'} \frac{\text{tr}[\text{B}_{\alpha'}^\dagger \Pi_E \text{B}_\alpha \Pi_{E'}]}{V_E V_{E'}}, \nonumber\\
&\gamma^{\alpha\alpha'}_{\text{rmt}}(E,E') = \frac{2\pi\lambda^2}{\delta} V_E V_{E'} (b_{\alpha'}^*(E,E') b_\alpha(E,E') + a^2\delta_{\alpha\alpha'}), \nonumber\\
&\gamma^{\alpha\alpha'}_{\text{eth}}(E,E') =\frac{2\pi\lambda^2}{\delta} V_E V_{E'} \frac{F^{\alpha\alpha'}(\bar{E},E-E')}{V_{\bar{E}}},
\end{align}
which makes the comparison easier. A connection can be found when the coupling has purely coarse grained components, namely, when the following approximation holds:
\begin{align}
\text{B}^\alpha &= \sum_{E_i E_j} \qav{E_i|\text{B}^\alpha|E_j} \ketbra{E_i}{E_j} \nonumber\\
&\approx\sum_{E E'} b^\alpha(E,E') \sum_{E_i \in E_\delta}\sum_{E_j\in E'_\delta} \ketbra{E_i}{E_j},\label{eq:b_connection}
\end{align}
where $b^\alpha(E,E')$ are functions of only the coarse grained energies.

\textit{Connection of the approaches} \ref{subsec:first} \textit{and} \ref{subsec:stochastic}:\\

Assuming the form in Eq.~\eqref{eq:b_connection}, the coarse bath dynamics method gives raise to the rates
\begin{align}
\gamma^{\alpha\alpha'}_{\text{heuristic}}(E,E') = \frac{2\pi\lambda^2}{\delta} V_E V_{E'} b^*_{\alpha'}(E,E') b_{\alpha}(E,E').\label{eq:rate_first}
\end{align}
On the other hand, in the limit of vanishing variance $a\to 0$, the random matrix coupling in Eq.~\eqref{eq:stochastic_coupling} reduces to \eqref{eq:b_connection}. Therefore, we find
\begin{align}
\gamma^{\alpha\alpha'}_{\text{rmt}}(E,E') &=\lim_{a\to 0} \frac{2\pi\lambda^2}{\delta} V_E V_{E'} ( b^*_{\alpha'}(E,E') b_{\alpha}(E,E')+a^2\delta_{\alpha\alpha'})\nonumber\\
&=\frac{2\pi\lambda^2}{\delta} V_E V_{E'} b^*_{\alpha'}(E,E') b_{\alpha}(E,E'),
\end{align}
obtaining, then, the same rates $\gamma_{\text{heuristic}}^{\alpha\alpha'}(E,E') = \gamma^{\alpha\alpha'}_{\text{rmt}}(E,E')$.\\

\textit{Connection of the approaches} \ref{subsec:first} \textit{and} \ref{subsec:eth}:\\

Using the ETH ansatz in Eq.~\eqref{eq:eth_ansatz} and assuming that the bath coupling operator has the coarse-grained structure in Eq.~\eqref{eq:b_connection}, it follows 
\begin{align}
\text{B}^{\alpha}_{ij} = \sqrt{\frac{1}{V_{\bar{E}}}}  f^{\alpha} (\bar{E},E-E') R^{\alpha}_{ij} \quad (E\neq E').\label{eq:eth_coarsed}
\end{align}
Using Eq.~\eqref{eq:eth_coarsed} into Eq.~\eqref{eq:first_rates}, and identifying the average $\overline{R^*_{\alpha',ij}R_{\alpha,ij}} = \sum_{ij} R^*_{\alpha',ij} R_{\alpha,ij}/(V_E V_{E'})$, we arrive at
\begin{align}
\gamma^{\alpha\alpha'}_{\text{heuristic}}(E,E') = \frac{2\pi\lambda^2}{\delta} V_E V_{E'} \frac{F^{\alpha\alpha'}(\bar{E},E-E')}{V_{\bar{E}}},
\end{align}
and, therefore, $\gamma_{\text{heuristic}}^{\alpha\alpha'}(E,E') = \gamma^{\alpha\alpha'}_{\text{eth}}(E,E')$.

\section{More details on the RMT approach}\label{app:time-integrals}

\Andreu{In this appendix, we give further details on the RMT approach. Our starting point is the finite-time Redfield version of the EMME which in the Schr\"odinger picture reads
\begin{align}
\partial_t \rho_S(E) =& -i [\text{H}_S,\rho_S(E)] \nonumber\\
 &\hspace{-1.5cm}+\sum_{\alpha\alpha'}\sum_{E'}\sum_{\omega\omega'} \int_0^t d\tau e^{-i\omega'\tau}\left(\mathbb{E}[C^{\alpha\alpha'}_B(E,E';-\tau)] \text{S}_\omega\rho_S(E')\text{S}_{\omega'}^\dagger \right.\nonumber\\
& \left.- \mathbb{E}[C^{\alpha\alpha'}_B(E',E;-\tau)]  \rho_S(E)\text{S}_{\omega'}^\dagger\text{S}_{\omega}\right)+\text{h.c.}. \label{eq:emme_second_order_3}
\end{align}
Our goal consists in evaluating more explicitly the integrals of the form
\begin{align}
\int_0^t d\tau\mathbb{E}[C^{\alpha\alpha'}_B(E,E';-\tau)]e^{i\omega \tau}.
\end{align}
that appear in Eq.~\eqref{eq:emme_second_order_3}. To that end, we use the results obtained in App.~\ref{app:details_stoch} to write the averaged microcanonical bath correlation function as
\begin{align}
\mathbb{E}[C_B^{\alpha\alpha'}(E,E';-\tau)] =\frac{\gamma^{\alpha\alpha'}_\text{rmt}(E,E')}{V_{E'}} h(\tau) e^{i(E-E')\tau}.
\end{align}
Then, the integrals of interest can be cast as
\begin{align}
\frac{\gamma^{\alpha\alpha'}_\text{rmt}(E,E')}{V_{E'}} \int_0^t d\tau h(\tau) e^{i(\omega+E-E')\tau}.\label{eq:time-integrals}
\end{align}
In the case where $\omega+E=E'$, the evaluation is simpler and can be done by conveniently introducing the function 
\begin{align}
\zeta(t) \coloneqq \frac{\delta}{\pi}\int_0^t d\tau \frac{\sin^2(\delta\tau/2)}{(\delta\tau/2)^2} = 2\int_0^t d\tau h(\tau).
\end{align}
Then, all the dissipation rates appearing in the EMME are multiplied by the time-dependent envelope $\zeta(t)$ as it is the case, for instance, of the example analyzed in Sec.~\ref{sec:case_study}.}

\begin{figure}
{\includegraphics[width=\textwidth]{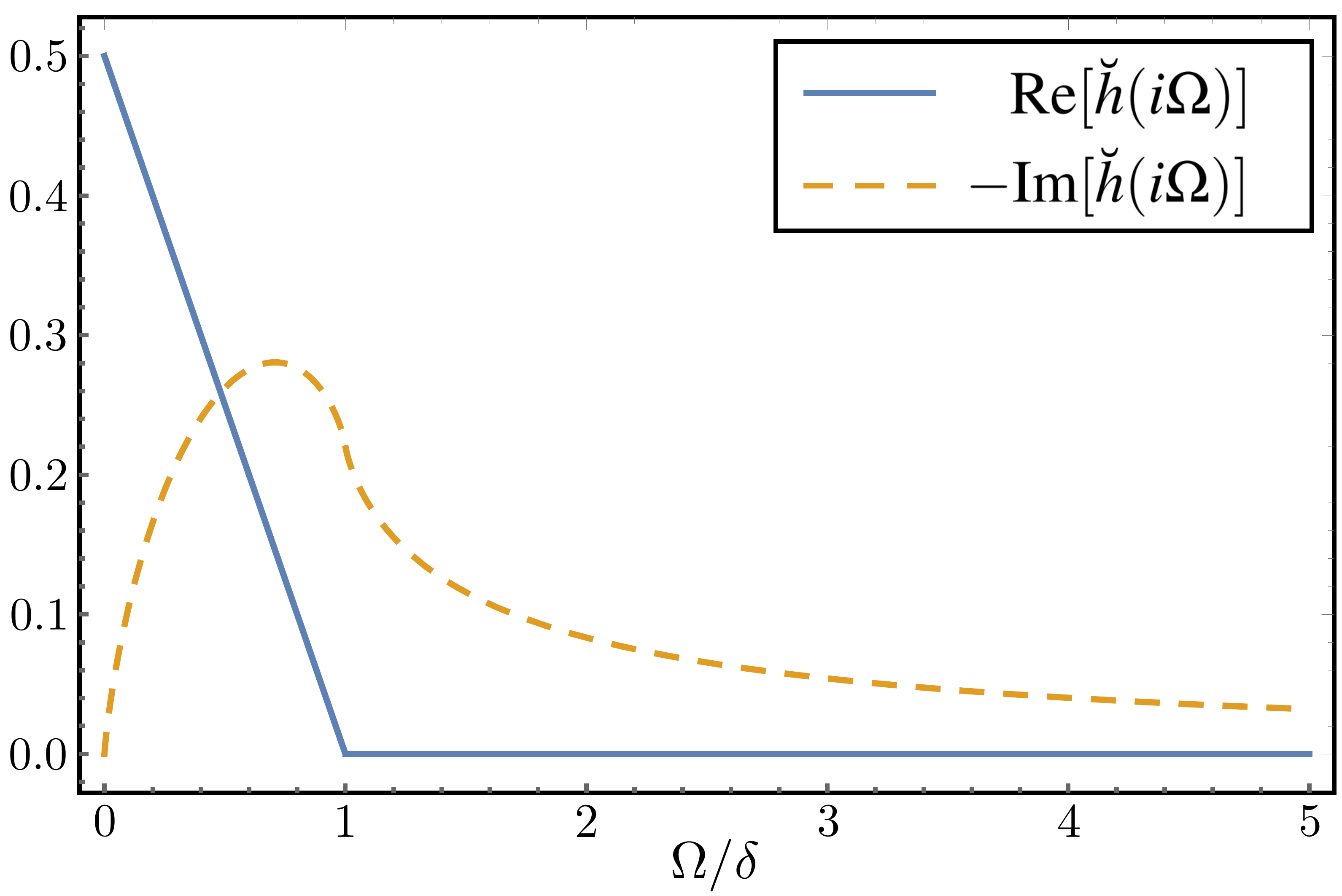}}
  \caption{Real (solid blue) and imaginary (dashed orange) parts of the function $\breve{h}(i\Omega)$. } \label{fig:wierdFunc}
\end{figure}

In general, however, there is no a priory reason why $E' = E+\omega$. Even though the general result for a finite $t$ is cumbersome, it is possible to evaluate exactly the time-integrals under the Markov approximation (i.e., for $t\to\infty$). Ultimately, our objective is to compute the integrals 
\begin{align}
I(1) \coloneqq \int_0^\infty d\tau h(\tau) e^{-i\Omega \tau} =\frac{1}{\pi} \int_0^\infty dx  \frac{\sin^2(x)}{x^2} e^{-i 2(\Omega/\delta) x },
\end{align}
that appear in Eq.~\eqref{eq:time-integrals}. Those integrals, can be regarded as the Laplace transform $\breve{h}(i\Omega) = \mathbb{L}[h(\tau)](s=i\Omega)$, with $\mathbb{L}[f(\tau)](s) = \int_0^\infty d\tau f(\tau) \exp(-s\tau)$. It is easy to see that the real and imaginary parts of $\breve{h}(i\Omega)$ correspond respectively to even and odd functions of $\Omega$. We introduce the frequency ratio $\xi = \Omega/\delta \in \mathbb{R}$ as well as the parameter-dependent integral
\begin{align}
I(a) &\coloneqq\frac{1}{\pi} \int_0^\infty dx \frac{\sin^2(ax)}{x^2} e^{-i2\xi x} \nonumber\\
&\Rightarrow I'(a) =\frac{1}{\pi} \int_0^\infty 2a dx\, \frac{ \sin(2ax)}{(2ax)} e^{-i2\xi x}.
\end{align}
Performing the change of variables $t = 2ax$, and noting the Laplace transform property
\begin{align}
&\mathbb{L}[f(t)/t] (s) = \int_s^\infty du \breve{f}(u)\nonumber\\ &\Rightarrow \mathbb{L}[\sin(t)/t] (s) = \int_s^\infty \frac{du}{1+u^2} = \frac{\pi}{2}-\arctan(s),
\end{align}
which brings to $I'(a) = 1/2 - \arctan( i \xi/ a))/\pi$. Now, the complex function $w(z) = \arctan(z)$ can be written in terms of logarithms using the following reasoning
\begin{align}
&z = \tan(w) = -i \frac{e^{i w}-e^{-i w}}{e^{i w}+e^{-i w}} \nonumber\\
&\Rightarrow 2iw = \log\frac{1+iz}{1-iz} +n 2\pi i \quad \text{with } n\in\mathbb{N}.
\end{align}
We choose the principal Riemann sheet $n=0$ to coincide with the real $\arctan$ function, i.e. $w(1) = \pi/4$. Noticing that $I(0) =0$, we can proceed to
\begin{align}
I(1)& = \int_0^1 \, da I'(a) = \int_0^1 da \left(\frac{1}{2} - \frac{\arctan(i\xi/2)}{\pi}\right) \nonumber\\
&=  \frac{1}{2}+ \frac{i}{2\pi } \int_0^1 da\left( \log(a -\xi) -\log(a+\xi)\right),
\end{align}
As we have discussed, the real and imaginary parts of the target integral $I(1)$ are respectively even and odd functions of $\xi$ and, then, we can restrict ourselves to $\xi>0$. Still, we have two different scenarios: 
\begin{align}
\int_0^1& da  ( \log(a+\xi) - \log(a-\xi) ) = \int_0^1 \log(a+\xi) \nonumber\\
&- \int_0^\xi da \log(a-\xi) - \int_\xi^1 (\log|a-\xi| + i\pi ) \quad (\text{for } \xi <1)\nonumber\\
\int_0^1 &da ( \log (a+\xi) - \log(a-\xi) ) = \int_0^1 \log(a+\xi) \nonumber\\
&+ \int_0^1 (\log|a-\xi| + i\pi ) \quad (\text{for }\xi >1),
\end{align}
which leads, for $\xi >0$, to the final result
\begin{align}
\text{Re}[I(1)] &= \left\{ \begin{matrix}
(1-\xi)/2  & \quad  \text{for } \xi < 1\\
0 & \quad \text{for } \xi > 1
\end{matrix}\right.,\nonumber\\
\text{Im}[I(1)] &= \frac{1}{2\pi}\left(
2\xi\log(\xi) -(1+\xi)\log(1+\xi)\right.\nonumber\\
&\left.+(1-\xi) \log|1-\xi|\right).
\end{align}
Therefore, there is no approximation in the $\text{Re}[\breve{h}(i\Omega)]$ by disregarding the non-resonant terms $\Omega>\delta$. For completeness we show the full behavior in Fig.~\ref{fig:wierdFunc}. Hence, the relation $\gamma^{\alpha\alpha'}_{\text{rmt}}(E,E';\omega) \propto \delta_{E',E+\omega}$ is obtained without any approximation in computing the time-integrals.

\floatsetup[figure]{style=plain,subcapbesideposition=top}

\section{Properties of the transtion rates}\label{app:general_props}

In this appendix, we consider the interaction $\text{H}_\text{int} = \lambda\sum_\alpha \text{S}^\alpha\otimes\text{B}^\alpha$ leading to the transition rates
\begin{align}
W_{kq}(E,E') \coloneqq\sum_{\alpha\alpha'} \langle q|\text{S}^{\alpha'\dagger} |k\rangle\langle k|\text{S}^\alpha|q\rangle \gamma^{\alpha\alpha'}(E,E').
\end{align}
With this definition, we proof the properties $W_{kq}(E,E')\geq 0$ and $W_{kq}(E,E') = W_{qk}(E',E)$ for the transition rates. The former is proven, as usual, through Bochner's theorem and the latter requires only basic algebraic manipulation. \\

\subsection{Positivity}

We start noting that the transition rates $W_{kq}(E,E')$ may be obtained from the scalar product
\begin{align}
(\mathbf{s}_{kq},\mathbf{\gamma}^T\, \mathbf{s}_{kq}) &= \sum_{\alpha\alpha'} s^{\alpha'*}_{kq} \gamma^{\alpha\alpha'}(E,E';\omega) s^\alpha_{kq}, \label{eq:inner_product}
\end{align}
where the arguments $E,E'$ and $\omega$ are implicit in the matrix $\gamma^T$. Defining $\text{V}_{kq}  = \lambda \sum_\alpha s^\alpha_{kq} \text{B}^\alpha$, Eq.~\eqref{eq:inner_product} is the Fourier transform
\begin{align}
(\mathbf{s}_{kq},\mathbf{\gamma}^T \, \mathbf{s}_{kq}) &= \int_\mathbb{R} d\tau \text{tr}_B [e^{-i\text{H}_B\tau}\text{V}^\dagger_{kq} e^{i\text{H}_B\tau} \Pi_{E} \text{V}_{kq}  \Pi_{E'}] e^{i\omega\tau}\nonumber\\
&= \int_\mathbb{R} d\tau f(\tau) e^{i\omega\tau} = \hat{f}(\omega).\label{eq:f_bochner}
\end{align}
Then, Bochner's theorem states that the Fourier transform $\hat{f}(\omega)$ of a function $f(\tau)$ is positive if $f(\tau)$ is of \textit{positive type}. A function is of positive type if for any set of times $\{\tau^\alpha\}$ the matrix $f^{\alpha\alpha'} \equiv f(\tau^\alpha-\tau^{\alpha'})$ is positive semidefinite. Then, we see that $f(\tau)$ is of positive type since taking a general vector $\mathbf{w}$ we have
\begin{align}
(\mathbf{w},\mathbf{f w}) &= \sum_{E_i E_j}\left|\sum_\alpha w^\alpha e^{-i(E_i -E_j)\tau^\alpha}\right|^2 |\qav{E_i|\text{V}_{kq}|E_j}|^2\nonumber\\ &\geq 0,\label{eq:transition_rates_positive}
\end{align}
where $E_i \in E_\delta$ and $E_j \in E'_\delta$. Finally, to obtain $W_{kq}(E,E')\geq 0$ from the inner product $(\mathbf{s}_{kq},\mathbf{\gamma}^T\, \mathbf{s}_{kq})\geq 0$ it is only left to assume the factorization condition which yields
\begin{align}
W_{kq}(E,E')\delta_{E',E+\omega} = (\mathbf{s}_{kq},\mathbf{\gamma}^T\, \mathbf{s}_{kq}) \geq 0.
\end{align}

\subsection{Symmetry}

The key observation is that, since the interaction $\text{V}$ is Hermitian, we have the property $\text{V}^\dagger_{kq} = \text{V}_{qk}$. Then, one can cast Eq.~\eqref{eq:f_bochner} in the alternative form
\begin{align}
(\mathbf{s}_{kq},\mathbf{\gamma}^T\, \mathbf{s}_{kq}) =\sum_{E_iE_j} |\qav{E_i|\text{V}_{kq}|E_j}|^2 \int_{\mathbb{R}} d\tau e^{i(\omega+E_i-E_j)\tau},
\end{align}
where $E_i \in E_\delta$ and $E_j \in E'_\delta$. From the equation above, one deduces
\begin{align}
(\mathbf{s}_{kq},\mathbf{\gamma}^T(E,E';\omega)\, \mathbf{s}_{kq})= (\mathbf{s}_{qk},\mathbf{\gamma}^T(E',E;-\omega)\, \mathbf{s}_{qk}).
\end{align}
Therefore, assuming the factorization condition in Eq.~\eqref{eq:factorization_condition} it follows
\begin{align}
W_{kq}(E,E') \delta_{E',E+\omega} = W_{qk}(E',E) \delta_{E,E'-\omega},
\end{align}
from where we deduce $W_{kq}(E,E') = W_{qk}(E',E)$.

\section{Mutual information inequality}\label{app:mutual_information}

In this appendix we proof the inequality $\mathcal{I}^{S:B}[\rho] \geq \mathcal{I}^{S:B}_\text{cg}(\mathbf{p})$. First, we note that the quantum mutual information can be cast as $\mathcal{I}^{S:B}[\rho] = \mathcal{D}[\rho||\rho_S\otimes\rho_B]$ where we have introduced the quantum relative entropy
\begin{align}
\mathcal{D}[\rho||\sigma] = \text{tr}[\rho(\log\rho-\log\sigma)]\geq 0,\label{eq:relative_entropy}
\end{align}
for $\rho$ and $\sigma$ two states. The relative entropy is contractive under the action of a  completely-positive and trace-preserving map $\mathcal{E}$, i.e., it fulfills the property $\mathcal{D}[\rho||\sigma] \geq \mathcal{D}[\mathcal{E}(\rho)||\mathcal{E}(\sigma)]$. Second, we note that the map 
\begin{align}
\mathcal{E}[\rho] &= \sum_{k,E}\text{tr}[\rho\ketbra{k}{k}\otimes \Pi_E] \ketbra{k}{k}\otimes\frac{\Pi_E}{V_E} \nonumber\\
&= \sum_{k,E}p(\varepsilon_k,E) \ketbra{k}{k}\otimes\frac{\Pi_E}{V_E}, 
\end{align}
is a valid completely-positive and trace-preserving map since its Kraus decomposition can be read from
\begin{align}
\mathcal{E}[\rho] = \sum_{k,E}\sum_{E_i\in E_\delta}\sum_{E_j\in E_\delta} \frac{\ketbra{k,E_i}{k,E_j}}{\sqrt{V_E}}\rho \frac{\ketbra{k,E_j}{k,E_i}}{\sqrt{V_E}}.
\end{align}
Therefore, we obtain
\begin{align}
\mathcal{I}^{S:B}[\rho] &= \mathcal{D}[\rho||\rho_S\otimes\rho_B] \nonumber\\
&\geq \mathcal{D}[\mathcal{E}[\rho]||\mathcal{E}[\rho_S\otimes\rho_B]] = \mathcal{I}^{S:B}_\text{cg}(\mathbf{p}).
\end{align}

\section{Analytic solution for the spin system}\label{app:spin}

Here we provide the analytic solution for the spin system studied in Sec.~\ref{sec:case_study}. 
Taking as a starting point Eq.~\eqref{eq:qubit_explicit}, we project into the system eigenstates from where it follows the population rate equations
\begin{align}
\partial_t p(\varepsilon_0,E) &=\zeta(t) \gamma(E,E-\Delta \varepsilon) \nonumber\\
&\times \left(\frac{p(\varepsilon_1,E - \Delta \varepsilon)}{V_{E-\Delta\varepsilon}}  -\frac{p(\varepsilon_0,E)}{V_E}\right),\nonumber\\
\partial_t p(\varepsilon_1,E)  &= \zeta(t) \gamma(E,E+\Delta \varepsilon)\nonumber \\
& \times \left(\frac{p(\varepsilon_0,E+\Delta \varepsilon)}{V_{E+\Delta\varepsilon}}  -\frac{p(\varepsilon_1,E)}{V_E}\right),\label{eq:qubit_pops}
\end{align}
and we have defined $\gamma(E,E')=0$ if either $E$ or $E'$ does not exist. Equivalently, we could have written Eq.~\eqref{eq:qubit_pops} in matrix form by gathering all the populations $p(\varepsilon_k, E)$ in the population vector $\mathbf{p}$ as
\begin{align}
\partial_t \mathbf{p} = \zeta(t) \Lambda \mathbf{p},\label{eq:me_matrix_spin}
\end{align}
where the entries of the matrix $\Lambda$ should be readed from Eq.~\eqref{eq:qubit_pops} (see below). We note that Eq.~\eqref{eq:qubit_pops} leads to the block-diagonal structure
\begin{align}
\Lambda = \bigoplus_{E_{\text{tot}}} \Lambda(E_{\text{tot}}),
\end{align}
where $\Lambda(E_{\text{tot}})$ acts on the subspace $\mathbf{p}(E_{\text{tot}}) = \{p(\varepsilon_1,E),p(\varepsilon_0,E+\Delta \varepsilon)\}$ of total energy $E_{\text{tot}}=\varepsilon_1 + E$. Explicitly, we can write down
\begin{align}
\Lambda(E_{\text{tot}}) = \gamma(E,E+\Delta \varepsilon)  \begin{pmatrix}
-1/V_E & 1/V_{E+\Delta\varepsilon}\\
1/V_E & -1/V_{E+\Delta\varepsilon}
\end{pmatrix},
\end{align}
which is a stochastic matrix. Now, we are at the position where it is possible to integrate Eq.~\eqref{eq:me_matrix_spin} to arrive at
\begin{align}
\mathbf{p}(t) = \bigoplus_{E_{\text{tot}}} e^{\Lambda(E_{\text{tot}}) \Xi(t)} \mathbf{p}(0),
\end{align}
where $\Xi(t) = \int_0^t dt' \zeta(t')$. Remarkably the matrix $\Lambda(E)$ has the property
\begin{align}
\Lambda(E_{\text{tot}})^2 &= - \gamma(E,E+\Delta\varepsilon) \left(\frac{1}{V_E}+\frac{1}{V_{E+\Delta\varepsilon}}\right) \Lambda(E_{\text{tot}})
\nonumber\\ &\equiv-2 \bar{\gamma}(E,E+\Delta \varepsilon)\Lambda(E_{\text{tot}}),
\end{align}
which leads to the final solution
\begin{align}
&e^{\Lambda(E_{\text{tot}})\Xi(t)} \mathbf{p}(E_{\text{tot}};0) =  \sum_{n=0}^\infty \frac{(\Lambda(E_{\text{tot}}) \Xi(t))^n}{n!} \mathbf{p}(E_{\text{tot}};0)\nonumber\\
&\qquad = \left(\text{1}+\frac{1-e^{-2\bar{\gamma}(E,E+\Delta\varepsilon)\, \Xi(t)}}{2\bar{\gamma}(E,E+\Delta\varepsilon)} \Lambda(E_{\text{tot}})\right)\mathbf{p}(E_{\text{tot}};0).
\end{align}
Since the function $\zeta(t)$ saturates rapidly to $\zeta(t\to\infty) = 1$, we expect $\Xi(t) \sim t$ at long times. Therefore, the steady-state can be computed
\begin{align}
\frac{p_\text{eq}(\varepsilon_1,E)}{p_\text{eq}(\varepsilon_0,E+\Delta\varepsilon)} = \frac{V_{E}}{V_{E+\Delta\varepsilon}},\label{eq:spin_steady_state}
\end{align}
while keeping constant at all times the probability of being in a certain energy shell of total energy $E_{\text{tot}} = \varepsilon_1 + E$, i.e., $p(\varepsilon_1,E) + p(\varepsilon_0,E+\Delta\varepsilon) = \text{constant}$. 

\section{Additional numerical results}\label{app:EMME_failure}
\floatsetup[figure]{style=plain,subcapbesideposition=top}
\begin{figure*}
{\includegraphics[width=\textwidth]{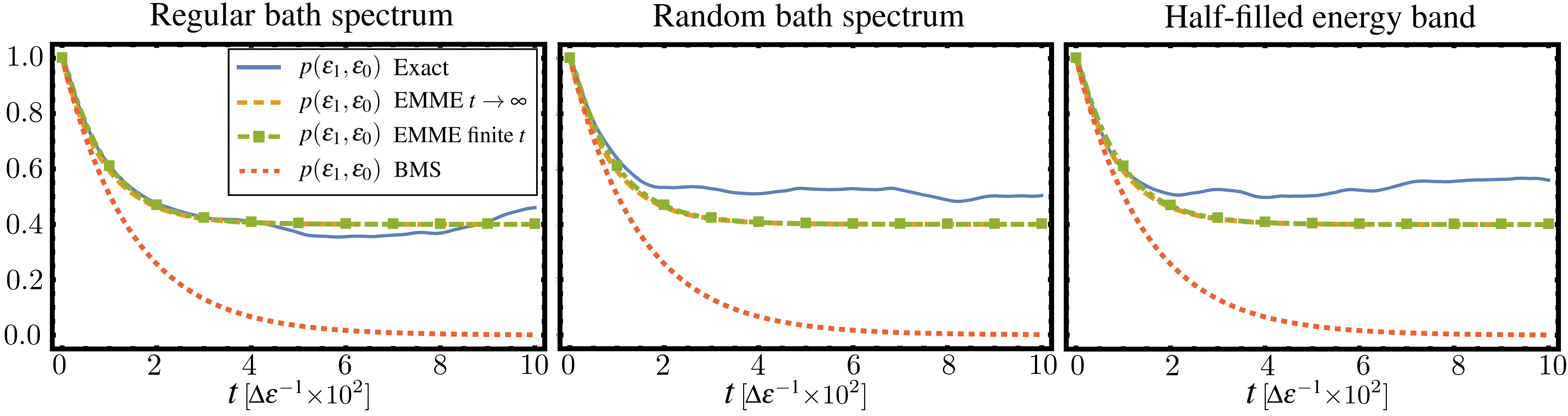}}\\
\caption{Comparison of the evolution of the joint probability $p(\varepsilon_1,E=\varepsilon_0)$ for a spin system coupled to a two bands bath (see Fig.~\ref{fig:twobands} for details). Parameters: $\lambda = 3\times 10^{-3}$, $\delta = 0.5$, $\varepsilon_0 = 0$, $\varepsilon_1 = 1$, $V_{\varepsilon_0} =  20$ and $V_{\varepsilon_1} = 40$, and $a=1$.\label{fig:comparative_dynamics_a} }
\end{figure*}
\floatsetup[figure]{style=plain,subcapbesideposition=top}
\begin{figure*}
{\includegraphics[width=\textwidth]{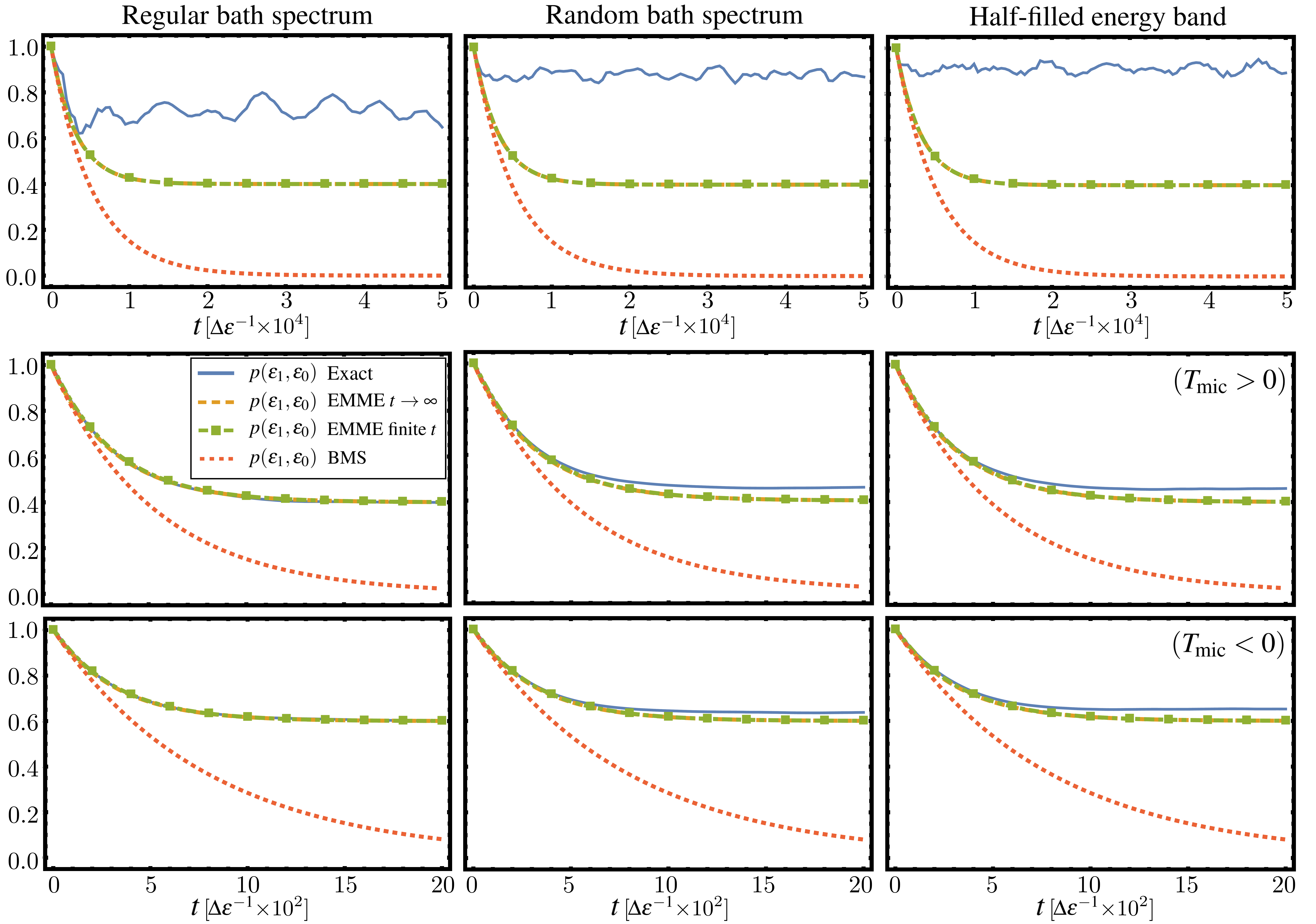}}\\
\caption{Comparison of the evolution of the joint probability $p(\varepsilon_1,E=\varepsilon_0)$ for a spin system coupled to a two bands bath (see Fig.~\ref{fig:twobands} for details). First row: $V_{\varepsilon_0} =  20$ and $V_{\varepsilon_1} = 40$. Second row: $V_{\varepsilon_0} =  400$ and $V_{\varepsilon_1} = 600$. Third row: $V_{\varepsilon_0} =  600$ and $V_{\varepsilon_1} = 400$. Parameters: $\lambda = 5\times 10^{-4}$, $\delta = 0.5$, $\varepsilon_0 = 0$, $\varepsilon_1 = 1$, and $a=1$. \label{fig:comparative_dynamics_b}}
\end{figure*}
\floatsetup[figure]{style=plain,subcapbesideposition=top}
\begin{figure*}
{\includegraphics[width=\textwidth]{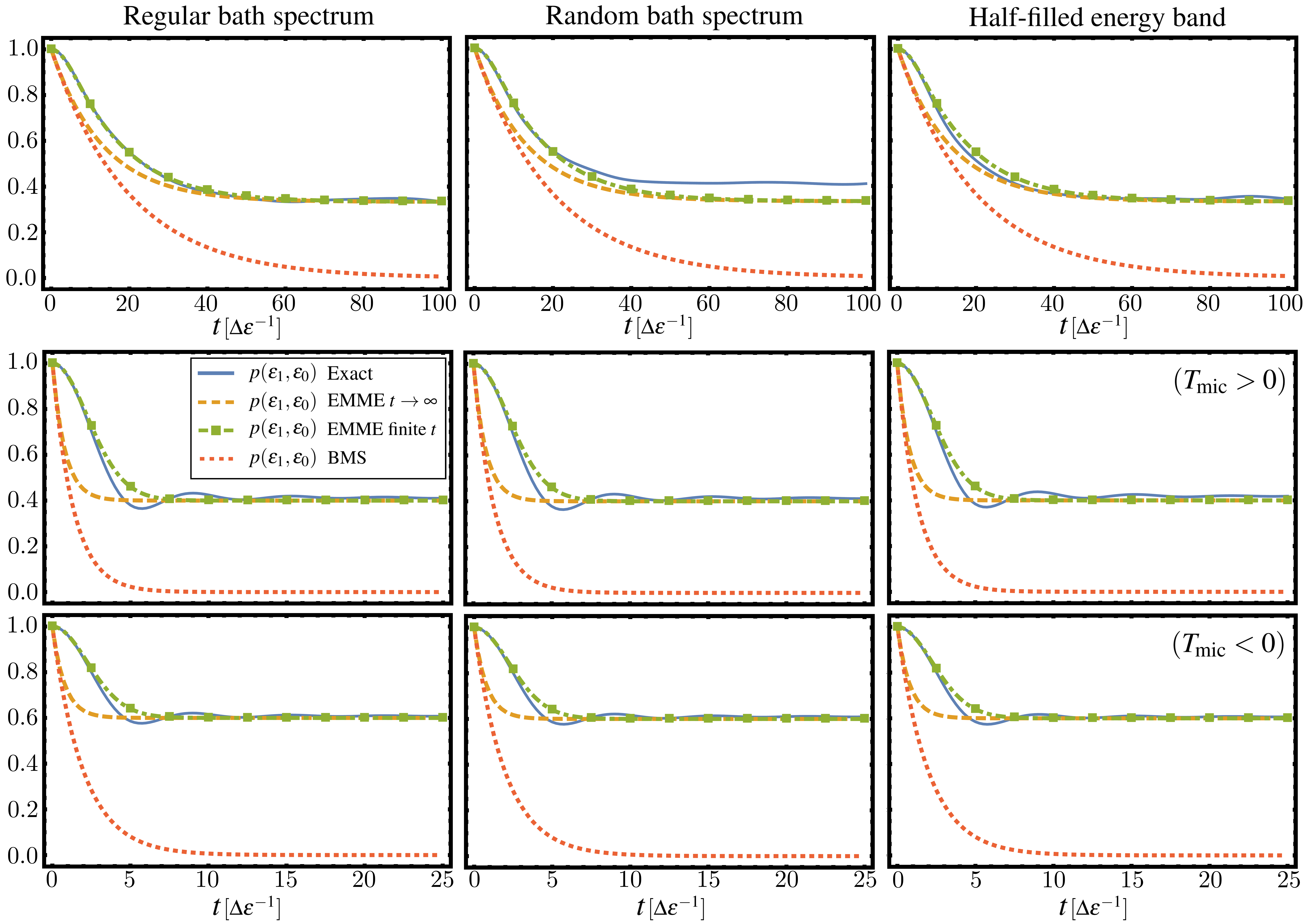}}\\
\caption{Comparison of the evolution of the joint probability $p(\varepsilon_1,E=\varepsilon_0)$ for a spin system coupled to a two bands bath (see Fig.~\ref{fig:twobands} for details). First row: $V_{\varepsilon_0} =  20$ and $V_{\varepsilon_1} = 40$. Second row: $V_{\varepsilon_0} =  400$ and $V_{\varepsilon_1} = 600$. Third row: $V_{\varepsilon_0} =  600$ and $V_{\varepsilon_1} = 400$. Parameters: $\lambda = 10^{-2}$, $\delta = 0.5$, $\varepsilon_0 = 0$, $\varepsilon_1 = 1$, and $a=1$. \label{fig:comparative_dynamics_c}}
\end{figure*}

We here show the evolution of the population for stronger and weaker coupling strengths as compared to the main text Fig.~\ref{fig:twobands}. First, in the first row of Fig.~\ref{fig:comparative_dynamics_a}, Fig.~\ref{fig:comparative_dynamics_b}, and Fig.~\ref{fig:comparative_dynamics_c}, we observe that reducing the volumes of the energy bands $V_E \lesssim 100$ typically leads to a disagreement of the exact dynamics with the prediction of the EMME. This expected behavior arises from the fact that the bath is too small: recurrences are unavoidable and no perturbative master equation approach can correctly capture the reduced system dynamics in this case.

Regarding the second and third rows of Fig.~\ref{fig:comparative_dynamics_b}, we observe that reducing the coupling strength $\lambda$ further can lead to imprecise results even though the second order approximation becomes more accurate. The underlying reason is that, as $\lambda\to 0$, the linewidth of a jump process becomes narrower and, then, the system can resolve the fine-structure of the bath energy bands. Typically, resolving the structure of the bath energy bands leads to a reduced effective volume of the band or, for a very small number of available levels, may avoid thermalization completely. 

In the opposite limit of a ``large'' coupling $\lambda$ (see Fig.~\ref{fig:comparative_dynamics_c}), the EMME fails to describe the transient behavior while the steady state is correctly predicted. Of course, for even larger $\lambda$, the second order approximation breaks down completely leading to an incorrect description of the dynamics. 

The aforementioned observations are in unison with the findings in \cite{Esposito2003b}, where upper and lower bounds for the coupling strength were found to the validity of their microcanonical master equation.

\section{Details on multiple environments}\label{app:multiple_baths}

In this appendix, we show in some detail the derivation of the EMME for the case of multiple baths and, also, we obtain the nonequilibrium thermodynamic description in analogous manner to the case of a single bath. We generalize the results shown in the main text to a coupling operator of the form $\text{H}_{\text{int},\nu} = \sum_{\alpha_\nu} \text{S}^{\alpha_\nu}_{\nu}\otimes \text{B}^{\alpha_\nu}_{\text{int},\nu}$.

\subsection{Additive structure}

Following the derivation for a single bath, we consider that each bath is coupled to the system via the interaction operator
\begin{align}
\text{H}_{\text{int},\nu}  = \lambda \sum_{\alpha_\nu} \text{S}^{\alpha_\nu}_{\nu} \otimes \text{B}_{\text{int},\nu}^{\alpha_\nu}.
\end{align}
The evolution in the interaction picture is then governed by the von Neumann equation
\begin{align}
\mathcal{L}(t)[\rho] = -i \sum_{\nu}[\tilde{\text{H}}_{\text{int},\nu}(t),\tilde{\rho}(t)].
\end{align}
Again, it will prove useful to decompose the interaction into the respective block-diagonal and off-diagonal parts  
\begin{align}
\sum_\mathbf{E} \delta \text{H}_{\nu}(\mathbf{E})\Pi_\mathbf{E} &= \lambda \sum_{\alpha_\nu} \text{S}^{\alpha_\nu}_\nu \otimes \sum_\mathbf{E} \langle \text{B}_{\text{int},\nu}^{\alpha_\nu}  \rangle_\mathbf{E} \Pi_\mathbf{E}, \nonumber\\
 \text{V}_\nu &= \lambda \sum_{\alpha_\nu} \text{S}^{\alpha_\nu}_\nu \otimes (\text{B}_{\text{int},\nu}^{\alpha_\nu}-\sum_\mathbf{E} \langle \text{B}_{\text{int},\nu}^{\alpha_\nu}  \rangle_\mathbf{E} \Pi_\mathbf{E}),
\end{align}
and further introduce $\delta\text{H}_\nu = \sum_\mathbf{E}\delta \text{H}_{\nu}(\mathbf{E})$ and $\text{B}^{\alpha_\nu}_\nu = \text{B}^{\alpha_\nu}_\text{int}-\sum_{\textbf{E}}\langle\text{B}_{\text{int},\nu}^{\alpha_\nu}\rangle_{\textbf{E}}\Pi_{\textbf{E}}$. We now proceed to compute the different terms appearing in Eq.~\eqref{eq:redfield_equation}.  We start with 
\begin{align}
\mathcal{PL}(t)\mathcal{P}\rho &= -i \sum_\nu [\delta\tilde{\text{H}}_\nu(t),\mathcal{P}\tilde{\rho}(t)],
\end{align}
and using the linearity of the commutator together with $\mathcal{Q} = \mathcal{I-P}$, one deduces
\begin{align}
\mathcal{QL}(t)\mathcal{P}\rho = -i \sum_{\nu} [\tilde{\text{V}}_\nu(t),\mathcal{P}\tilde{\rho}(t)].
\end{align}
Therefore, the second order term is found to be
\begin{align}
&\mathcal{PL}(t)\mathcal{QL}(t')\mathcal{P}\rho\nonumber\\
& = \sum_{\nu\nu'}\sum_{\mathbf{E}} \text{tr}_B[\Pi_\mathbf{E}[\delta \tilde{\text{H}}_{\nu}(t) +\tilde{\text{V}}_\nu(t),[ \mathcal{P}\tilde{\rho}(t'),\text{V}_{\nu'}(t')]]]\frac{\Pi_\mathbf{E}}{V_\mathbf{E}}.
\end{align}
After a lengthy but straightforward manipulation, one obtains that the term involving $\delta \text{H}_\nu$ vanishes since it ends up being proportional to $\langle\text{B}^{\alpha_\nu}_\nu\rangle_\mathbf{E} = 0$ and one is left with 
\begin{align}
\mathcal{PL}(t)&\mathcal{QL}(t')\mathcal{P}\rho \nonumber\\
&= \sum_{\nu\nu'}\sum_{\mathbf{E}} \text{tr}_B[\Pi_\mathbf{E}[\tilde{\text{V}}_\nu(t), [\mathcal{P}\tilde{\rho}(t'),\tilde{\text{V}}_{\nu'}(t')]]]\frac{\Pi_\mathbf{E}}{V_\mathbf{E}}.
\end{align}
Using the explicit expression of $\text{V}_\nu$, one obtains that only the terms with $\nu=\nu'$ survive since, again, the case $\nu\neq\nu'$ lead to factors $\langle\text{B}^{\alpha_\nu}_\nu\rangle_\mathbf{E} = 0$. Therefore, we get an additive structure for the EMME. Then, one can proceed for each $\nu$ as we did for the single bath case. Performing the standard Markov and secular approximations leads to Eq.~\eqref{eq:emme_multiple_bath} of the main text.

\subsection{Rate equation}

The rate equation for the multiple environment case is found to be
\begin{align}
\partial_t p(\varepsilon_k,\textbf{E}) &= \sum_{\nu,q} \left( \frac{W_{\nu,kq}(E_\nu,E_\nu+\omega_{kq})}{V_{E_\nu+\omega_{kq}}} p(\varepsilon_q,\textbf{E}+\omega_{kq}\textbf{e}_\nu)\right. \nonumber\\
&\left. - \frac{W_{\nu,qk}(E_\nu+\omega_{kq},E_\nu)}{V_{E_\nu}} p(\varepsilon_k,\textbf{E}) \right),\label{eq:emme_populations_multiple}
\end{align}
where, for each bath $\nu$, we introduce the transition rates $W_{\nu,kq}$ analogously to the case of a single bath. Again, we find the properties $W_{\nu,kq}(E_\nu,E'_\nu) \geq 0$ and $W_{\nu,kq}(E_\nu,E'_\nu) = W_{\nu,qk}(E'_\nu,E_\nu)$.

\subsection{Energy conservation}

Using Eq.~\eqref{eq:emme_populations_multiple}, one can prove that $P(E_{\text{tot}}) = \sum_{k,\mathbf{E}} p(\varepsilon_k, \mathbf{E}) \delta_{E_{\text{tot}},\varepsilon_k +\sum_\nu E_\nu}$ is preserved. To this end we introduce, given $E_\text{tot}$, the energy of the bath $\nu$ given the system energy $\varepsilon_k$
\begin{align}
E_{\nu|k} =  E_{\text{tot}}-\sum_{\nu'\neq\nu} E_{\nu'} -\varepsilon_k,
\end{align}
with the property $E_{\nu|k} +\omega_{kq} = E_{\nu|q}$. We note that, under the action of the Kronecker $\delta$-function in the definition of $P(E_\text{tot})$, we have
\begin{align}
\partial_t& P(E_{\text{tot}}) = \nonumber\\
& \sum_\mathbf{E} \sum_{\nu} \sum_{kq} \left( \frac{W_{\nu,kq}(E_{\nu|k},E_{\nu|q})}{V_{E_{\nu|q}}} p(\varepsilon_q,E_1,\cdots,E_{\nu|q},\cdots,E_n)\right. \nonumber\\
\quad &\left. - \frac{W_{\nu,qk}(E_{\nu|q},E_{\nu|k})}{V_{E_{\nu|k}}} p(\varepsilon_k,E_1,\cdots,E_{\nu|k},\cdots,E_n) \right),
\end{align}
which vanishes after renaming the dummy variables $k\leftrightarrow q$ in the second line.

\subsection{Steady-state}

Imposing detailed balance into Eq.~\eqref{eq:emme_populations_multiple}, it follows the steady-state condition
\begin{align}
\frac{p(\varepsilon_k,E_1,\cdots,E_{\nu|k},\cdots, E_n)}{p(\varepsilon_q,E_1,\cdots,E_{\nu|q},\cdots, E_n)} = \frac{V_{E_{\nu|k}}}{V_{E_{\nu|q}}},
\end{align}
which is analogous to the one found for the single environment case in Eq.~\eqref{eq:steady_state}.

\bibliographystyle{apsrev4-1}
\bibliography{Refs}

\begin{thebibliography}{62}%
\makeatletter
\providecommand \@ifxundefined [1]{%
 \@ifx{#1\undefined}
}%
\providecommand \@ifnum [1]{%
 \ifnum #1\expandafter \@firstoftwo
 \else \expandafter \@secondoftwo
 \fi
}%
\providecommand \@ifx [1]{%
 \ifx #1\expandafter \@firstoftwo
 \else \expandafter \@secondoftwo
 \fi
}%
\providecommand \natexlab [1]{#1}%
\providecommand \enquote  [1]{``#1''}%
\providecommand \bibnamefont  [1]{#1}%
\providecommand \bibfnamefont [1]{#1}%
\providecommand \citenamefont [1]{#1}%
\providecommand \href@noop [0]{\@secondoftwo}%
\providecommand \href [0]{\begingroup \@sanitize@url \@href}%
\providecommand \@href[1]{\@@startlink{#1}\@@href}%
\providecommand \@@href[1]{\endgroup#1\@@endlink}%
\providecommand \@sanitize@url [0]{\catcode `\\12\catcode `\$12\catcode
  `\&12\catcode `\#12\catcode `\^12\catcode `\_12\catcode `\%12\relax}%
\providecommand \@@startlink[1]{}%
\providecommand \@@endlink[0]{}%
\providecommand \url  [0]{\begingroup\@sanitize@url \@url }%
\providecommand \@url [1]{\endgroup\@href {#1}{\urlprefix }}%
\providecommand \urlprefix  [0]{URL }%
\providecommand \Eprint [0]{\href }%
\providecommand \doibase [0]{http://dx.doi.org/}%
\providecommand \selectlanguage [0]{\@gobble}%
\providecommand \bibinfo  [0]{\@secondoftwo}%
\providecommand \bibfield  [0]{\@secondoftwo}%
\providecommand \translation [1]{[#1]}%
\providecommand \BibitemOpen [0]{}%
\providecommand \bibitemStop [0]{}%
\providecommand \bibitemNoStop [0]{.\EOS\space}%
\providecommand \EOS [0]{\spacefactor3000\relax}%
\providecommand \BibitemShut  [1]{\csname bibitem#1\endcsname}%
\let\auto@bib@innerbib\@empty
\bibitem [{\citenamefont {Breuer}\ and\ \citenamefont
  {Petruccione}(2002)}]{Breuer2002}%
  \BibitemOpen
  \bibfield  {author} {\bibinfo {author} {\bibfnamefont {H.-P.}\ \bibnamefont
  {Breuer}}\ and\ \bibinfo {author} {\bibfnamefont {F.}~\bibnamefont
  {Petruccione}},\ }\href@noop {} {\emph {\bibinfo {title} {The theory of open
  quantum systems}}}\ (\bibinfo  {publisher} {Oxford University Press},\
  \bibinfo {year} {2002})\BibitemShut {NoStop}%
\bibitem [{\citenamefont {Schaller}(2014)}]{Schaller2014}%
  \BibitemOpen
  \bibfield  {author} {\bibinfo {author} {\bibfnamefont {G.}~\bibnamefont
  {Schaller}},\ }\href {\doibase 10.1007/978-3-319-03877-3} {\emph {\bibinfo
  {title} {Open quantum systems far from equilibrium}}},\ Vol.\ \bibinfo
  {volume} {881}\ (\bibinfo  {publisher} {Springer},\ \bibinfo {year}
  {2014})\BibitemShut {NoStop}%
\bibitem [{\citenamefont {de~Vega}\ and\ \citenamefont
  {Alonso}(2017)}]{deVega2017}%
  \BibitemOpen
  \bibfield  {author} {\bibinfo {author} {\bibfnamefont {I.}~\bibnamefont
  {de~Vega}}\ and\ \bibinfo {author} {\bibfnamefont {D.}~\bibnamefont
  {Alonso}},\ }\href {\doibase 10.1103/RevModPhys.89.015001} {\bibfield
  {journal} {\bibinfo  {journal} {Rev. Mod. Phys.}\ }\textbf {\bibinfo {volume}
  {89}},\ \bibinfo {pages} {015001} (\bibinfo {year} {2017})}\BibitemShut
  {NoStop}%
\bibitem [{\citenamefont {Esposito}\ and\ \citenamefont
  {Gaspard}(2003{\natexlab{a}})}]{Esposito2003a}%
  \BibitemOpen
  \bibfield  {author} {\bibinfo {author} {\bibfnamefont {M.}~\bibnamefont
  {Esposito}}\ and\ \bibinfo {author} {\bibfnamefont {P.}~\bibnamefont
  {Gaspard}},\ }\href {\doibase 10.1103/PhysRevE.68.066112} {\bibfield
  {journal} {\bibinfo  {journal} {Phys. Rev. E}\ }\textbf {\bibinfo {volume}
  {68}},\ \bibinfo {pages} {066112} (\bibinfo {year}
  {2003}{\natexlab{a}})}\BibitemShut {NoStop}%
\bibitem [{\citenamefont {Budini}(2005)}]{Budini2005}%
  \BibitemOpen
  \bibfield  {author} {\bibinfo {author} {\bibfnamefont {A.~A.}\ \bibnamefont
  {Budini}},\ }\href {\doibase 10.1103/PhysRevE.72.056106} {\bibfield
  {journal} {\bibinfo  {journal} {Phys. Rev. E}\ }\textbf {\bibinfo {volume}
  {72}},\ \bibinfo {pages} {056106} (\bibinfo {year} {2005})}\BibitemShut
  {NoStop}%
\bibitem [{\citenamefont {Breuer}\ \emph {et~al.}(2006)\citenamefont {Breuer},
  \citenamefont {Gemmer},\ and\ \citenamefont {Michel}}]{Breuer2006}%
  \BibitemOpen
  \bibfield  {author} {\bibinfo {author} {\bibfnamefont {H.-P.}\ \bibnamefont
  {Breuer}}, \bibinfo {author} {\bibfnamefont {J.}~\bibnamefont {Gemmer}}, \
  and\ \bibinfo {author} {\bibfnamefont {M.}~\bibnamefont {Michel}},\ }\href
  {\doibase 10.1103/PhysRevE.73.016139} {\bibfield  {journal} {\bibinfo
  {journal} {Phys. Rev. E}\ }\textbf {\bibinfo {volume} {73}},\ \bibinfo
  {pages} {016139} (\bibinfo {year} {2006})}\BibitemShut {NoStop}%
\bibitem [{\citenamefont {Budini}(2006)}]{Budini2006}%
  \BibitemOpen
  \bibfield  {author} {\bibinfo {author} {\bibfnamefont {A.~A.}\ \bibnamefont
  {Budini}},\ }\href {\doibase 10.1103/PhysRevA.74.053815} {\bibfield
  {journal} {\bibinfo  {journal} {Phys. Rev. A}\ }\textbf {\bibinfo {volume}
  {74}},\ \bibinfo {pages} {053815} (\bibinfo {year} {2006})}\BibitemShut
  {NoStop}%
\bibitem [{\citenamefont {Breuer}(2007)}]{Breuer2007}%
  \BibitemOpen
  \bibfield  {author} {\bibinfo {author} {\bibfnamefont {H.-P.}\ \bibnamefont
  {Breuer}},\ }\href {\doibase 10.1103/PhysRevA.75.022103} {\bibfield
  {journal} {\bibinfo  {journal} {Phys. Rev. A}\ }\textbf {\bibinfo {volume}
  {75}},\ \bibinfo {pages} {022103} (\bibinfo {year} {2007})}\BibitemShut
  {NoStop}%
\bibitem [{\citenamefont {Esposito}\ and\ \citenamefont
  {Gaspard}(2003{\natexlab{b}})}]{Esposito2003b}%
  \BibitemOpen
  \bibfield  {author} {\bibinfo {author} {\bibfnamefont {M.}~\bibnamefont
  {Esposito}}\ and\ \bibinfo {author} {\bibfnamefont {P.}~\bibnamefont
  {Gaspard}},\ }\href {\doibase 10.1103/PhysRevE.68.066113} {\bibfield
  {journal} {\bibinfo  {journal} {Phys. Rev. E}\ }\textbf {\bibinfo {volume}
  {68}},\ \bibinfo {pages} {066113} (\bibinfo {year}
  {2003}{\natexlab{b}})}\BibitemShut {NoStop}%
\bibitem [{\citenamefont {Fischer}\ and\ \citenamefont
  {Breuer}(2007)}]{Fischer2007}%
  \BibitemOpen
  \bibfield  {author} {\bibinfo {author} {\bibfnamefont {J.}~\bibnamefont
  {Fischer}}\ and\ \bibinfo {author} {\bibfnamefont {H.-P.}\ \bibnamefont
  {Breuer}},\ }\href {\doibase 10.1103/PhysRevA.76.052119} {\bibfield
  {journal} {\bibinfo  {journal} {Phys. Rev. A}\ }\textbf {\bibinfo {volume}
  {76}},\ \bibinfo {pages} {052119} (\bibinfo {year} {2007})}\BibitemShut
  {NoStop}%
\bibitem [{\citenamefont {von Neumann}(1929)}]{vonNeumann1929_ger}%
  \BibitemOpen
  \bibfield  {author} {\bibinfo {author} {\bibfnamefont {J.}~\bibnamefont {von
  Neumann}},\ }\href {\doibase 10.1007/BF01339852} {\bibfield  {journal}
  {\bibinfo  {journal} {Z. Phys.}\ }\textbf {\bibinfo {volume} {57}},\ \bibinfo
  {pages} {30} (\bibinfo {year} {1929})}\BibitemShut {NoStop}%
\bibitem [{\citenamefont {von Neumann}(2010)}]{vonNeumann1929_en}%
  \BibitemOpen
  \bibfield  {author} {\bibinfo {author} {\bibfnamefont {J.}~\bibnamefont {von
  Neumann}},\ }\href {\doibase 10.1140/epjh/e2010-00008-5} {\bibfield
  {journal} {\bibinfo  {journal} {{E}ur. {P}hys. {J}. {H}}\ }\textbf {\bibinfo
  {volume} {35}},\ \bibinfo {pages} {201} (\bibinfo {year} {2010})}\BibitemShut
  {NoStop}%
\bibitem [{\citenamefont {Reimann}(2008)}]{Reimann2008}%
  \BibitemOpen
  \bibfield  {author} {\bibinfo {author} {\bibfnamefont {P.}~\bibnamefont
  {Reimann}},\ }\href {\doibase 10.1103/PhysRevLett.101.190403} {\bibfield
  {journal} {\bibinfo  {journal} {Phys. Rev. Lett.}\ }\textbf {\bibinfo
  {volume} {101}},\ \bibinfo {pages} {190403} (\bibinfo {year}
  {2008})}\BibitemShut {NoStop}%
\bibitem [{\citenamefont {Venuti}(2015)}]{Venuti2015}%
  \BibitemOpen
  \bibfield  {author} {\bibinfo {author} {\bibfnamefont {L.~C.}\ \bibnamefont
  {Venuti}},\ }\href {https://arxiv.org/abs/1509.04352} {\bibfield  {journal}
  {\bibinfo  {journal} {arXiv:1509.04352}\ } (\bibinfo {year}
  {2015})}\BibitemShut {NoStop}%
\bibitem [{\citenamefont {Strasberg}(2019)}]{Strasberg2019}%
  \BibitemOpen
  \bibfield  {author} {\bibinfo {author} {\bibfnamefont {P.}~\bibnamefont
  {Strasberg}},\ }\href {https://arxiv.org/abs/1906.09933} {\bibfield
  {journal} {\bibinfo  {journal} {arXiv:1906.09933}\ } (\bibinfo {year}
  {2019})}\BibitemShut {NoStop}%
\bibitem [{\citenamefont {Esposito}\ and\ \citenamefont
  {Gaspard}(2007)}]{Esposito2007}%
  \BibitemOpen
  \bibfield  {author} {\bibinfo {author} {\bibfnamefont {M.}~\bibnamefont
  {Esposito}}\ and\ \bibinfo {author} {\bibfnamefont {P.}~\bibnamefont
  {Gaspard}},\ }\href {\doibase 10.1103/PhysRevE.76.041134} {\bibfield
  {journal} {\bibinfo  {journal} {Phys. Rev. E}\ }\textbf {\bibinfo {volume}
  {76}},\ \bibinfo {pages} {041134} (\bibinfo {year} {2007})}\BibitemShut
  {NoStop}%
\bibitem [{\citenamefont {Beenakker}(1997)}]{Beenakker1997}%
  \BibitemOpen
  \bibfield  {author} {\bibinfo {author} {\bibfnamefont {C.~W.~J.}\
  \bibnamefont {Beenakker}},\ }\href {\doibase 10.1103/RevModPhys.69.731}
  {\bibfield  {journal} {\bibinfo  {journal} {Rev. Mod. Phys.}\ }\textbf
  {\bibinfo {volume} {69}},\ \bibinfo {pages} {731} (\bibinfo {year}
  {1997})}\BibitemShut {NoStop}%
\bibitem [{\citenamefont {Gelbart}\ \emph {et~al.}(1972)\citenamefont
  {Gelbart}, \citenamefont {Rice},\ and\ \citenamefont {Freed}}]{Gelbart1972}%
  \BibitemOpen
  \bibfield  {author} {\bibinfo {author} {\bibfnamefont {W.~M.}\ \bibnamefont
  {Gelbart}}, \bibinfo {author} {\bibfnamefont {S.~A.}\ \bibnamefont {Rice}}, \
  and\ \bibinfo {author} {\bibfnamefont {K.~F.}\ \bibnamefont {Freed}},\ }\href
  {\doibase https://doi.org/10.1063/1.1678139} {\bibfield  {journal} {\bibinfo
  {journal} {{J}. {C}hem. {P}hys.}\ }\textbf {\bibinfo {volume} {57}},\
  \bibinfo {pages} {4699} (\bibinfo {year} {1972})}\BibitemShut {NoStop}%
\bibitem [{\citenamefont {Mello}\ \emph {et~al.}(1988)\citenamefont {Mello},
  \citenamefont {Pereyra},\ and\ \citenamefont {Kumar}}]{Mello1988}%
  \BibitemOpen
  \bibfield  {author} {\bibinfo {author} {\bibfnamefont {P.~A.}\ \bibnamefont
  {Mello}}, \bibinfo {author} {\bibfnamefont {P.}~\bibnamefont {Pereyra}}, \
  and\ \bibinfo {author} {\bibfnamefont {N.}~\bibnamefont {Kumar}},\ }\href
  {\doibase 10.1007/BF01015321} {\bibfield  {journal} {\bibinfo  {journal} {J.
  Stat. Phys.}\ }\textbf {\bibinfo {volume} {51}},\ \bibinfo {pages} {77}
  (\bibinfo {year} {1988})}\BibitemShut {NoStop}%
\bibitem [{\citenamefont {Pereyra}(1991)}]{Pereyra1991}%
  \BibitemOpen
  \bibfield  {author} {\bibinfo {author} {\bibfnamefont {P.}~\bibnamefont
  {Pereyra}},\ }\href {\doibase https://doi.org/10.1007/BF01053754} {\bibfield
  {journal} {\bibinfo  {journal} {J. Stat. Phys.}\ }\textbf {\bibinfo {volume}
  {65}},\ \bibinfo {pages} {773} (\bibinfo {year} {1991})}\BibitemShut
  {NoStop}%
\bibitem [{\citenamefont {Cohen}\ \emph {et~al.}(2000)\citenamefont {Cohen},
  \citenamefont {Izrailev},\ and\ \citenamefont {Kottos}}]{Cohen2000}%
  \BibitemOpen
  \bibfield  {author} {\bibinfo {author} {\bibfnamefont {D.}~\bibnamefont
  {Cohen}}, \bibinfo {author} {\bibfnamefont {F.~M.}\ \bibnamefont {Izrailev}},
  \ and\ \bibinfo {author} {\bibfnamefont {T.}~\bibnamefont {Kottos}},\ }\href
  {\doibase 10.1103/PhysRevLett.84.2052} {\bibfield  {journal} {\bibinfo
  {journal} {Phys. Rev. Lett.}\ }\textbf {\bibinfo {volume} {84}},\ \bibinfo
  {pages} {2052} (\bibinfo {year} {2000})}\BibitemShut {NoStop}%
\bibitem [{\citenamefont {Lebowitz}\ and\ \citenamefont
  {Pastur}(2004)}]{Lebowitz2004}%
  \BibitemOpen
  \bibfield  {author} {\bibinfo {author} {\bibfnamefont {J.~L.}\ \bibnamefont
  {Lebowitz}}\ and\ \bibinfo {author} {\bibfnamefont {L.}~\bibnamefont
  {Pastur}},\ }\href {\doibase 10.1088/0305-4470/37/5/004} {\bibfield
  {journal} {\bibinfo  {journal} {J. Phys. A}\ }\textbf {\bibinfo {volume}
  {37}},\ \bibinfo {pages} {1517} (\bibinfo {year} {2004})}\BibitemShut
  {NoStop}%
\bibitem [{\citenamefont {Cohen}\ and\ \citenamefont
  {Kottos}(2004)}]{Cohen2004}%
  \BibitemOpen
  \bibfield  {author} {\bibinfo {author} {\bibfnamefont {D.}~\bibnamefont
  {Cohen}}\ and\ \bibinfo {author} {\bibfnamefont {T.}~\bibnamefont {Kottos}},\
  }\href {\doibase 10.1103/PhysRevE.69.055201} {\bibfield  {journal} {\bibinfo
  {journal} {Phys. Rev. E}\ }\textbf {\bibinfo {volume} {69}},\ \bibinfo
  {pages} {055201(R)} (\bibinfo {year} {2004})}\BibitemShut {NoStop}%
\bibitem [{\citenamefont {Nation}\ and\ \citenamefont
  {Porras}(2019)}]{Nation2019}%
  \BibitemOpen
  \bibfield  {author} {\bibinfo {author} {\bibfnamefont {C.}~\bibnamefont
  {Nation}}\ and\ \bibinfo {author} {\bibfnamefont {D.}~\bibnamefont
  {Porras}},\ }\href {\doibase 10.1103/PhysRevE.99.052139} {\bibfield
  {journal} {\bibinfo  {journal} {Phys. Rev. E}\ }\textbf {\bibinfo {volume}
  {99}},\ \bibinfo {pages} {052139} (\bibinfo {year} {2019})}\BibitemShut
  {NoStop}%
\bibitem [{\citenamefont {Reimann}(2015)}]{Reimann2015}%
  \BibitemOpen
  \bibfield  {author} {\bibinfo {author} {\bibfnamefont {P.}~\bibnamefont
  {Reimann}},\ }\href {\doibase 10.1088/1367-2630/17/5/055025} {\bibfield
  {journal} {\bibinfo  {journal} {New J. Phys.}\ }\textbf {\bibinfo {volume}
  {17}},\ \bibinfo {pages} {055025} (\bibinfo {year} {2015})}\BibitemShut
  {NoStop}%
\bibitem [{\citenamefont {Deutsch}(1991)}]{Deutsch1991}%
  \BibitemOpen
  \bibfield  {author} {\bibinfo {author} {\bibfnamefont {J.~M.}\ \bibnamefont
  {Deutsch}},\ }\href {\doibase 10.1103/PhysRevA.43.2046} {\bibfield  {journal}
  {\bibinfo  {journal} {Phys. Rev. A}\ }\textbf {\bibinfo {volume} {43}},\
  \bibinfo {pages} {2046} (\bibinfo {year} {1991})}\BibitemShut {NoStop}%
\bibitem [{\citenamefont {Srednicki}(1994)}]{Srednicki1994}%
  \BibitemOpen
  \bibfield  {author} {\bibinfo {author} {\bibfnamefont {M.}~\bibnamefont
  {Srednicki}},\ }\href {\doibase 10.1103/PhysRevE.50.888} {\bibfield
  {journal} {\bibinfo  {journal} {Phys. Rev. E}\ }\textbf {\bibinfo {volume}
  {50}},\ \bibinfo {pages} {888} (\bibinfo {year} {1994})}\BibitemShut
  {NoStop}%
\bibitem [{\citenamefont {D'Alessio}\ \emph {et~al.}(2016)\citenamefont
  {D'Alessio}, \citenamefont {Kafri}, \citenamefont {Polkovnikov},\ and\
  \citenamefont {Rigol}}]{dAlessio2016}%
  \BibitemOpen
  \bibfield  {author} {\bibinfo {author} {\bibfnamefont {L.}~\bibnamefont
  {D'Alessio}}, \bibinfo {author} {\bibfnamefont {Y.}~\bibnamefont {Kafri}},
  \bibinfo {author} {\bibfnamefont {A.}~\bibnamefont {Polkovnikov}}, \ and\
  \bibinfo {author} {\bibfnamefont {M.}~\bibnamefont {Rigol}},\ }\href
  {\doibase 10.1080/00018732.2016.1198134} {\bibfield  {journal} {\bibinfo
  {journal} {{A}dv. {P}hys.}\ }\textbf {\bibinfo {volume} {65}},\ \bibinfo
  {pages} {239} (\bibinfo {year} {2016})}\BibitemShut {NoStop}%
\bibitem [{\citenamefont {Deutsch}(2018)}]{Deutsch2018}%
  \BibitemOpen
  \bibfield  {author} {\bibinfo {author} {\bibfnamefont {J.~M.}\ \bibnamefont
  {Deutsch}},\ }\href {\doibase 10.1088/1361-6633/aac9f1} {\bibfield  {journal}
  {\bibinfo  {journal} {{R}ep. {P}rog. {P}hys.}\ }\textbf {\bibinfo {volume}
  {81}},\ \bibinfo {pages} {082001} (\bibinfo {year} {2018})}\BibitemShut
  {NoStop}%
\bibitem [{\citenamefont {Maes}\ and\ \citenamefont
  {Neto{\v{c}}n{\`y}}(2003)}]{Maes2003}%
  \BibitemOpen
  \bibfield  {author} {\bibinfo {author} {\bibfnamefont {C.}~\bibnamefont
  {Maes}}\ and\ \bibinfo {author} {\bibfnamefont {K.}~\bibnamefont
  {Neto{\v{c}}n{\`y}}},\ }\href {\doibase 10.1023/A:1021026930129} {\bibfield
  {journal} {\bibinfo  {journal} {J. {S}tat. {P}hys.}\ }\textbf {\bibinfo
  {volume} {110}},\ \bibinfo {pages} {269} (\bibinfo {year}
  {2003})}\BibitemShut {NoStop}%
\bibitem [{\citenamefont {Braun}\ \emph {et~al.}(2013)\citenamefont {Braun},
  \citenamefont {Ronzheimer}, \citenamefont {Schreiber}, \citenamefont
  {Hodgman}, \citenamefont {Rom}, \citenamefont {Bloch},\ and\ \citenamefont
  {Schneider}}]{Braun2013}%
  \BibitemOpen
  \bibfield  {author} {\bibinfo {author} {\bibfnamefont {S.}~\bibnamefont
  {Braun}}, \bibinfo {author} {\bibfnamefont {J.~P.}\ \bibnamefont
  {Ronzheimer}}, \bibinfo {author} {\bibfnamefont {M.}~\bibnamefont
  {Schreiber}}, \bibinfo {author} {\bibfnamefont {S.~S.}\ \bibnamefont
  {Hodgman}}, \bibinfo {author} {\bibfnamefont {T.}~\bibnamefont {Rom}},
  \bibinfo {author} {\bibfnamefont {I.}~\bibnamefont {Bloch}}, \ and\ \bibinfo
  {author} {\bibfnamefont {U.}~\bibnamefont {Schneider}},\ }\href {\doibase
  10.1126/science.1227831} {\bibfield  {journal} {\bibinfo  {journal}
  {Science}\ }\textbf {\bibinfo {volume} {339}},\ \bibinfo {pages} {52}
  (\bibinfo {year} {2013})}\BibitemShut {NoStop}%
\bibitem [{\citenamefont {Romero-Roch\'{\i}n}(2013)}]{RomeroRochin2013}%
  \BibitemOpen
  \bibfield  {author} {\bibinfo {author} {\bibfnamefont {V.}~\bibnamefont
  {Romero-Roch\'{\i}n}},\ }\href {\doibase 10.1103/PhysRevE.88.022144}
  {\bibfield  {journal} {\bibinfo  {journal} {Phys. Rev. E}\ }\textbf {\bibinfo
  {volume} {88}},\ \bibinfo {pages} {022144} (\bibinfo {year}
  {2013})}\BibitemShut {NoStop}%
\bibitem [{\citenamefont {Dunkel}\ and\ \citenamefont
  {Hilbert}(2014)}]{Dunkel2014}%
  \BibitemOpen
  \bibfield  {author} {\bibinfo {author} {\bibfnamefont {J.}~\bibnamefont
  {Dunkel}}\ and\ \bibinfo {author} {\bibfnamefont {S.}~\bibnamefont
  {Hilbert}},\ }\href {\doibase https://doi.org/10.1038/nphys2815} {\bibfield
  {journal} {\bibinfo  {journal} {{N}at. {P}hys.}\ }\textbf {\bibinfo {volume}
  {10}},\ \bibinfo {pages} {67} (\bibinfo {year} {2014})}\BibitemShut {NoStop}%
\bibitem [{\citenamefont {Campisi}(2015)}]{Campisi2015}%
  \BibitemOpen
  \bibfield  {author} {\bibinfo {author} {\bibfnamefont {M.}~\bibnamefont
  {Campisi}},\ }\href {\doibase 10.1103/PhysRevE.91.052147} {\bibfield
  {journal} {\bibinfo  {journal} {Phys. Rev. E}\ }\textbf {\bibinfo {volume}
  {91}},\ \bibinfo {pages} {052147} (\bibinfo {year} {2015})}\BibitemShut
  {NoStop}%
\bibitem [{\citenamefont {Abraham}\ and\ \citenamefont
  {Penrose}(2017)}]{Eitan2017}%
  \BibitemOpen
  \bibfield  {author} {\bibinfo {author} {\bibfnamefont {E.}~\bibnamefont
  {Abraham}}\ and\ \bibinfo {author} {\bibfnamefont {O.}~\bibnamefont
  {Penrose}},\ }\href {\doibase 10.1103/PhysRevE.95.012125} {\bibfield
  {journal} {\bibinfo  {journal} {Phys. Rev. E}\ }\textbf {\bibinfo {volume}
  {95}},\ \bibinfo {pages} {012125} (\bibinfo {year} {2017})}\BibitemShut
  {NoStop}%
\bibitem [{\citenamefont {Swendsen}(2018)}]{Swendsen2018}%
  \BibitemOpen
  \bibfield  {author} {\bibinfo {author} {\bibfnamefont {R.~H.}\ \bibnamefont
  {Swendsen}},\ }\href {\doibase 10.1088/1361-6633/aac18c} {\bibfield
  {journal} {\bibinfo  {journal} {{R}ep. {P}rog. {P}hys.}\ }\textbf {\bibinfo
  {volume} {81}},\ \bibinfo {pages} {072001} (\bibinfo {year}
  {2018})}\BibitemShut {NoStop}%
\bibitem [{\citenamefont {Schneider}\ \emph {et~al.}(2014)\citenamefont
  {Schneider}, \citenamefont {Mandt}, \citenamefont {Rapp}, \citenamefont
  {Braun}, \citenamefont {Weimer}, \citenamefont {Bloch},\ and\ \citenamefont
  {Rosch}}]{Schneider2014}%
  \BibitemOpen
  \bibfield  {author} {\bibinfo {author} {\bibfnamefont {U.}~\bibnamefont
  {Schneider}}, \bibinfo {author} {\bibfnamefont {S.}~\bibnamefont {Mandt}},
  \bibinfo {author} {\bibfnamefont {A.}~\bibnamefont {Rapp}}, \bibinfo {author}
  {\bibfnamefont {S.}~\bibnamefont {Braun}}, \bibinfo {author} {\bibfnamefont
  {H.}~\bibnamefont {Weimer}}, \bibinfo {author} {\bibfnamefont
  {I.}~\bibnamefont {Bloch}}, \ and\ \bibinfo {author} {\bibfnamefont
  {A.}~\bibnamefont {Rosch}},\ }\href {https://arxiv.org/abs/1407.4127}
  {\bibfield  {journal} {\bibinfo  {journal} {arXiv:1407.4127}\ } (\bibinfo
  {year} {2014})}\BibitemShut {NoStop}%
\bibitem [{\citenamefont {{M}itchison}\ and\ \citenamefont
  {{P}lenio}(2018)}]{Mitchison2018}%
  \BibitemOpen
  \bibfield  {author} {\bibinfo {author} {\bibfnamefont {M.~T.}\ \bibnamefont
  {{M}itchison}}\ and\ \bibinfo {author} {\bibfnamefont {M.~B.}\ \bibnamefont
  {{P}lenio}},\ }\href {\doibase 10.1088/1367-2630/aa9f70} {\bibfield
  {journal} {\bibinfo  {journal} {New {J}. {P}hys.}\ }\textbf {\bibinfo
  {volume} {20}},\ \bibinfo {pages} {033005} (\bibinfo {year}
  {2018})}\BibitemShut {NoStop}%
\bibitem [{\citenamefont {Braun}\ \emph {et~al.}(2001)\citenamefont {Braun},
  \citenamefont {Haake},\ and\ \citenamefont {Strunz}}]{Braun2001}%
  \BibitemOpen
  \bibfield  {author} {\bibinfo {author} {\bibfnamefont {D.}~\bibnamefont
  {Braun}}, \bibinfo {author} {\bibfnamefont {F.}~\bibnamefont {Haake}}, \ and\
  \bibinfo {author} {\bibfnamefont {W.~T.}\ \bibnamefont {Strunz}},\ }\href
  {\doibase 10.1103/PhysRevLett.86.2913} {\bibfield  {journal} {\bibinfo
  {journal} {Phys. Rev. Lett.}\ }\textbf {\bibinfo {volume} {86}},\ \bibinfo
  {pages} {2913} (\bibinfo {year} {2001})}\BibitemShut {NoStop}%
\bibitem [{\citenamefont {\ifmmode~\check{S}\else \v{S}\fi{}afr\'anek}\ \emph
  {et~al.}(2019{\natexlab{a}})\citenamefont {\ifmmode~\check{S}\else
  \v{S}\fi{}afr\'anek}, \citenamefont {Deutsch},\ and\ \citenamefont
  {Aguirre}}]{Safranek2019_1}%
  \BibitemOpen
  \bibfield  {author} {\bibinfo {author} {\bibfnamefont {D.}~\bibnamefont
  {\ifmmode~\check{S}\else \v{S}\fi{}afr\'anek}}, \bibinfo {author}
  {\bibfnamefont {J.~M.}\ \bibnamefont {Deutsch}}, \ and\ \bibinfo {author}
  {\bibfnamefont {A.}~\bibnamefont {Aguirre}},\ }\href {\doibase
  10.1103/PhysRevA.99.010101} {\bibfield  {journal} {\bibinfo  {journal} {Phys.
  Rev. A}\ }\textbf {\bibinfo {volume} {99}},\ \bibinfo {pages} {010101(R)}
  (\bibinfo {year} {2019}{\natexlab{a}})}\BibitemShut {NoStop}%
\bibitem [{\citenamefont {\ifmmode~\check{S}\else \v{S}\fi{}afr\'anek}\ \emph
  {et~al.}(2019{\natexlab{b}})\citenamefont {\ifmmode~\check{S}\else
  \v{S}\fi{}afr\'anek}, \citenamefont {Deutsch},\ and\ \citenamefont
  {Aguirre}}]{Safranek2019_2}%
  \BibitemOpen
  \bibfield  {author} {\bibinfo {author} {\bibfnamefont {D.}~\bibnamefont
  {\ifmmode~\check{S}\else \v{S}\fi{}afr\'anek}}, \bibinfo {author}
  {\bibfnamefont {J.~M.}\ \bibnamefont {Deutsch}}, \ and\ \bibinfo {author}
  {\bibfnamefont {A.}~\bibnamefont {Aguirre}},\ }\href {\doibase
  10.1103/PhysRevA.99.012103} {\bibfield  {journal} {\bibinfo  {journal} {Phys.
  Rev. A}\ }\textbf {\bibinfo {volume} {99}},\ \bibinfo {pages} {012103}
  (\bibinfo {year} {2019}{\natexlab{b}})}\BibitemShut {NoStop}%
\bibitem [{\citenamefont {Schindler}\ \emph {et~al.}(2020)\citenamefont
  {Schindler}, \citenamefont {{\v{S}}afr{\'a}nek},\ and\ \citenamefont
  {Aguirre}}]{Schindler2020}%
  \BibitemOpen
  \bibfield  {author} {\bibinfo {author} {\bibfnamefont {J.}~\bibnamefont
  {Schindler}}, \bibinfo {author} {\bibfnamefont {D.}~\bibnamefont
  {{\v{S}}afr{\'a}nek}}, \ and\ \bibinfo {author} {\bibfnamefont
  {A.}~\bibnamefont {Aguirre}},\ }\href {https://arxiv.org/abs/2005.05408}
  {\bibfield  {journal} {\bibinfo  {journal} {ar{X}iv:2005.05408}\ } (\bibinfo
  {year} {2020})}\BibitemShut {NoStop}%
\bibitem [{\citenamefont {Strasberg}\ and\ \citenamefont
  {Winter}(2020)}]{Strasberg2020}%
  \BibitemOpen
  \bibfield  {author} {\bibinfo {author} {\bibfnamefont {P.}~\bibnamefont
  {Strasberg}}\ and\ \bibinfo {author} {\bibfnamefont {A.}~\bibnamefont
  {Winter}},\ }\href {https://arxiv.org/abs/2002.08817} {\bibfield  {journal}
  {\bibinfo  {journal} {arXiv:2002.08817}\ } (\bibinfo {year}
  {2020})}\BibitemShut {NoStop}%
\bibitem [{\citenamefont {Strasberg}\ and\ \citenamefont
  {Esposito}(2019)}]{Strasberg2019_b}%
  \BibitemOpen
  \bibfield  {author} {\bibinfo {author} {\bibfnamefont {P.}~\bibnamefont
  {Strasberg}}\ and\ \bibinfo {author} {\bibfnamefont {M.}~\bibnamefont
  {Esposito}},\ }\href {\doibase 10.1103/PhysRevE.99.012120} {\bibfield
  {journal} {\bibinfo  {journal} {Phys. Rev. E}\ }\textbf {\bibinfo {volume}
  {99}},\ \bibinfo {pages} {012120} (\bibinfo {year} {2019})}\BibitemShut
  {NoStop}%
\bibitem [{\citenamefont {Esposito}\ \emph {et~al.}(2009)\citenamefont
  {Esposito}, \citenamefont {Harbola},\ and\ \citenamefont
  {Mukamel}}]{Esposito2009}%
  \BibitemOpen
  \bibfield  {author} {\bibinfo {author} {\bibfnamefont {M.}~\bibnamefont
  {Esposito}}, \bibinfo {author} {\bibfnamefont {U.}~\bibnamefont {Harbola}}, \
  and\ \bibinfo {author} {\bibfnamefont {S.}~\bibnamefont {Mukamel}},\ }\href
  {\doibase 10.1103/RevModPhys.81.1665} {\bibfield  {journal} {\bibinfo
  {journal} {Rev. Mod. Phys.}\ }\textbf {\bibinfo {volume} {81}},\ \bibinfo
  {pages} {1665} (\bibinfo {year} {2009})}\BibitemShut {NoStop}%
\bibitem [{\citenamefont {Kelly}\ and\ \citenamefont
  {Markland}(2013)}]{Kelly2013}%
  \BibitemOpen
  \bibfield  {author} {\bibinfo {author} {\bibfnamefont {A.}~\bibnamefont
  {Kelly}}\ and\ \bibinfo {author} {\bibfnamefont {T.~E.}\ \bibnamefont
  {Markland}},\ }\href {\doibase 10.1063/1.4812355} {\bibfield  {journal}
  {\bibinfo  {journal} {{J}. {C}hem. {P}hys.}\ }\textbf {\bibinfo {volume}
  {139}},\ \bibinfo {pages} {014104} (\bibinfo {year} {2013})}\BibitemShut
  {NoStop}%
\bibitem [{\citenamefont {Kelly}\ \emph {et~al.}(2015)\citenamefont {Kelly},
  \citenamefont {Brackbill},\ and\ \citenamefont {Markland}}]{Kelly2015}%
  \BibitemOpen
  \bibfield  {author} {\bibinfo {author} {\bibfnamefont {A.}~\bibnamefont
  {Kelly}}, \bibinfo {author} {\bibfnamefont {N.}~\bibnamefont {Brackbill}}, \
  and\ \bibinfo {author} {\bibfnamefont {T.~E.}\ \bibnamefont {Markland}},\
  }\href {\doibase 10.1063/1.4913686} {\bibfield  {journal} {\bibinfo
  {journal} {{J}. {C}hem. {P}hys.}\ }\textbf {\bibinfo {volume} {142}},\
  \bibinfo {pages} {094110} (\bibinfo {year} {2015})}\BibitemShut {NoStop}%
\bibitem [{\citenamefont {Brandes}(2005)}]{Brandes2005}%
  \BibitemOpen
  \bibfield  {author} {\bibinfo {author} {\bibfnamefont {T.}~\bibnamefont
  {Brandes}},\ }\href {\doibase 10.1016/j.physrep.2004.12.002} {\bibfield
  {journal} {\bibinfo  {journal} {{P}hys. {R}ep.}\ }\textbf {\bibinfo {volume}
  {408}},\ \bibinfo {pages} {315} (\bibinfo {year} {2005})}\BibitemShut
  {NoStop}%
\bibitem [{\citenamefont {Segal}(2006)}]{Segal2006}%
  \BibitemOpen
  \bibfield  {author} {\bibinfo {author} {\bibfnamefont {D.}~\bibnamefont
  {Segal}},\ }\href {\doibase 10.1103/PhysRevB.73.205415} {\bibfield  {journal}
  {\bibinfo  {journal} {Phys. Rev. B}\ }\textbf {\bibinfo {volume} {73}},\
  \bibinfo {pages} {205415} (\bibinfo {year} {2006})}\BibitemShut {NoStop}%
\bibitem [{\citenamefont {Schaller}\ \emph {et~al.}(2013)\citenamefont
  {Schaller}, \citenamefont {Krause}, \citenamefont {Brandes},\ and\
  \citenamefont {Esposito}}]{Schaller2013}%
  \BibitemOpen
  \bibfield  {author} {\bibinfo {author} {\bibfnamefont {G.}~\bibnamefont
  {Schaller}}, \bibinfo {author} {\bibfnamefont {T.}~\bibnamefont {Krause}},
  \bibinfo {author} {\bibfnamefont {T.}~\bibnamefont {Brandes}}, \ and\
  \bibinfo {author} {\bibfnamefont {M.}~\bibnamefont {Esposito}},\ }\href
  {\doibase 10.1088/1367-2630/15/3/033032} {\bibfield  {journal} {\bibinfo
  {journal} {{N}ew {J}. {P}hys.}\ }\textbf {\bibinfo {volume} {15}},\ \bibinfo
  {pages} {033032} (\bibinfo {year} {2013})}\BibitemShut {NoStop}%
\bibitem [{\citenamefont {Gelbwaser-Klimovsky}\ and\ \citenamefont
  {Aspuru-Guzik}(2015)}]{Gelbwaser2015}%
  \BibitemOpen
  \bibfield  {author} {\bibinfo {author} {\bibfnamefont {D.}~\bibnamefont
  {Gelbwaser-Klimovsky}}\ and\ \bibinfo {author} {\bibfnamefont
  {A.}~\bibnamefont {Aspuru-Guzik}},\ }\href {\doibase
  10.1021/acs.jpclett.5b01404} {\bibfield  {journal} {\bibinfo  {journal} {J.
  Phys. Chem. Lett.}\ }\textbf {\bibinfo {volume} {6}},\ \bibinfo {pages}
  {3477} (\bibinfo {year} {2015})}\BibitemShut {NoStop}%
\bibitem [{\citenamefont {Wang}\ \emph {et~al.}(2015)\citenamefont {Wang},
  \citenamefont {Ren},\ and\ \citenamefont {Cao}}]{Wang2015}%
  \BibitemOpen
  \bibfield  {author} {\bibinfo {author} {\bibfnamefont {C.}~\bibnamefont
  {Wang}}, \bibinfo {author} {\bibfnamefont {J.}~\bibnamefont {Ren}}, \ and\
  \bibinfo {author} {\bibfnamefont {J.}~\bibnamefont {Cao}},\ }\href {\doibase
  10.1038/srep11787} {\bibfield  {journal} {\bibinfo  {journal} {{S}ci.
  {R}ep.}\ }\textbf {\bibinfo {volume} {5}},\ \bibinfo {pages} {11787}
  (\bibinfo {year} {2015})}\BibitemShut {NoStop}%
\bibitem [{\citenamefont {Hughes}\ \emph
  {et~al.}(2009{\natexlab{a}})\citenamefont {Hughes}, \citenamefont {Christ},\
  and\ \citenamefont {Burghardt}}]{Hughes2009a}%
  \BibitemOpen
  \bibfield  {author} {\bibinfo {author} {\bibfnamefont {K.~H.}\ \bibnamefont
  {Hughes}}, \bibinfo {author} {\bibfnamefont {C.~D.}\ \bibnamefont {Christ}},
  \ and\ \bibinfo {author} {\bibfnamefont {I.}~\bibnamefont {Burghardt}},\
  }\href {\doibase 10.1063/1.3159671} {\bibfield  {journal} {\bibinfo
  {journal} {J. {C}hem. {P}hys}\ }\textbf {\bibinfo {volume} {131}},\ \bibinfo
  {pages} {024109} (\bibinfo {year} {2009}{\natexlab{a}})}\BibitemShut
  {NoStop}%
\bibitem [{\citenamefont {Martinazzo}\ \emph {et~al.}(2011)\citenamefont
  {Martinazzo}, \citenamefont {Vacchini}, \citenamefont {Hughes},\ and\
  \citenamefont {Burghardt}}]{Martinazzo2011}%
  \BibitemOpen
  \bibfield  {author} {\bibinfo {author} {\bibfnamefont {R.}~\bibnamefont
  {Martinazzo}}, \bibinfo {author} {\bibfnamefont {B.}~\bibnamefont
  {Vacchini}}, \bibinfo {author} {\bibfnamefont {K.~H.}\ \bibnamefont
  {Hughes}}, \ and\ \bibinfo {author} {\bibfnamefont {I.}~\bibnamefont
  {Burghardt}},\ }\href {\doibase 10.1063/1.3532408} {\bibfield  {journal}
  {\bibinfo  {journal} {J. {C}hem. {P}hys.}\ }\textbf {\bibinfo {volume}
  {134}},\ \bibinfo {pages} {011101} (\bibinfo {year} {2011})}\BibitemShut
  {NoStop}%
\bibitem [{\citenamefont {Woods}\ \emph {et~al.}(2014)\citenamefont {Woods},
  \citenamefont {Groux}, \citenamefont {Chin}, \citenamefont {Huelga},\ and\
  \citenamefont {Plenio}}]{Woods2014}%
  \BibitemOpen
  \bibfield  {author} {\bibinfo {author} {\bibfnamefont {M.}~\bibnamefont
  {Woods}}, \bibinfo {author} {\bibfnamefont {R.}~\bibnamefont {Groux}},
  \bibinfo {author} {\bibfnamefont {A.}~\bibnamefont {Chin}}, \bibinfo {author}
  {\bibfnamefont {S.~F.}\ \bibnamefont {Huelga}}, \ and\ \bibinfo {author}
  {\bibfnamefont {M.~B.}\ \bibnamefont {Plenio}},\ }\href {\doibase
  10.1063/1.4866769} {\bibfield  {journal} {\bibinfo  {journal} {{J}. {M}ath.
  {P}hys.}\ }\textbf {\bibinfo {volume} {55}},\ \bibinfo {pages} {032101}
  (\bibinfo {year} {2014})}\BibitemShut {NoStop}%
\bibitem [{\citenamefont {Hughes}\ \emph
  {et~al.}(2009{\natexlab{b}})\citenamefont {Hughes}, \citenamefont {Christ},\
  and\ \citenamefont {Burghardt}}]{Hughes2009b}%
  \BibitemOpen
  \bibfield  {author} {\bibinfo {author} {\bibfnamefont {K.~H.}\ \bibnamefont
  {Hughes}}, \bibinfo {author} {\bibfnamefont {C.~D.}\ \bibnamefont {Christ}},
  \ and\ \bibinfo {author} {\bibfnamefont {I.}~\bibnamefont {Burghardt}},\
  }\href {\doibase 10.1063/1.3226343} {\bibfield  {journal} {\bibinfo
  {journal} {J. {C}hem. {P}hys}\ }\textbf {\bibinfo {volume} {131}},\ \bibinfo
  {pages} {124108} (\bibinfo {year} {2009}{\natexlab{b}})}\BibitemShut
  {NoStop}%
\bibitem [{\citenamefont {Iles-Smith}\ \emph {et~al.}(2014)\citenamefont
  {Iles-Smith}, \citenamefont {Lambert},\ and\ \citenamefont
  {Nazir}}]{IlesSmith2014}%
  \BibitemOpen
  \bibfield  {author} {\bibinfo {author} {\bibfnamefont {J.}~\bibnamefont
  {Iles-Smith}}, \bibinfo {author} {\bibfnamefont {N.}~\bibnamefont {Lambert}},
  \ and\ \bibinfo {author} {\bibfnamefont {A.}~\bibnamefont {Nazir}},\ }\href
  {\doibase 10.1103/PhysRevA.90.032114} {\bibfield  {journal} {\bibinfo
  {journal} {Phys. Rev. A}\ }\textbf {\bibinfo {volume} {90}},\ \bibinfo
  {pages} {032114} (\bibinfo {year} {2014})}\BibitemShut {NoStop}%
\bibitem [{\citenamefont {Strasberg}\ \emph {et~al.}(2016)\citenamefont
  {Strasberg}, \citenamefont {Schaller}, \citenamefont {Lambert},\ and\
  \citenamefont {Brandes}}]{Strasberg2016}%
  \BibitemOpen
  \bibfield  {author} {\bibinfo {author} {\bibfnamefont {P.}~\bibnamefont
  {Strasberg}}, \bibinfo {author} {\bibfnamefont {G.}~\bibnamefont {Schaller}},
  \bibinfo {author} {\bibfnamefont {N.}~\bibnamefont {Lambert}}, \ and\
  \bibinfo {author} {\bibfnamefont {T.}~\bibnamefont {Brandes}},\ }\href
  {\doibase 10.1088/1367-2630/18/7/073007} {\bibfield  {journal} {\bibinfo
  {journal} {{N}ew {J}. {P}hys.}\ }\textbf {\bibinfo {volume} {18}},\ \bibinfo
  {pages} {073007} (\bibinfo {year} {2016})}\BibitemShut {NoStop}%
\bibitem [{\citenamefont {Newman}\ \emph {et~al.}(2017)\citenamefont {Newman},
  \citenamefont {Mintert},\ and\ \citenamefont {Nazir}}]{Newman2017}%
  \BibitemOpen
  \bibfield  {author} {\bibinfo {author} {\bibfnamefont {D.}~\bibnamefont
  {Newman}}, \bibinfo {author} {\bibfnamefont {F.}~\bibnamefont {Mintert}}, \
  and\ \bibinfo {author} {\bibfnamefont {A.}~\bibnamefont {Nazir}},\ }\href
  {\doibase 10.1103/PhysRevE.95.032139} {\bibfield  {journal} {\bibinfo
  {journal} {Phys. Rev. E}\ }\textbf {\bibinfo {volume} {95}},\ \bibinfo
  {pages} {032139} (\bibinfo {year} {2017})}\BibitemShut {NoStop}%
\bibitem [{\citenamefont {Strasberg}\ \emph {et~al.}(2018)\citenamefont
  {Strasberg}, \citenamefont {Schaller}, \citenamefont {Schmidt},\ and\
  \citenamefont {Esposito}}]{Strasberg2018}%
  \BibitemOpen
  \bibfield  {author} {\bibinfo {author} {\bibfnamefont {P.}~\bibnamefont
  {Strasberg}}, \bibinfo {author} {\bibfnamefont {G.}~\bibnamefont {Schaller}},
  \bibinfo {author} {\bibfnamefont {T.~L.}\ \bibnamefont {Schmidt}}, \ and\
  \bibinfo {author} {\bibfnamefont {M.}~\bibnamefont {Esposito}},\ }\href
  {\doibase 10.1103/PhysRevB.97.205405} {\bibfield  {journal} {\bibinfo
  {journal} {Phys. Rev. B}\ }\textbf {\bibinfo {volume} {97}},\ \bibinfo
  {pages} {205405} (\bibinfo {year} {2018})}\BibitemShut {NoStop}%
\bibitem [{\citenamefont {Tamascelli}\ \emph {et~al.}(2018)\citenamefont
  {Tamascelli}, \citenamefont {Smirne}, \citenamefont {Huelga},\ and\
  \citenamefont {Plenio}}]{Tamascelli2018}%
  \BibitemOpen
  \bibfield  {author} {\bibinfo {author} {\bibfnamefont {D.}~\bibnamefont
  {Tamascelli}}, \bibinfo {author} {\bibfnamefont {A.}~\bibnamefont {Smirne}},
  \bibinfo {author} {\bibfnamefont {S.~F.}\ \bibnamefont {Huelga}}, \ and\
  \bibinfo {author} {\bibfnamefont {M.~B.}\ \bibnamefont {Plenio}},\ }\href
  {\doibase 10.1103/PhysRevLett.120.030402} {\bibfield  {journal} {\bibinfo
  {journal} {Phys. Rev. Lett.}\ }\textbf {\bibinfo {volume} {120}},\ \bibinfo
  {pages} {030402} (\bibinfo {year} {2018})}\BibitemShut {NoStop}%
\bibitem [{\citenamefont {Brenes}\ \emph {et~al.}(2020)\citenamefont {Brenes},
  \citenamefont {Mendoza-Arenas}, \citenamefont {Purkayastha}, \citenamefont
  {Mitchison}, \citenamefont {Clark},\ and\ \citenamefont
  {Goold}}]{Brenes2020}%
  \BibitemOpen
  \bibfield  {author} {\bibinfo {author} {\bibfnamefont {M.}~\bibnamefont
  {Brenes}}, \bibinfo {author} {\bibfnamefont {J.~J.}\ \bibnamefont
  {Mendoza-Arenas}}, \bibinfo {author} {\bibfnamefont {A.}~\bibnamefont
  {Purkayastha}}, \bibinfo {author} {\bibfnamefont {M.~T.}\ \bibnamefont
  {Mitchison}}, \bibinfo {author} {\bibfnamefont {S.~R.}\ \bibnamefont
  {Clark}}, \ and\ \bibinfo {author} {\bibfnamefont {J.}~\bibnamefont
  {Goold}},\ }\href {\doibase 10.1103/PhysRevX.10.031040} {\bibfield  {journal}
  {\bibinfo  {journal} {Phys. Rev. X}\ }\textbf {\bibinfo {volume} {10}},\
  \bibinfo {pages} {031040} (\bibinfo {year} {2020})}\BibitemShut {NoStop}%
\end{thebibliography}%

\end{document}